\documentclass[aps, prd, reprint, nofootinbib, longbibliography]{revtex4-1}

\usepackage{graphics,graphicx,epsfig}
\usepackage{amsmath, amssymb, color}
\usepackage[dvipsnames]{xcolor}
\usepackage[colorlinks=true, linkcolor=blue, citecolor=magenta]{hyperref}
\usepackage{url}

\usepackage{float, placeins}
\usepackage{mwe}
\usepackage[caption=false]{subfig}
\usepackage{morefloats}
\usepackage{empheq}
\usepackage{bm}
\usepackage{booktabs} 
\usepackage{multirow}
\usepackage{makecell}

\newcommand\Mpl{M_{\textrm{pl}}}

\begin{document}

\title{Flows into inflation: An effective field theory approach}

\author{Feraz Azhar}
\email[Email address: {feraz\_azhar@fas.harvard.edu}]{}
\affiliation{Black Hole Initiative, Harvard University, Cambridge, MA 02138, USA}

\author{David I.~Kaiser}
\email[Email address: dikaiser@mit.edu]{}
\affiliation{Department of Physics, Massachusetts Institute of Technology, Cambridge, MA 02139, USA}

\date{\today}

\begin{abstract}
We analyze the flow into inflation for generic ``single-clock" systems, by combining an effective field theory approach with a dynamical-systems analysis. In this approach, we construct an expansion for the potential-like term in the effective action as a function of time, rather than specifying a particular functional dependence on a scalar field. We may then identify fixed points in the effective phase space for such systems, order-by-order, as various constraints are placed on the $M$th time derivative of the potential-like function. For relatively simple systems, we find significant probability for the background spacetime to flow into an inflationary state, and for inflation to persist for at least 60 efolds. Moreover, for systems that are compatible with single-scalar-field realizations, we find a single, universal functional form for the effective potential, $V (\phi)$, which is similar to the well-studied potential for power-law inflation. We discuss the compatibility of such dynamical systems with observational constraints.
\end{abstract}

\maketitle

\section{Introduction\label{SEC:Introduction}}

Early-universe inflation remains the leading explanation for several observable features of our universe today, such as its large-scale homogeneity and spatial flatness, as well as the specific pattern of primordial perturbations visible in the cosmic microwave background radiation. (For reviews, see Refs.~\cite{guthkaiser_05,bassett_05,liddle+lyth_09,martin+al_14b,guth+al_14,linde_15,baumann+mcallister_15}.) An important question has been whether the onset of inflation itself may be considered generic, or whether inflation requires fine-tuned initial conditions. (For reviews, see Refs.~\cite{goldwirth_92,brandenberger_16}.) Recent work, including numerical studies that implement full $(3+1)$-dimensional numerical relativity \cite{east+al_16,clough+al_16} and 
topological arguments \cite{kleban_16}, suggests that the onset of early-universe inflation may be rather generic, even amid inhomogeneous and anisotropic initial conditions. 

In the light of these recent results, it is of interest to explore generic characteristics of the onset (or otherwise) of inflation. Are there common features of the dynamical flow into inflation that one may identify, without needing to consider many distinct models, one at a time? 

In this paper, we combine recent work on effective field theory (EFT) approaches to inflation \cite{cheung+al_08,weinberg_08} (for a review, see Ref.~\cite{baumann+mcallister_15}) with a dynamical-systems analysis originally formulated to characterize late-universe acceleration \cite{frusciante+al_14}. Our goal is to develop tools with which to address the flow into inflationary states for general ``single-clock" descriptions of inflation --- a formulation that includes, but is not limited to, single-scalar-field (SSF) models of inflation. 

In order to develop the formalism we restrict attention here to background spacetimes that are (already) homogeneous, isotropic, and spatially flat, and aim to relax these assumptions in future work. We focus on the dynamical flow into inflationary states for initial conditions that are not expressly geared to trigger inflation, and develop heuristic measures over such initial conditions with which to estimate the probability that inflation will begin and persist for at least 60 efolds. 

Given our focus on the dynamics of single-clock systems, we construct an expansion for the potential-like term in the effective action as a function of {\it time}, rather than specifying a particular functional dependence on a scalar field, $\phi$. We may then study the dynamics of such systems, order by order, as various constraints are placed on the $M$th time derivative of that function. Our approach complements 
the techniques developed in Refs.~\cite{SalopekBond90,LiddleParsonsBarrow94,HoffmanTurner01,Kinney02,Liddle03,RamirezLiddle05,Chongchitnan05,remmen_13,remmen+carroll_14,VenninHorizonFlow14,VenninStochastic,MarshBarrow,VenninWandsUltraSlowRoll} to study attractor behavior for inflationary models, either by specifying a particular form for the scalar field's potential, $V (\phi)$, or by adopting the Hamilton-Jacobi formalism to study the evolution of the Hubble parameter as a function of the scalar field, $H (\phi)$.

Within the effective phase space for the systems we consider, we identify at most two hyperbolic inflationary fixed points at any order $M$ of the dynamical system. Each fixed point can be mapped onto one of two types of behavior: either pure de Sitter evolution of the background spacetime, or evolution in accord with a particular solution to power-law inflation. Moreover, by setting down heuristic probability distributions over initial conditions within the phase spaces for the two simplest orders of the dynamical system, we demonstrate that the probability for the system to flow into an inflationary state 
can be significant. In fact, when we consider initial conditions that are not compatible with SSF realizations, the probability of flowing into inflation can be enhanced. Lastly, for trajectories through phase space that are compatible with SSF realizations, we find that each such trajectory is compatible with a single, universal functional form for the effective potential, $V (\phi)$, a form which is similar to the well-studied potential for power-law inflation \cite{abbott+wise_84, lucchin_85,SalopekBond90,mukhanov_13,geng_15,geng_17}.

In Section \ref{SEC:DynamicalEquations}, we introduce the effective action and equations of motion for the relevant degrees of freedom, and introduce the variables in terms of which we parameterize the dynamical system for any order $M$. In Section \ref{SEC:DynamicalTrajectories} we consider dynamical trajectories for the zeroth- and first-order systems, identify fixed points in the effective phase space, describe various types of flows for the dynamical system, and estimate the probability that inflation will begin and persist for at least $60$ efolds.
In Section \ref{SEC:GenPotential} we demonstrate that trajectories through the phase space for zeroth- and first-order systems 
that correspond to an SSF realization may each be fit with a single functional form for the effective potential, $V (\phi)$, and in Section \ref{SEC:observables} we discuss how this inferred form for $V (\phi)$ may be constrained by recent observations. Concluding remarks follow in Section \ref{SEC:Discussion}. We explore aspects of the second-order system in Appendix~\ref{APP:secondorder}, and consider aspects of the $M$th order system (for $M \geq 3$) in Appendix~\ref{APP:Mthorder}.

\section{Dynamical equations for the background\label{SEC:DynamicalEquations}}

In this section we introduce the relevant degrees of freedom and dynamical equations that govern the system of interest. We build upon the effective field theory (EFT) of inflation pioneered in Refs.~\cite{cheung+al_08,weinberg_08}, combined with complementary EFT techniques from Ref.~\cite{frusciante+al_14}, which were originally designed to address late-universe acceleration. We work in units with $\hbar=c= 1$, so that the reduced Planck mass may be written $\Mpl=(8\pi G)^{-1/2}\simeq 2.4\times 10^{18}\,\textrm{GeV}$. We restrict attention to four spacetime dimensions and adopt the metric signature $(-,+,+,+)$. Lower-case Greek letters $\alpha,\beta,\dots = 0, 1, 2, 3$ label spacetime indices.

\subsection{The effective action}
\label{SEC:Ch2EFT}

Inflation may be described as a period of accelerated expansion of space, during which the universe evolves in a quasi-de Sitter state. The inflationary phase cannot be an exact de Sitter state, because the accelerated expansion must end. Hence the time-translation invariance of the action describing the relevant degrees of freedom during inflation must be broken: the action should be symmetric under time-dependent, 3-dimensional spatial diffeomorphisms, rather than under 4-dimensional spacetime diffeomorphisms. In other words, there must exist a clock that counts down the time until inflation ends. Though it is typical to model early-universe inflation in terms of the dynamics of one or more scalar fields, the clock need not correspond to a scalar field \cite{cheung+al_08}. 

The selection of a gauge effects a $(3+1)$-dimensional decomposition of the underlying spacetime, foliating it with $3$-dimensional spatial hypersurfaces. (See, e.g., Ref.~\cite{liddle+lyth_09}.) The quasi-de Sitter background of inflation has a privileged spatial slicing, determined by the symmetries of the (physical) clock. One may select a slicing (or gauge) in which fluctuations in the clock at different spatial locations vanish (to first order), leaving only perturbations in the spacetime metric. This choice of time slicing is known as ``unitary gauge" \cite{cheung+al_08}. (We will see below how to implement unitary gauge for the familiar case of an inflationary model involving a single scalar field.) 

Following Ref.~\cite{cheung+al_08}, we adopt unitary gauge and consider the most general effective action that respects time-dependent spatial diffeomorphisms, expanding around a spatially flat Friedmann-Lema\^{i}tre-Robertson-Walker (FLRW) metric. The action may then be written
\begin{equation}\label{EQN:ActionFO}
S = S_0 + \Delta S,
\end{equation}
where
\begin{equation}\label{EQN:S0}
S_0 =\int d^{4}x\sqrt{-g}\left[\frac{\Mpl^{2}}{2}R-L(t)-c(t) g^{00}\right] .
\end{equation}
Here $R$ is the spacetime Ricci scalar, $g^{00}$ is the `time-time' component of the (inverse) metric tensor, and $L(t)$ and $c(t)$ are (as yet unspecified) functions of time. The term $\Delta S$ includes terms that are quadratic (and higher) in the fluctuations of the metric, 
as well as terms that contain higher-order derivatives of the metric (which we assume are suppressed in the low-energy effective theory). Because we are interested in the dynamics of the background spacetime, we will focus on $S_0$ in the remainder of our analysis.

Varying $S_0$ with respect to $g^{\mu\nu}$ yields the Friedmann equations,
\begin{align}
H^{2}&=\frac{1}{3 \Mpl^{2}}\left[c(t)+L(t)\right], \label{EQN:EE1}\\
\dot{H}+H^{2}&=-\frac{1}{3 \Mpl^{2}}\left[2 c(t)-L(t)\right], \label{EQN:EE2}
\end{align}
where $H \equiv \dot{a} / a$ is the Hubble parameter, and overdots denote derivatives with respect to time. Solving Eqs.~(\ref{EQN:EE1}) and (\ref{EQN:EE2}) for $c(t)$ and $L(t)$ yields 
\begin{align}
c(t)&=-\Mpl^{2}\dot{H},\label{EQN:cSolve}\\
L(t)&=\Mpl^{2}\left(3H^2+\dot{H}\right)\label{EQN:LSolve}.
\end{align}
We may then substitute Eqs.~(\ref{EQN:cSolve}) and (\ref{EQN:LSolve}) back into Eq.~(\ref{EQN:S0}) to find 
\begin{equation}\label{EQN:ActionFOfinal}
S_0 =  \int d^{4}x\sqrt{-g}\, \Mpl^2 \left[\frac{R}{2}-3H^2+\dot{H}\left( g^{00} - 1\right) \right] .
\end{equation}
The term $S_0$ is known as the ``universal" part of the action, since this contribution to $S$ is fixed by the history of the background. (See, e.g., Appendix B of Ref.~\cite{baumann+mcallister_15}.)

The relationships in Eqs.~(\ref{EQN:EE1}) and (\ref{EQN:EE2}) enable us to identify the energy density $\rho = c + L$ and pressure $p = c - L$ for the matter degrees of freedom filling the FLRW spacetime.
Then we may specify the various point-wise energy conditions \cite{hawking+ellis_73,wald_84,visser_97} in terms of $c(t)$ and $L(t)$. The null energy condition (NEC) may be written
\begin{equation}\label{EQN:nullenergy}
{\rm NEC}:  \rho + p \geq 0 \rightarrow c (t) \geq 0 .
\end{equation}
The weak energy condition (WEC) becomes
\begin{align}\label{EQN:weakenergy}
 {\rm WEC}:\>\> &\rho \geq 0 \>\> {\rm and} \>\> \rho + p \geq 0 \\
\nonumber &\rightarrow c (t) + L (t) \geq 0 \>\> {\rm and} \>\> c (t) \geq 0 .
\end{align}
The dominant energy condition (DEC) may be written
\begin{align}\label{EQN:dominantenergy}
 {\rm DEC}:\>\> &\rho \geq 0 \>\> {\rm and} \>\> \rho \pm p \geq 0 \\
\nonumber &\rightarrow c (t) + L (t) \geq 0 \>, \> c (t) \geq 0 \> , \> {\rm and} \>\> L (t) \geq 0 .
\end{align}
And the strong energy condition (SEC) takes the form
\begin{align}\label{EQN:strongenergy}
 {\rm SEC}:\>\> &\rho + 3 p \geq 0 \>\> {\rm and} \>\> \rho + p \geq 0 \\
\nonumber &\rightarrow 2c (t) - L(t) \geq 0 \>\> {\rm and} \>\> c (t) \geq 0 .
\end{align}
As usual, we expect the strong energy condition to be violated during an inflationary phase. Moreover, it is possible that an effective field theory may violate the other (point-wise) energy conditions without yielding unphysical instabilities \cite{creminelli_06}; in such cases, appropriately averaged versions of the energy conditions may still be satisfied.

In single-scalar-field (SSF) models of inflation, the evolution of the scalar field $\phi$ plays the role of the physical clock. As usual, we may decompose the scalar field as
\begin{equation}\label{EQN:scalarPert}
\phi(x^\mu)\equiv \phi_{0}(t) + \delta\phi(x^\mu),
\end{equation}
where $\vert \delta\phi(x^\mu) \vert$ is considered to be small compared to $\phi_{0}(t)$. The field fluctuations $\delta \phi$ are gauge dependent. Hence we may choose a spatial slicing such that the scalar field is homogeneous across space but evolves over time, $\phi (x^\mu) \rightarrow \phi_0 (t)$ with $\delta \phi (x^\mu) = 0$, leaving only perturbations in the spacetime metric. In particular, if we perform a shift of the time coordinate,
\begin{equation}\label{EQN:timeDIFF}
t\to t+\xi^{0}(x^\mu), 
\end{equation}
where $\vert \xi^{0}(x^\mu) \vert$ is also considered to be small, then the field fluctuation transforms as 
\begin{equation}
\delta\phi(x^\mu) \to \delta\phi(x^\mu) + \dot{\phi}_{0}(t)\, \xi^{0}(x^\mu),
\end{equation}
to first order in $\xi^0$. (See, e.g., Refs.~\cite{liddle+lyth_09,cheung+al_08}.) We may choose 
\begin{equation}
\xi^{0}(x^\mu)\equiv -\frac{\delta\phi(x^\mu)}{\dot{\phi}_{0}(t)},
\end{equation}
so that
\begin{equation}
\phi(x^\mu) \to \phi_{0}(t),
\end{equation}
thereby implementing unitary gauge. In this way, the fluctuations of the scalar field have been gauged away, and a new time coordinate has been defined to track the value of the field $\phi_{0}(t)$ \cite{cheung+al_08}.

For an SSF model of inflation involving a minimally coupled scalar field subject to a potential $V (\phi)$, we may write the action as
\begin{equation}\label{EQN:ActionSFI}
S = \int d^{4}x\sqrt{-g}\left[\frac{\Mpl^{2}}{2}R-\frac{1}{2}g^{\mu\nu}\partial_{\mu}\phi\partial_{\nu}\phi -V(\phi)\right].
\end{equation}
In unitary gauge, $\phi\to \phi_{0}(t)$, so Eq.~(\ref{EQN:ActionSFI}) becomes
\begin{equation}\label{EQN:ActionSFIUG}
S = \int d^{4}x\sqrt{-g}\left[\frac{\Mpl^{2}}{2}R-\frac{1}{2}g^{00}\dot{\phi}_{0}^2(t)-V(\phi_{0}(t))\right].
\end{equation}
Eq.~(\ref{EQN:ActionSFIUG}) has the same form as Eq.~(\ref{EQN:S0}), and hence for an SSF model in unitary gauge we may identify
\begin{equation}\label{EQN:VLphidotc}
V (\phi_0 (t)) \leftrightarrow L (t) \> , \>\> \frac{1}{2} \dot{\phi}_0^2 (t) \leftrightarrow c (t) ,
\end{equation}
and similarly recognize Eqs.~(\ref{EQN:EE1}) and (\ref{EQN:EE2}) as the usual background-order relations
\begin{align}
\frac{1}{2} \dot{\phi}_0^2 (t) &= - \Mpl^2 \dot{H} , \label{EQN:Friedmann1} \\
V (\phi_0 (t)) &= \Mpl^2 \left( 3 H^2 + \dot{H} \right) \label{EQN:Friedmann2} .
\end{align}
 Substituting these relations into Eq.~(\ref{EQN:ActionSFIUG}) yields the expression for $S_0$ in Eq.~(\ref{EQN:ActionFOfinal}). Thus the usual action for an SSF model with a minimally coupled scalar field, in unitary gauge, corresponds to the universal part of the action displayed in Eq.~(\ref{EQN:ActionFOfinal}), though the action in Eq.~(\ref{EQN:ActionFOfinal}) is not limited to the case of an SSF model \cite{cheung+al_08}.

\subsection{Dynamical equations of motion}\label{SEC:DynEqMotion}

If we neglect the higher-order terms contained in $\Delta S$, then the dynamics of a system described by the action in Eq.~(\ref{EQN:S0}) depends on only two functions of time, $c(t)$ and $L(t)$. To study the dynamics of this system, one could solve for the evolution of $a(t)$ and thereby derive the behavior of $c(t)$ and $L(t)$. Or, adopting a dynamical-systems point of view, one may leave $c(t)$ and $L(t)$ free and study what forms of these functions yield viable expansion histories $a(t)$. Following Ref.~\cite{frusciante+al_14}, we adopt the latter approach. (See also, e.g., Ref.~\cite{Odintsov}.)

First we note that we may take a time derivative of Eq.~(\ref{EQN:EE1}) and use Eq.~(\ref{EQN:cSolve}) to find an analogue of the continuity equation:
\begin{equation}\label{EQN:cdot}
\dot{c}(t)+\dot{L}(t) = -6Hc(t).
\end{equation}
Naturally Eq.~(\ref{EQN:cdot}) is not independent of Eqs.~(\ref{EQN:EE1}) and (\ref{EQN:EE2}), though it is convenient to consider all three of these equations. Then, following Ref.~\cite{frusciante+al_14}, we may define the dimensionless variables (suppressing the explicit time dependences for now):
\begin{align}
x&\equiv \frac{c}{3 \Mpl^{2} H^2},\label{EQN:xF}\\
y&\equiv \frac{L}{3 \Mpl^{2} H^2},\label{EQN:yF}\\
\lambda_{m}&\equiv-\frac{L^{(m+1)}}{H L^{(m)}},\label{EQN:lF}
\end{align}
for $m=0,1,2,\dots$. In Eq.~(\ref{EQN:lF}), $(m)$ represents the $m$th derivative with respect to time. Eq.~(\ref{EQN:lF}) introduces an infinite tower of dimensionless variables that encode implicit choices for the functional form of $L (t)$, though in practice we will only consider a finite number of these terms for a given phase-space analysis.  

Our next task is to derive a set of coupled, ordinary differential equations (ODEs) with which we may construct a dynamical-systems analysis. Making use of Eq.~(\ref{EQN:cdot}) as well as the definitions in Eqs.~(\ref{EQN:xF})--(\ref{EQN:lF}), we find
\begin{align}
\frac{dx}{d\ln a} &=\lambda_{0} y -6 x -2 x \frac{\dot{H}}{H^2},\label{EQN:xFtime}\\
\frac{dy}{d\ln a} &=\left(-\lambda_{0} -2  \frac{\dot{H}}{H^2}\right)y,\label{EQN:yFtime}\\
\frac{d\lambda_{m-1}}{d\ln a} &= \left(-\lambda_{m}+\lambda_{m-1}- \frac{\dot{H}}{H^2}\right)\lambda_{m-1},\label{EQN:lFtime}
\end{align}
for $m \geq 1$. Similarly, the first Friedmann equation, Eq.~(\ref{EQN:EE1}), is equivalent to the constraint
\begin{equation}
x+y=1\label{EQN:Constraint}.
\end{equation}
Furthermore, we may use Eqs.~(\ref{EQN:cSolve}) and (\ref{EQN:LSolve}) to find an expression for the slow-roll parameter, $\epsilon$:
\begin{equation}
\epsilon\equiv-\frac{\dot{H}}{H^2} = \frac{3}{2}(1+x-y) = 3x ,\label{EQN:Ep}
\end{equation}
where the final expression comes from applying the constraint of Eq.~(\ref{EQN:Constraint}). Eqs.~(\ref{EQN:xFtime})--(\ref{EQN:Constraint}) are derived from Eqs.~(\ref{EQN:EE1}), (\ref{EQN:EE2}), and (\ref{EQN:cdot}).

We only consider scenarios in which $H(t)$ is real and hence $H^2 \geq 0$. Therefore the null energy condition (NEC) in Eq.~(\ref{EQN:nullenergy}) corresponds to $x \geq 0$. Since we always impose the constraint of Eq.~(\ref{EQN:Constraint}), any trajectory through the effective phase space that satisfies $x \geq 0$ will also satisfy the weak energy condition (WEC) of Eq.~(\ref{EQN:weakenergy}). On the other hand, only trajectories for which $x \geq 0$ {\it and} $y \geq 0$ will satisfy the dominant energy condition (DEC) of Eq.~(\ref{EQN:dominantenergy}).

Eqs.~(\ref{EQN:xFtime})--(\ref{EQN:Constraint}) do not form a closed system, because of the infinite tower in Eq.~(\ref{EQN:lFtime}), but we can make them so by fixing $\lambda_{M}$ to be a constant for some $M\geq 0$. We will refer to the dynamical system so obtained as the \emph{Mth-order system}. In this case, the dynamics are controlled by a constrained system of $2+M$ ODEs (for $x$, $y$, and $\lambda_{m}$, where $m=0,1,\dots,M-1$), where the constraint is given by Eq.~(\ref{EQN:Constraint}). (Solving this closed and constrained system allows one to determine all $\lambda_m$ for $m > M$.) Thus one may study the dynamics of such a system order-by-order in $M$.

Upon setting $\lambda_M =\textrm{constant}$ for some $M \geq 0$, Eqs.~(\ref{EQN:xFtime})--(\ref{EQN:Constraint}) take the form:
\begin{subequations}\label{EQN:MthorderLISTa}
\begin{align}
\frac{dx}{d\ln a} &=\lambda_{0} y -3 x +3 x^2-3 x y\label{EQN:xFtimeM},\\
\frac{dy}{d\ln a} &=-\lambda_{0} y +3 y +3 x y-3 y^2\label{EQN:yFtimeM},\\
\frac{d\lambda_{0}}{d\ln a} &= \left[-\lambda_{1}+\lambda_{0}+\frac{3}{2}(1+x-y)\right]\lambda_{0}\label{EQN:lzerotimeM},\\
\frac{d\lambda_{1}}{d\ln a} &= \left[-\lambda_{2}+\lambda_{1}+\frac{3}{2}(1+x-y)\right]\lambda_{1}\label{EQN:lonetimeM},\\
&\vdots\nonumber\\
\frac{d\lambda_{M-1}}{d\ln a} &= \left[-\lambda_{M}+\lambda_{M-1}+\frac{3}{2}(1+x-y)\right]\lambda_{M-1}\label{EQN:lMminusonetimeM},\\
1&=x+y, \label{EQN:ConstraintM}
\end{align}
\end{subequations}
and the slow-roll parameter $\epsilon$ is given by Eq.~(\ref{EQN:Ep}). Hence one may determine whether the system is in an inflationary state simply by monitoring the value of $x$. In particular,
\begin{equation}\label{EQN:InflationEpCond}
\emph{inflation occurs \emph{if and only if}}\; x < \frac{1}{3} ,
\end{equation}
which corresponds to $\epsilon < 1$ and hence $\ddot{a} > 0$.

The effective phase space of this dynamical system is $(1+M)$-dimensional, stemming from the $(2+M)$ ODEs in Eqs.~(\ref{EQN:xFtimeM})--(\ref{EQN:lMminusonetimeM}), subject to the constraint of Eq.~(\ref{EQN:ConstraintM}).\footnote{Note that the structure of the equations described in Eq.~(\ref{EQN:MthorderLISTa}) is somewhat simpler than it first appears, because there exist invariant manifolds at any order. For example, it is straightforward to show that for the $M$th-order system, $d (x + y - 1) / d \ln a = 3 (x - y)(x + y - 1)$. Thus one does not move off the constraint surface if one begins on it. By the same reasoning, the surfaces $y=0$, and $\lambda_{m}=0$ for $m=0,1,\dots M-1$, are also invariant manifolds.} 
The phase space is naturally described in terms of the coordinates $(x,y,\lambda_{0}, \dots, \lambda_{M-1})$.

Two (related) time coordinates prove to be especially convenient: cosmic time, $t$, and 
\begin{equation}
\tau \equiv \ln a (t) .
\label{EQN:taudef}
\end{equation}
One may study Eqs.~(\ref{EQN:xFtimeM})--(\ref{EQN:lMminusonetimeM}) in terms of $t$ rather than $\tau$, but then an explicit factor of the Hubble parameter $H$ will appear in each equation (since $d \ln a / dt = H$), and one must then also use Eq.~(\ref{EQN:Ep}) when solving the coupled system of equations.

Given the definition of $\lambda_m$ in Eq.~(\ref{EQN:lF}), we see that fixing $\lambda_M$ to be a constant for some $M \geq 0$ yields
\begin{equation}\label{EQN:clamM}
L^{(M)}(t) = L^{(M)}(t_{i})\left[\frac{a(t_{i})}{a(t)}\right]^{\lambda_{M}}, 
\end{equation}
where $t_{i}$ is some fixed initial time. Thus, setting $\lambda_{M}$ to be a constant corresponds to assuming that the $M$th time-derivative of  $L(t)$ scales as a power law in the scale factor $a(t)$, with power $-\lambda_{M}$. Moreover, following Ref.~\cite{frusciante+al_14}, we note that possessing an expression for the $M$th time-derivative of $L(t)$ allows us to expand $L(t)$ as a Taylor series about $t_{i}$:
\begin{widetext}
\begin{align}
L(t) &= \sum_{m=0}^{M-1}\frac{L^{(m)}(t_{i})}{m!}(t-t_{i})^{m}+\int_{t_{i}}^{t}d\tilde{t}\, \frac{(t-\tilde{t})^{M-1}}{(M-1)!}L^{(M)}(\tilde{t})\nonumber\\
&=\sum_{m=0}^{M-1}\frac{L^{(m)}(t_{i})}{m!}(t-t_{i})^{m}+L^{(M)}(t_{i})\int_{t_{i}}^{t}d\tilde{t} \, \frac{(t-\tilde{t})^{M-1}}{(M-1)!}\left[\frac{a(t_{i})}{a(\tilde{t} )}\right]^{\lambda_{M}}\label{EQN:Taylor}.
\end{align}
\end{widetext}
As such, $\lambda_{M}$ parameterizes the remainder term in the Taylor expansion. Thus, a higher-order system allows for more terms in the Taylor expansion in Eq.~(\ref{EQN:Taylor}). The significance of Eqs.~(\ref{EQN:clamM}) and (\ref{EQN:Taylor}) becomes more clear when we map the $M$th-order system onto SSF realizations.

For an SSF model involving a minimally coupled scalar field in unitary gauge, we may combine Eq.~(\ref{EQN:VLphidotc}) with Eqs.~(\ref{EQN:xF})--(\ref{EQN:lF}) to write
\begin{align}
x&=\frac{\dot{\phi}^{2}}{6 \Mpl^{2}H^2},\label{EQN:xSF}\\
y&=\frac{V(\phi)}{3 \Mpl^{2} H^2},\label{EQN:ySF}\\
\lambda_{m}&=-\frac{[V(\phi)]^{(m+1)}}{H [V(\phi)]^{(m)}},\label{EQN:lSF}
\end{align}
for $m=0,1,\dots M-1$. The mapping onto (standard) SSF models of inflation thus restricts $x\geq 0$ and $y\geq 0$. There is no analogous constraint on the $\lambda_{m}$'s.

Given that $x$ and $y$ must satisfy the constraint of Eq.~(\ref{EQN:ConstraintM}), we see that for an SSF realization, $x$ represents the fractional kinetic-energy density of the field and $y$ represents the fractional potential-energy density of the field. Furthermore, Eq.~(\ref{EQN:clamM}) in combination with the identification in Eq.~(\ref{EQN:VLphidotc}) yields:
\begin{equation}\label{EQN:VM}
V^{(M)}(t) = V^{(M)}(t_{i})\left[\frac{a(t_{i})}{a(t)}\right]^{\lambda_{M}}.
\end{equation}
For SSF realizations, in other words, fixing the order of the dynamical system (by setting $\lambda_M =$ constant for some $M \geq 0$) means that dynamical trajectories for the $M$th-order system correspond to scenarios in which the $M$th time-derivative of the potential-energy density, $V^{(M)}(t)$, scales as $[a(t)]^{-\lambda_M}$.

\section{Dynamical trajectories\label{SEC:DynamicalTrajectories}}

Using the effective action and the parameterized equations of motion from Sec.~\ref{SEC:DynamicalEquations}, in this section we identify important features of the resulting phase space for zeroth- and first-order systems, and compute the probabilities that such systems will flow into inflation. (We discuss second-order systems in Appendix \ref{APP:secondorder}, and identify interesting features of $M$th-order systems, with $M \geq 3$ in Appendix \ref{APP:Mthorder}.) For both the zeroth- and first-order systems, we first identify relevant fixed points and then consider representative trajectories for the system through the phase space, before considering measures for the flow into inflation.

\subsection{Zeroth-order system}

The zeroth-order system is the simplest dynamical system, and arises when we set $\lambda_{0}=\textrm{constant}$. Under this assumption, the equations governing the dynamics, Eqs.~(\ref{EQN:xFtimeM})--(\ref{EQN:ConstraintM}), simplify to an effectively one-dimensional system (in a phase space coordinatized by $x$ and $y$):
\begin{subequations}
\begin{align}
\frac{dx}{d\ln a} &=\lambda_{0} y -3 x +3 x^{2}-3 x y,\label{EQN:xFtimeZERO}\\
\frac{dy}{d\ln a} &=-\lambda_{0} y +3 y +3 x y-3 y^2,\label{EQN:yFtimeZERO}\\
1&=x+y, \label{EQN:ConstraintZERO}
\end{align}
\end{subequations}
with the 
slow-roll parameter $\epsilon$ given by Eq.~(\ref{EQN:Ep}).

The fixed points for the zeroth-order system are simply found by setting the right-hand sides of Eqs.~(\ref{EQN:xFtimeZERO}) and (\ref{EQN:yFtimeZERO}) to zero, subject to the constraint of Eq.~(\ref{EQN:ConstraintZERO}). Since we focus only on hyperbolic fixed points, their stability properties can be established by analyzing the eigenvalues of the Jacobian matrix evaluated at each fixed point. (See, e.g.,  Sec.~3.1 of Ref.~\cite{frusciante+al_14}.) One finds that there are at most two hyperbolic fixed points for the system, whose stability properties depend on the value of $\lambda_{0}$. We summarize these findings in Table~\ref{TAB:ZO}. Although any dynamical trajectory that begins on the constraint surface $x + y = 1$ will remain there, we analyze stability properties for fixed points considering the entire effective phase space, rather than limiting attention only to the constraint surface.
%
\begin{table*}
\begin{center}
\begin{tabular}{ c|c|c|l }
  \emph{Fixed point} & \emph{Inflationary?}  & \emph{Eigenvalues} & \emph{Stability properties} \\
 $(x, y)$ &$(\epsilon= 3x<1?)$ &[\emph{Hyperbolic iff}]&\\
\Xhline{1.5pt}    
   {\bf FP0a} 	& No	& $\{3, 6-\lambda_{0}\}$ 	& $\lambda_{0}<6$: Unstable \\
  $(1,0)$		&			& $[\lambda_{0}\neq 6]$	& $\lambda_{0}>6$: Saddle point\\
  \hline
   {\bf FP0b}		& Yes  $(\lambda_{0} < 2)$ & $\{-6+\lambda_{0}, -3+\lambda_{0}\}$ & $\lambda_{0}<3$: Attractor \\
    $\left(\frac{\lambda_{0}}{6}, 1-\frac{\lambda_{0}}{6}\right)$	& & $[\lambda_{0}\neq 3,6]$ & $3<\lambda_{0}<6$: Saddle point\\
    									     	& & & $\lambda_{0}>6$: Unstable\\
										\end{tabular}
\caption{Fixed points for the zeroth-order system and their stability properties. There are at most two hyperbolic fixed points for $M = 0$. Fixed point {\bf FP0a} does not correspond to an inflationary state, whereas {\bf FP0b} is inflationary if and only if $\lambda_{0}<2$, in which case it is an inflationary attractor.}\label{TAB:ZO}									
\end{center}
\end{table*}

The fixed points for the zeroth-order system display these features:
\begin{itemize}
\item[(i)] There is (at most) one inflationary fixed point, {\bf FP0b}, which is inflationary (with $x< 1/ 3$ and hence $\epsilon < 1$) if and only if $\lambda_{0}<2$. If we choose $0 < \lambda_{0} < 2$, {\bf FP0b} has, via Eq.~(\ref{EQN:Ep}), a Hubble parameter whose time derivative is nonzero---thus a trajectory that starts out (and indeed remains) at this fixed point describes the time evolution of a quasi-de Sitter background. If $\lambda_{0}=0$, this fixed point corresponds to a pure de Sitter background.
\item[(ii)] For {\bf FP0b}, $\lambda_{0}<0$ corresponds to $\dot{H} > 0$. We will therefore exclude cases with $\lambda_{0}<0$ from our analysis, since under these conditions a background spacetime that is initially expanding will develop a singularity in the scale factor in a finite time, akin to ``big rip" scenarios \cite{starobinsky_00, caldwell_02, caldwell+al_03}.
\end{itemize}

\begin{figure*}
\begin{minipage}{.49\linewidth}
\centering
\subfloat[]{\includegraphics[scale=0.5]{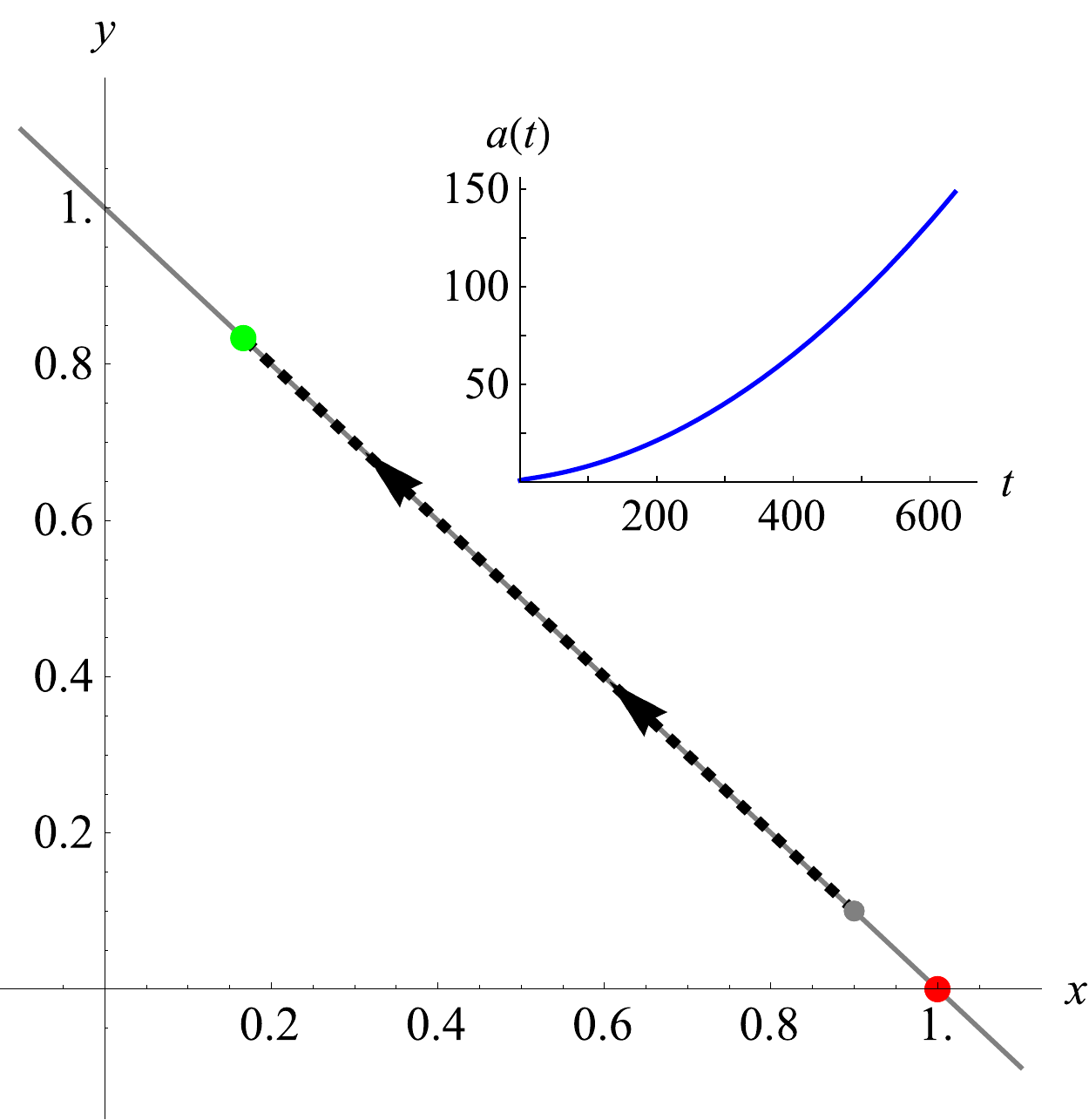}}
\end{minipage}
\begin{minipage}{.49\linewidth}
\centering
\subfloat[]{\includegraphics[scale=0.5]{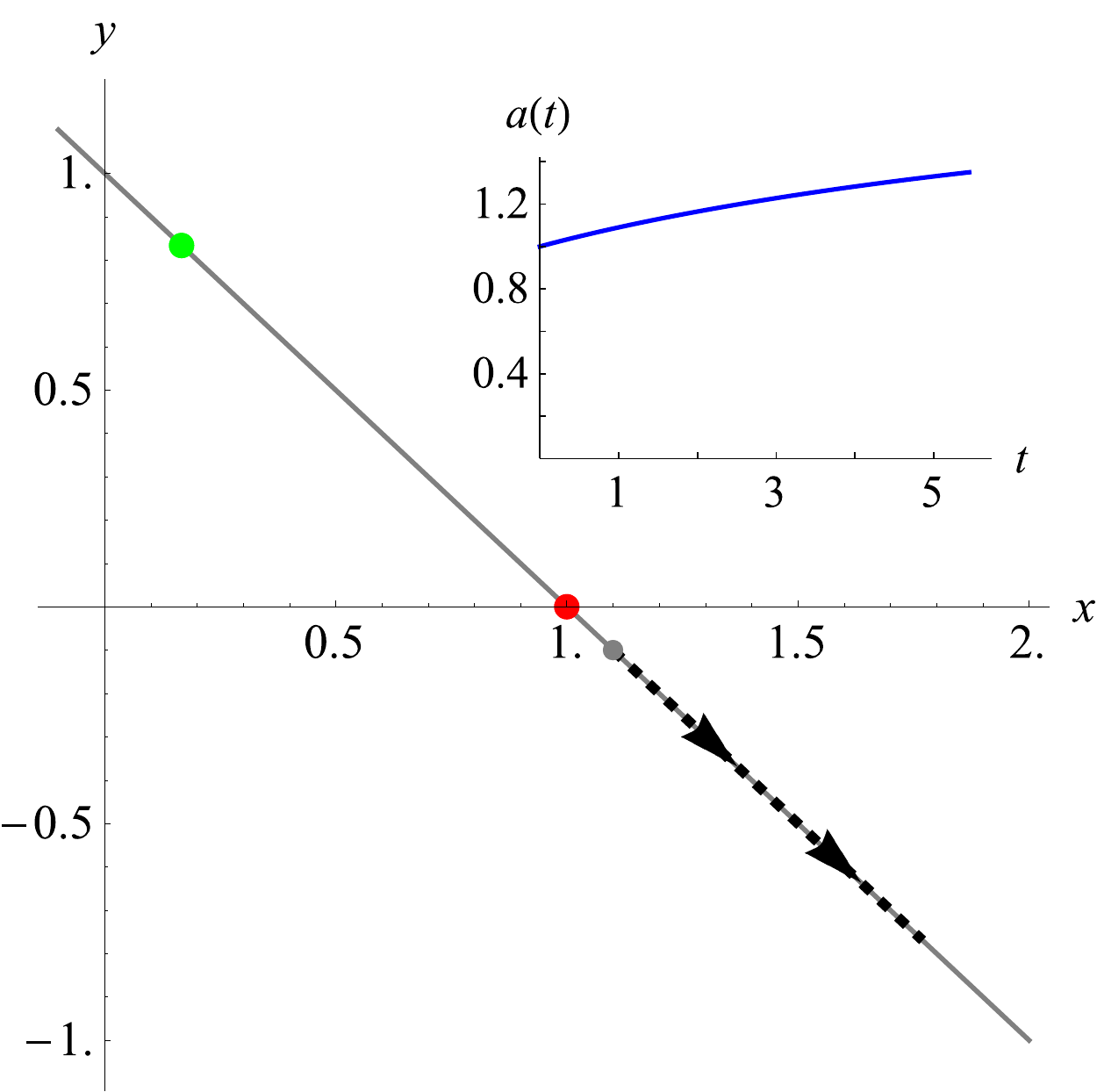}}
\end{minipage}
\caption{Phase-space plots of trajectories for the zeroth-order system with $\lambda_{0}=1$. In each plot, the green dot corresponds to the inflationary attractor (fixed point {\bf FP0b} in Table~\ref{TAB:ZO}), and the red dot corresponds to an unstable, non-inflationary fixed point ({\bf FP0a} in Table~\ref{TAB:ZO}). Each trajectory (dashed black line) begins at a gray dot and moves along the constraint surface, $x + y = 1$ (gray line), as a function of $\tau \equiv \ln a(t)$. The inset plots show the evolution of the scale factor $a(t)$ with cosmic time $t$, as the background follows its respective trajectory. Initial conditions were selected with $H (\tau_i) = 0.1$ (in units of $\Mpl$) and $a (t_i) = 1$, and with $(x (\tau_i), y (\tau_i) )$ equal to 
$(0.9,0.1)$ (case a), and $(1.1,-0.1)$ (case b). 
}
\label{FIG:TrajZero}
\end{figure*}

In Fig.~\ref{FIG:TrajZero}, we display some illustrative trajectories that arise from solving the equations of motion for the zeroth-order system, Eqs.~(\ref{EQN:xFtimeZERO})--(\ref{EQN:ConstraintZERO}), with $\lambda_{0}=1$. In this case, fixed point {\bf FP0b} in Table~\ref{TAB:ZO} is an inflationary attractor, marked by a green dot at $(x,y) = (\frac{1}{6},\frac{5}{6})$. The other fixed point, {\bf FP0a}, is a non-inflationary, unstable fixed point, marked by the red dot at $(x,y) = (1,0)$. For each trajectory, initial conditions at time $\tau_i$ are chosen such that the background spacetime is not initially inflating, with $x (\tau_i) \geq 1/3$; the starting point for each trajectory is denoted by a gray dot. 

Fig.~\ref{FIG:TrajZero}a shows the system evolving toward the inflationary attractor (fixed point {\bf FP0b}), whereas Fig.~\ref{FIG:TrajZero}b shows the system flowing away from the inflationary fixed point. Case (a) may be represented by an SSF model in which the field's kinetic energy initially dominates its potential energy. Case (b), on the other hand, begins with $L(\tau_i) < 0$, and hence violates the dominant energy condition (DEC) of Eq.~(\ref{EQN:dominantenergy}), although it satisfies both the null (NEC) and weak (WEC) energy conditions of Eqs.~(\ref{EQN:nullenergy}) and (\ref{EQN:weakenergy}) respectively. Given the identifications in Eq.~(\ref{EQN:VLphidotc}), we see that no trajectory with $L (\tau_i) < 0$ can be represented by an SSF model in which the field's potential energy is positive-definite, since $L \leftrightarrow V (\phi)$.

Next we may estimate the probability that a zeroth-order system will flow into an inflationary state that persists long enough to address the usual shortcomings of the standard big bang scenario, producing at least 60 efolds of inflation. We are particularly interested in situations like that shown in Fig.~\ref{FIG:TrajZero}a, in which there exists an inflationary fixed point toward which the system will flow, even for initial conditions dominated by kinetic (rather than potential) energy. It is straightforward to demonstrate that scenarios like Fig.~\ref{FIG:TrajZero}a generically produce sufficient inflation for zeroth-order systems.

Consider a vector field $\vec{v}_{0}$ along the constraint surface ($x + y = 1$) in the effective phase space. We may consider the conditions under which $\vec{v}_0$ will point toward the inflationary fixed point. Using the right-hand sides of Eqs.~(\ref{EQN:xFtimeZERO}) and (\ref{EQN:yFtimeZERO}), we may write:
\begin{align}\label{EQN:v0}
\vec{v}_{0} &\equiv (\lambda_{0} y -3 x +3 x^2-3 x y, -\lambda_{0} y +3 y +3 x y-3 y^2)\nonumber\\
&=(\mathcal{A}(x),-\mathcal{A}(x)),
\end{align}
where
\begin{equation}
\mathcal{A}(x)\equiv(\lambda_{0}-6x)(1-x),
\end{equation}
and the second line of Eq.~(\ref{EQN:v0}) follows upon using the constraint of Eq.~(\ref{EQN:ConstraintZERO}).

We may now consider various values of $\lambda_{0}$. Recall that the two fixed points at zeroth order occur at {\bf FP0a}: $(x,y)=(1,0)$ and {\bf FP0b}: $(x,y)= \left(\frac{\lambda_{0}}{6},1-\frac{\lambda_{0}}{6}\right)$, and that {\bf FP0b} is an attractor for $\lambda_{0}<3$ (see Table~\ref{TAB:ZO}). If we take $\lambda_0 < 2$ (as in Fig.~\ref{FIG:TrajZero}), then {\bf FP0b} is an inflationary attractor. For kinetic-energy-dominated initial conditions that are consistent with SSF realizations, 
the system starts with a value of $x$ that is greater than the $x$-value of {\bf FP0b} but less than the $x$-value of {\bf FP0a},
and will remain in that position relative to both fixed points throughout the ensuing evolution, i.e., $\frac{\lambda_{0}}{6}<x (\tau) <1$. In that case,
\begin{equation}
\mathcal{A}(x)\equiv\underbrace{(\lambda_{0}-6x)}_{<0}\underbrace{(1-x)}_{>0} <0.
\end{equation}
Thus the vector field $\vec{v}_{0} $ points along the constraint surface towards the attractor. \emph{Any} initially kinetic-energy-dominated trajectory that has an SSF realization (that is, with $x (\tau_i) \lesssim 1$) will flow into (and remain in) an inflationary state. 

On the other hand, if $x(\tau_{i})=1$, then $\mathcal{A}(x)=0$, so that $\vec{v}_{0}=(0,0)$, reflecting the fact that the system is positioned at {\bf FP0a} and will remain there for all time. Moreover, if $x(\tau_{i})>1$, then $\mathcal{A}(x)>0$, and the system gets driven away from {\bf FP0a}, deeper into the lower-right quadrant of the EFT phase space, as in Fig.~\ref{FIG:TrajZero}(d) --- though, as noted above, such an initial condition (which requires $L(\tau_i) < 0$) violates the dominant energy condition (DEC) of Eq.~(\ref{EQN:dominantenergy}) and is not consistent with SSF realizations. 

These results imply that for a zeroth-order system starting from kinetic-energy-dominated initial conditions that are consistent with SSF realizations, with $0 < \lambda_0 < 2$, the probability that the system will flow through sufficient inflation is unity. Any (normalized) probability distribution defined only over such kinetic-energy-dominated initial conditions, integrated over the subset of initial conditions that yield sufficient inflation, will yield unity.

The results in this subsection 
are easy to understand in terms of corresponding SSF models. For a zeroth-order system, the potential energy $V (\phi)$ will redshift as in Eq.~(\ref{EQN:VM}) with $M = 0$, and hence $V (t) \propto V (t_i) [ a(t)]^{-\lambda_0}$. Clearly, for any such system with $V (t_i) > 0$ and $0 < \lambda_0 < 2$, the potential energy will redshift more gradually than 
the kinetic energy of the field 
and will eventually dominate the system's dynamics. As we will see in Sec.~\ref{SEC:firstorder} and the Appendices, these relationships become considerably less trivial for systems with $M \geq 1$.

\subsection{First-order system}\label{SEC:firstorder}

To obtain the first-order system, we set $\lambda_{1}$ = constant. Under this assumption, the equations governing the dynamics, Eqs.~(\ref{EQN:xFtimeM})--(\ref{EQN:ConstraintM}), take the form
\begin{subequations}
\begin{align}
\frac{dx}{d\ln a} &=\lambda_{0} y -3 x +3 x^2-3 x y,\label{EQN:xFtimeONE}\\
\frac{dy}{d\ln a} &=-\lambda_{0} y +3 y +3 x y-3 y^2,\label{EQN:yFtimeONE}\\
\frac{d\lambda_{0}}{d\ln a} &= \left[-\lambda_{1}+\lambda_{0}+\frac{3}{2}(1+x-y)\right]\lambda_{0},\label{EQN:lzerotimeONE}\\
1&=x+y, \label{EQN:ConstraintONE}
\end{align}
\end{subequations}
and the 
slow-roll parameter is again given by $\epsilon = 3x$, as in Eq.~(\ref{EQN:Ep}). For the first-order system, our general path to computing probabilities for inflation will mirror that adopted for the zeroth-order system. Thus, we will first describe first-order hyperbolic fixed points, after which we exhibit a number of example trajectories in the corresponding first-order EFT phase space. Finally, we describe a way to make probabilistic statements about inflation at first order.

To find the fixed points of the system, we set the right-hand sides of Eqs.~(\ref{EQN:xFtimeONE})--(\ref{EQN:lzerotimeONE}) to zero, subject to the constraint of Eq.~(\ref{EQN:ConstraintONE}). One finds that there are at most four hyperbolic fixed points for the system, whose stability properties depend on the value of $\lambda_{1}$. The fixed points, together with some relevant properties, are given in Table~\ref{TAB:FO}. For the eigenvalues related to {\bf FP1d}, we define the constants
\begin{equation}
\alpha_{\pm}\equiv\frac{1}{3}\left[-9+2\lambda_{1}\pm\sqrt{81-2\lambda_{1}(-9+\lambda_{1})}\right] .
\label{EQN:alphapmdef}
\end{equation}
\begin{table*}
\begin{center}
\begin{tabular}{  c | c | c | l }
  \emph{Fixed point} & \emph{Inflationary?}  &\emph{Eigenvalues} & \emph{Stability properties} \\
  $(x, y, \lambda_{0})$ & $(\epsilon= 3x<1?)$&[\emph{Hyperbolic iff}]&\\
  \Xhline{2pt}
  {\bf FP1a} 														&  Yes							& 	$\{-6, -3,-\lambda_{1}\}$ 				& $\lambda_{1}<0$: {\small Saddle} \\$(0,1,0)$
  																	&			   								&		$[\lambda_{1}\neq 0]$							& $\lambda_{1}>0$: {\small Attractor}\\
  \hline
  {\bf FP1b}  														&  No 									&  	$\{6,3,3-\lambda_{1}\}$ 				& $\lambda_{1}<3$: {\small Unstable} \\$(1,0,0)$
    																	&											&		$[\lambda_{1}\neq 3]$			        				& $\lambda_{1}>3$: {\small Saddle}\\
  \hline
  {\bf FP1c}  											&  No 									&  	$\{3,9-\lambda_{1},-3+\lambda_{1}\}$ 	& $\lambda_{1}<3$: {\small Saddle} \\$(1,0,-3+\lambda_{1})$
    																	&											&					     $[\lambda_{1}\neq 3,9]$   				& $3<\lambda_{1}<9$: {\small Unstable}\\
																	&											&									& $\lambda_{1}>9$: {\small Saddle}\\
   \hline
    {\bf FP1d}  		&  Yes  $(\lambda_{1}<3)$ 			&  	$\{-3+\frac{2\lambda_{1}}{3},\alpha_{-},\alpha_{+}\}$ 			& $\lambda_{1}<\frac{9}{2}(1-\sqrt{3})$: {\small Stable focus} \\$(\frac{\lambda_{1}}{9},1-\frac{\lambda_{1}}{9},\frac{2\lambda_{1}}{3})$
    																	&											&			$[\lambda_{1}\neq 0,\frac{9}{2},9]$		        				& $\frac{9}{2}(1-\sqrt{3})\leq\lambda_{1}<0$: {\small Attractor}\\
																							&					&									& $0<\lambda_{1}<\frac{9}{2}$: {\small Saddle}\\
																							& & &  $\frac{9}{2}<\lambda_{1}<9$: {\small Saddle}\\
																							& & & $9<\lambda_{1}\leq\frac{9}{2}(1+\sqrt{3})$: {\small Unstable}\\
																							& & & $\lambda_{1}>\frac{9}{2}(1+\sqrt{3})$: {\small Unstable focus}\\ 
\end{tabular}
\caption{Fixed points for the first-order system and their stability properties. There are at most four hyperbolic fixed points for $M = 1$, two of which may yield inflationary states. The constants $\alpha_{\pm}$ that appear in the eigenvalues corresponding to {\bf FP1d} are defined in Eq.~(\ref{EQN:alphapmdef}). } 
\label{TAB:FO}
\end{center}
\end{table*}
These solutions have some interesting features:
\begin{itemize}
\item[(i)] There are at most two inflationary fixed points: {\bf FP1a} and {\bf FP1d}. {\bf FP1a} corresponds to a background that is (exactly) de Sitter; {\bf FP1d} has a Hubble parameter that varies with time, giving a quasi-de Sitter inflating background for $0<\lambda_{1}<3$. 
\item[(ii)] For {\bf FP1d}, $\lambda_{1}<0$ corresponds to
$\dot{H} > 0$. Analogously to the case discussed for fixed-point {\bf FP0b} for a zeroth-order system,
we will exclude from our analysis cases in which $\lambda_{1}<0$, as under these conditions a background that is initially expanding will develop a
singularity in the scale factor in a finite time, akin to ``big rip" scenarios \cite{starobinsky_00, caldwell_02, caldwell+al_03}.
\end{itemize}

\begin{figure*}
\begin{minipage}{.55\linewidth}
\centering
\subfloat[]{\includegraphics[scale=0.5]{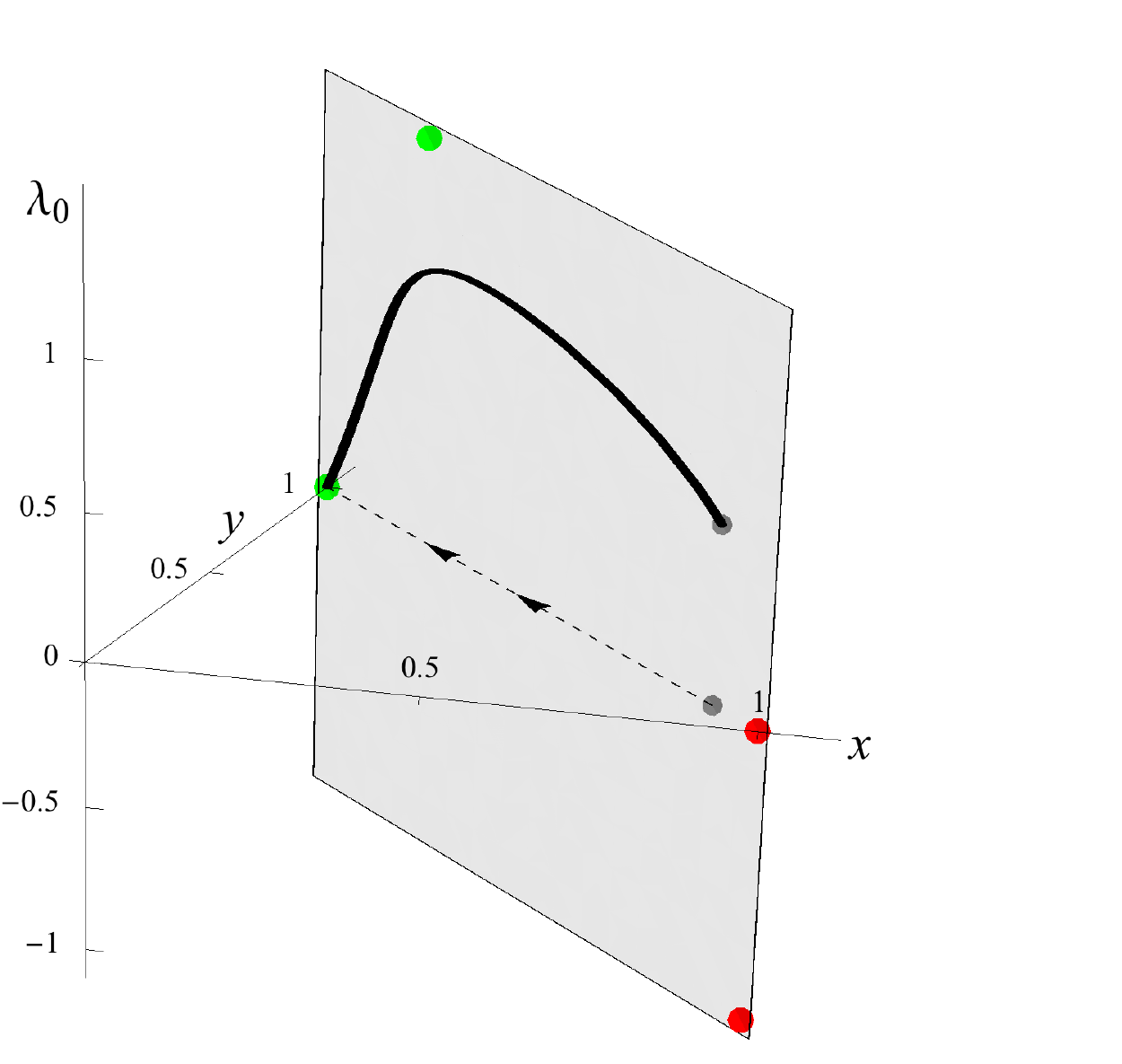}}
\end{minipage}
\begin{minipage}{.44\linewidth}
\centering
\subfloat[]{\includegraphics[scale=0.4]{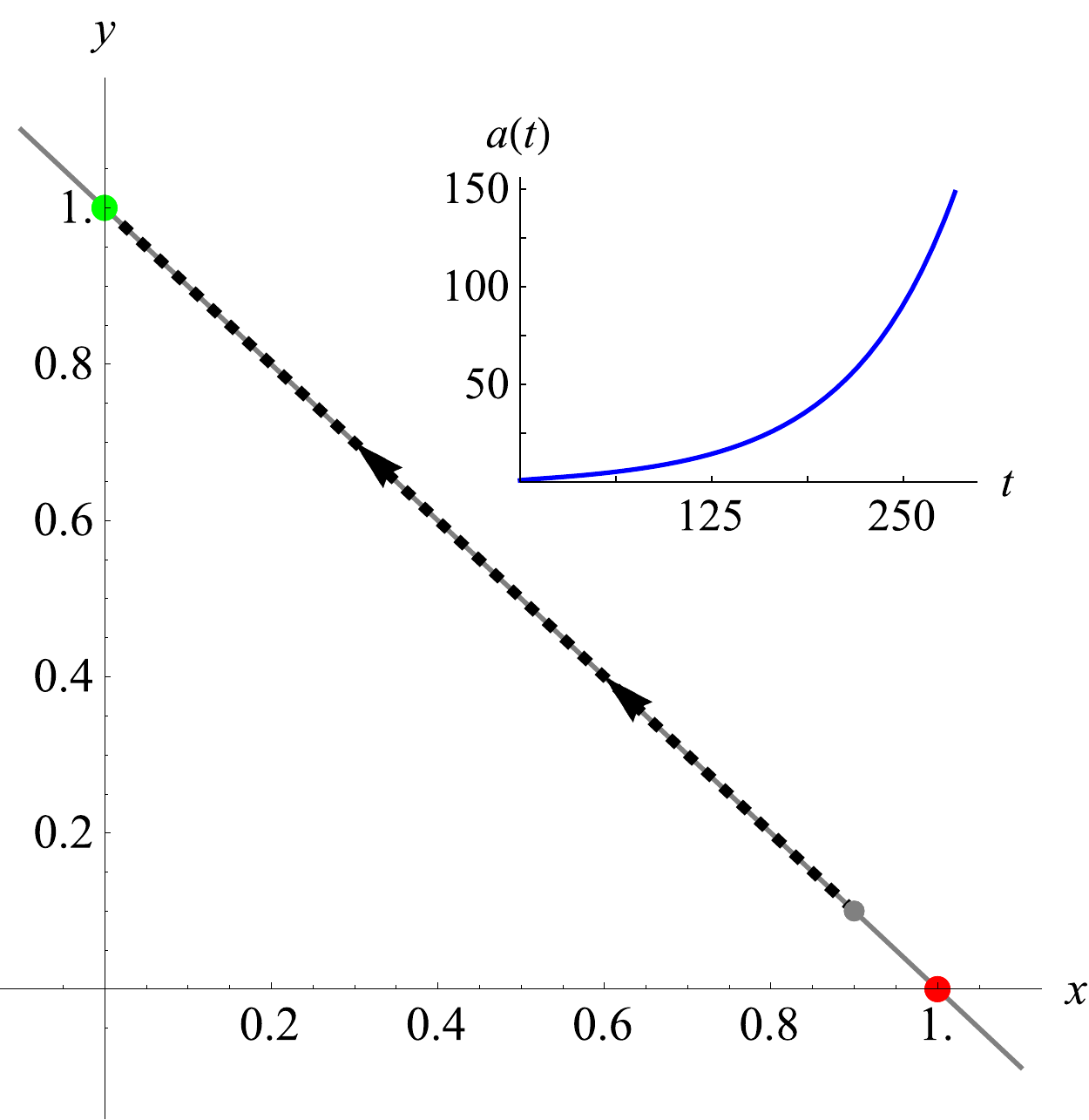}}
\end{minipage}
\begin{minipage}{.55\linewidth}
\centering
\subfloat[]{\includegraphics[scale=0.5]{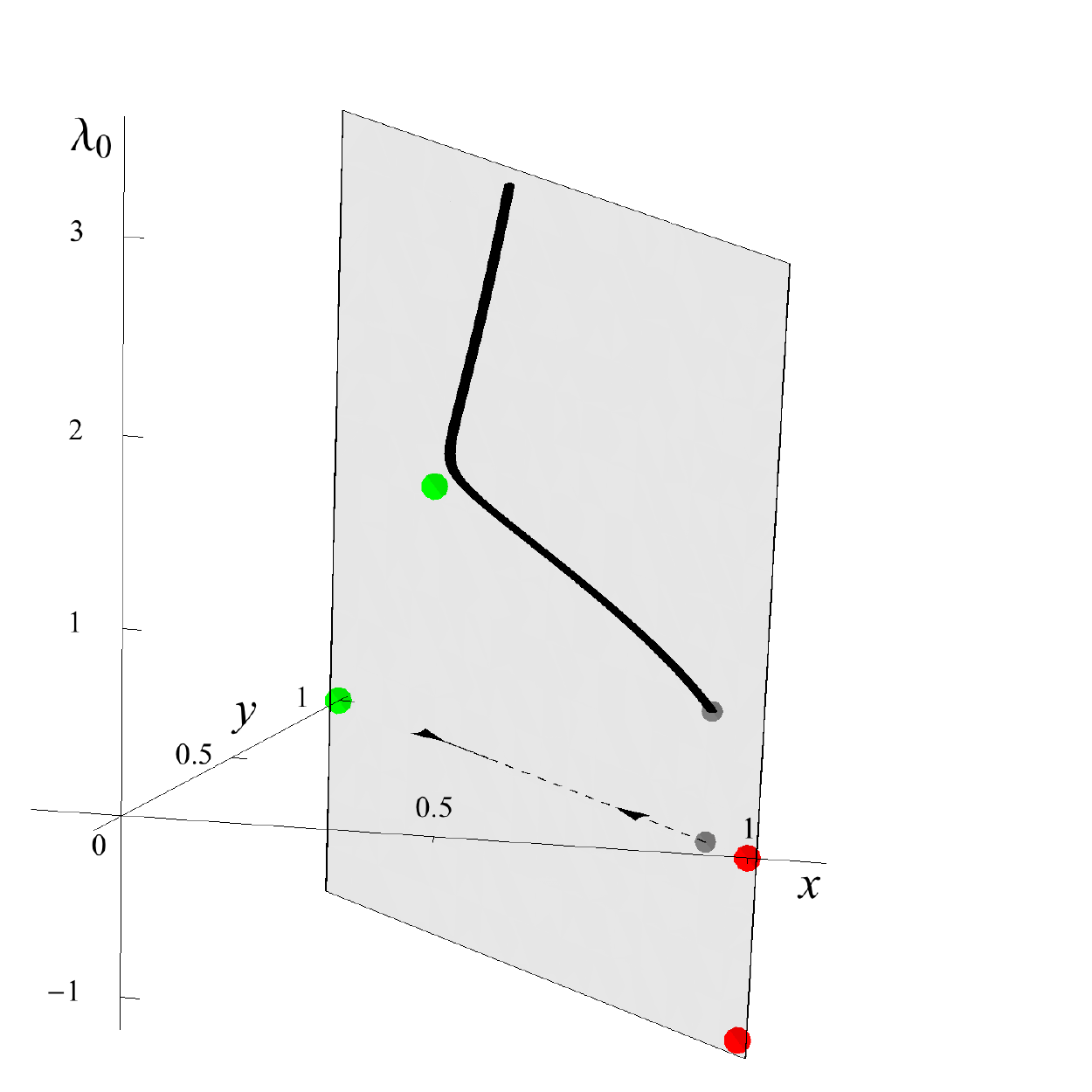}}
\end{minipage}
\begin{minipage}{.44\linewidth}
\centering
\subfloat[]{\includegraphics[scale=0.4]{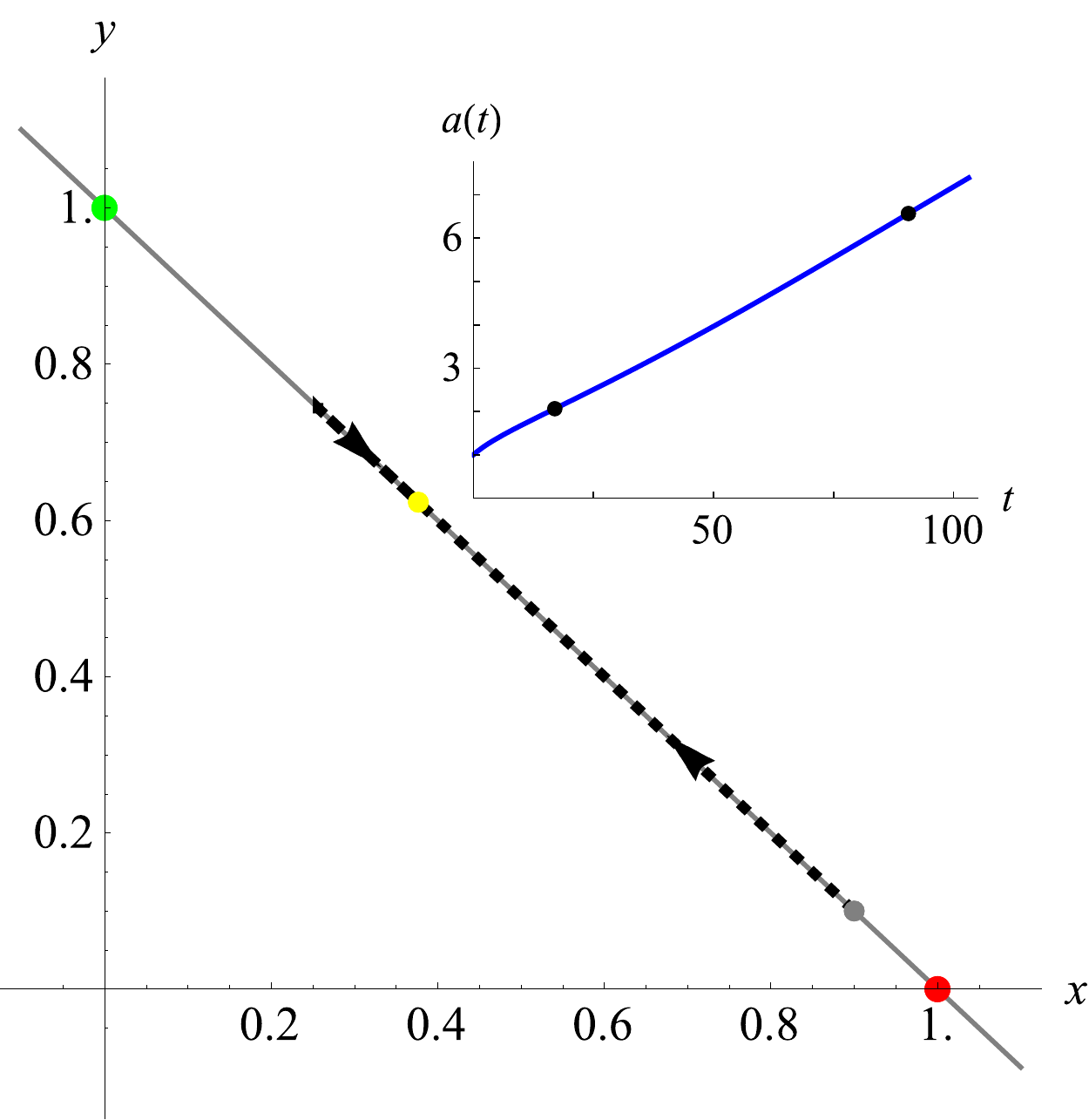}}
\end{minipage}
\begin{minipage}{.55\linewidth}
\centering
\subfloat[]{\includegraphics[scale=0.5]{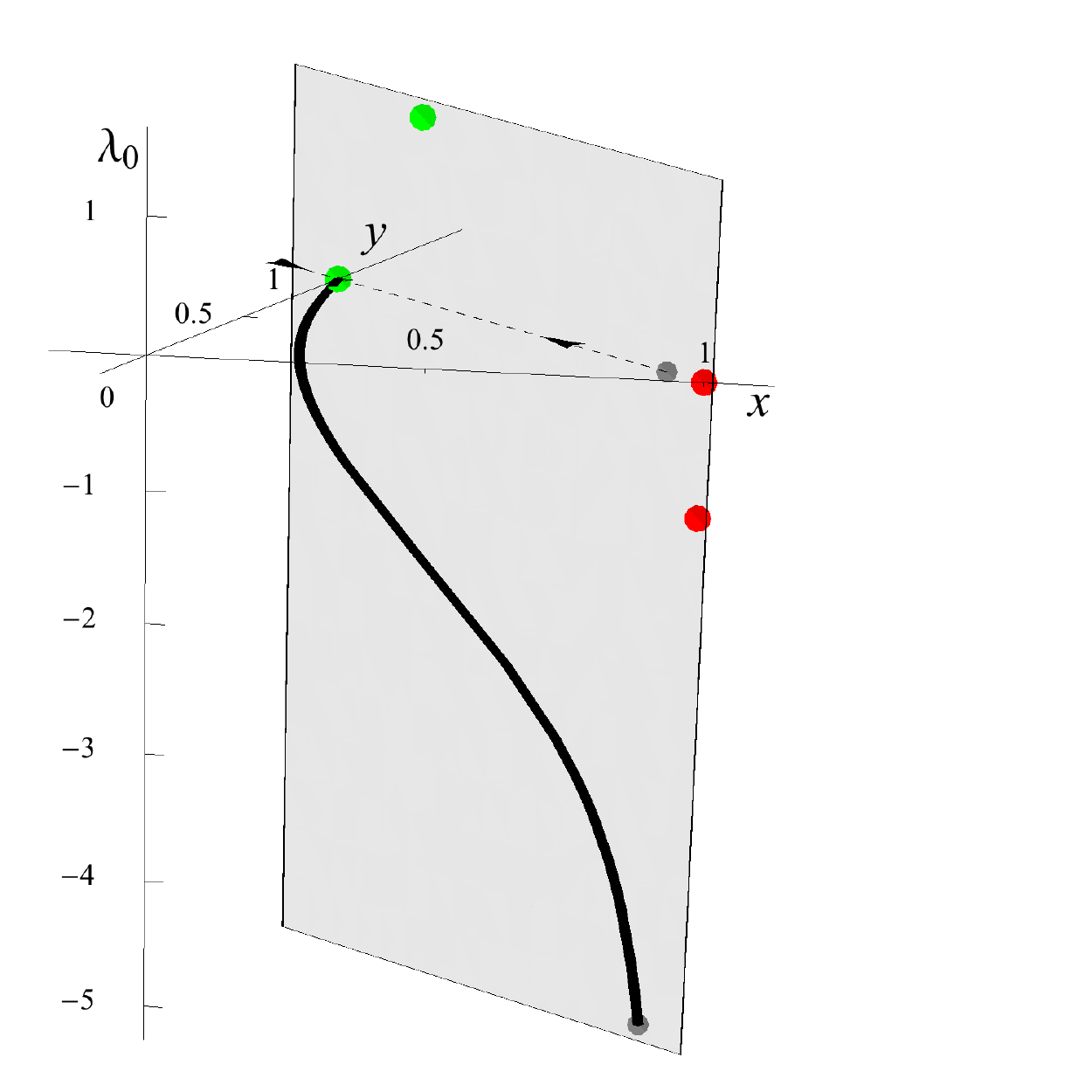}}
\end{minipage}
\begin{minipage}{.44\linewidth}
\centering
\subfloat[]{\includegraphics[scale=0.4]{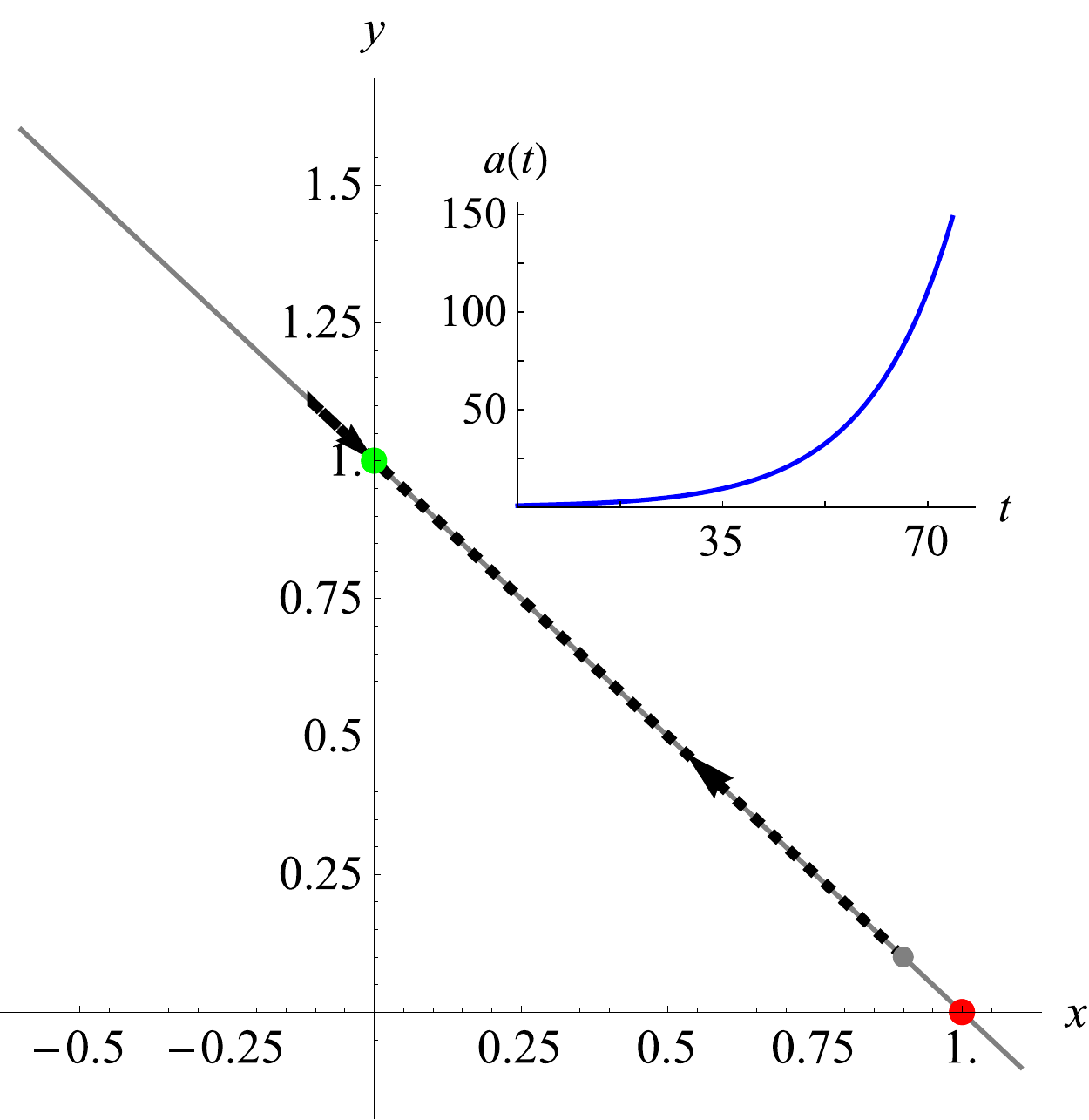}}
\end{minipage}
\caption{Phase-space plots of trajectories for the first-order system for $\lambda_{1}=2$, with $(x(\tau_{i}),y(\tau_{i}))=(0.9,0.1)$, $H(\tau_{i})=0.1$ (in units of $M_{\rm pl}$), and different values of $\lambda_{0}(\tau_{i})$. Each trajectory is constrained to the plane $x + y = 1$. The figures on the left show the full 3-dimensional trajectory (solid black line), together with a projection of that trajectory onto the $x\textrm{-}y$ plane (dashed line). The figures on the right show the projected trajectory, together with an inset displaying the evolution of $a(t)$. Green dots correspond to inflationary fixed points and red dots to non-inflationary fixed points (as given in Table~\ref{TAB:FO}). Grey dots correspond to the starting point of the trajectory. The plots show trajectories for $\lambda_0 (t_i) = 0.6$ (top row), $\lambda_0 (\tau_i) = 0.7$ (middle row), and $\lambda_0 (\tau_i) = -5$ (bottom row). For $\lambda (\tau_i) = 0.6$ and $-5$, the system evolves toward the inflationary attractor {\bf FP1a}. For $\lambda (\tau_i) = 0.7$, the system is deflected upwards by the inflationary saddle point {\bf FP1d}, and inflation only occurs between the black dots on the inset plot of $a(t)$. The yellow dot in the main panel of plot (d) corresponds to the projection of the end point of the trajectory, which was chosen arbitrarily but before a suspected finite-time singularity.}
\label{FIG:TrajOne_09}
\end{figure*}

The first-order system, as defined by Eqs.~(\ref{EQN:xFtimeONE})--(\ref{EQN:ConstraintONE}), corresponds to an effectively two-dimensional system in the EFT phase space. In Fig.~\ref{FIG:TrajOne_09} we display example trajectories that arise from solving these equations, setting $\lambda_{1}=2$ but varying the initial value of $\lambda_0 (\tau_i)$. Under these conditions, fixed point {\bf FP1d} in Table~\ref{TAB:FO} is an inflationary saddle point, whereas fixed point {\bf FP1a} is an inflationary attractor. For each trajectory, we begin with initial conditions such that the background spacetime is not inflating, with $(x (\tau_i) , y (\tau_i)) = (0.9,0.1)$ and $H (\tau_i) = 0.1$ (in units of $M_{\rm pl}$). 

As we vary $\lambda_0 (\tau_i)$, we find qualitatively different behavior for the resulting trajectories. Figs.~\ref{FIG:TrajOne_09}a,b correspond to the case $\lambda_0 (\tau_i) = 0.6$, for which the system is deflected downwards by the inflationary saddle point {\bf FP1d} (upper green dot) and evolves toward the inflationary attractor {\bf FP1a} (lower green dot). Increasing $\lambda_0 (\tau_i)$ to $\lambda_0 (\tau_i) = 0.7$, we find qualitatively different behavior in Figs.~\ref{FIG:TrajOne_09}c,d: the trajectory is deflected upwards by the inflationary saddle point {\bf FP1d}, such that inflation occurs for a brief period of time
($\sim 1$ efold).\footnote{The subsequent evolution of the background in this case reveals what we suspect is a finite-time singularity \cite{getz+jacobson_77,goriely+hyde_00}. For the first-order system, this corresponds to the magnitude of the vector of phase-space coordinates $||(x(t),y(t), \lambda_{0}(t))|| \to\infty$ as $t\to t_{\textrm{f}}$ for some $t_{\textrm{f}}<\infty$, where $||\dots||$ represents the norm. Such finite-time singularities are not particularly problematic at either zeroth or first order. At zeroth order, the singularities do not occur for the types of trajectories considered in Fig.~\ref{FIG:TrajZero}a. At first order, the singularities seem only to occur either after an inflationary phase or for trajectories that do not inflate at all. See Ref.~\cite{barrow+graham_15} and references therein for a recent discussion of various types of cosmological singularities.} For values of $\lambda_0 (\tau_i) < 0.6$, meanwhile --- including negative values, such as $\lambda_0 (\tau_i) = -5$ in Figs.~\ref{FIG:TrajOne_09}e,f --- the system again flows toward the inflationary attractor {\bf FP1a}. 

We find the same qualitative behavior for trajectories as we vary initial conditions $(x (\tau_i), y (\tau_i))$, with $0.5 \leq x (\tau_i) < 1$: the particular values of $\lambda_0 (\tau_i)$ separating the types of trajectories change, but the presence of these three types of trajectories remains common. Likewise, we find that initial conditions with $x (\tau_i) > 1$ (which violate the dominant energy condition, DEC) generically do not inflate, akin to the behavior shown in Fig.~\ref{FIG:TrajZero}b.

Estimating the probability of inflation is more subtle for first-order systems than for the zeroth-order case. First we note that for $\lambda_1 > 0$, there always exists an inflationary attractor at first order, viz., fixed point {\bf FP1a} at $(x,y, \lambda_0) = (0, 1, 0)$. For $\lambda_1 < 3$, there exists a second inflationary fixed point, {\bf FP1d}, which is never an attractor for $\lambda_1 > 0$. Hence we consider two distinct cases: $0 < \lambda_1 < 3$ and $\lambda_1 > 3$. For concreteness, we study examples with $\lambda_1 = 2$ (case 1) and $\lambda_1 = 4$ (case 2). 

For each case, we estimate the probability of inflation in three steps: first we fix $\lambda_1$ (as required at first order) and numerically find that portion of phase space that (i) corresponds to kinetic-energy-dominated initial conditions, 
i.e., $0.5 \leq x (\tau_i) \leq 1$, and (ii) flows through at least 60 efolds of inflation. (We denote this region of phase space the ``basin of sufficient inflation," ${\cal R}$.) Next we set down a specific, heuristic probability distribution, $P_{\rm KE} (x, \lambda_0)$, over {\it all} possible kinetic-energy-dominated initial conditions. (Given the constraint of 
Eq.~(\ref{EQN:ConstraintONE})
we may always parameterize the phase space for first-order systems by $\{x, \lambda_0\}$.) Finally, we integrate the probability distribution over ${\cal R}$ to find the probability that a first-order system will flow through at least 60 efolds of inflation, having started from kinetic-energy-dominated initial conditions:
\begin{equation}
{\rm Pr}_{\rm Inf} \equiv \int_{\cal R} dx \, d\lambda_0 \, P_{\rm KE} (x, \lambda_0) .
\label{EQN:1ProbInf}
\end{equation}

For each case that we consider ($\lambda_1 = 2$ and $\lambda_1 = 4$), we first focus on systems in which $\lambda_0 (\tau_i) \geq 0$ before considering the unrestricted case. We do so because for first-order systems that can be represented by SSF realizations, from Eq.~(\ref{EQN:lSF}) we have
\begin{equation}
\lambda_{0}= - \frac{\dot{V}}{H V}\sim -\frac{{\delta V}/V}{{\delta t}/t_{H}},
\end{equation}
where $t_{H}\equiv H^{-1}$ is the Hubble time. That is, for SSF systems, $\lambda_{0}$ can be interpreted as (minus) the fractional change in the potential-energy density per unit Hubble time. Put another way, for an SSF system at initial time $t_i$ we have $\dot{V} (t_i) = - \lambda_0 (t_i) H (t_i) V (t_i)$ and $V (t_i) = 3 M_{\rm pl}^2 H^2 (t_i) y (t_i)$. In all SSF realizations, $H (t_i) > 0$ and $y (t_i) > 0$, and hence $\lambda_0 (t_i) < 0$ corresponds to $\dot{V} (t_i) > 0$, a scenario that would presumably favor the onset of inflation. Since our aim is to consider initial conditions that do not expressly favor inflation, we first consider non-negative initial values of $\lambda_0$.

\begin{figure*}
\begin{minipage}{.49\linewidth}
\centering
\subfloat[]{\includegraphics[scale=0.45]{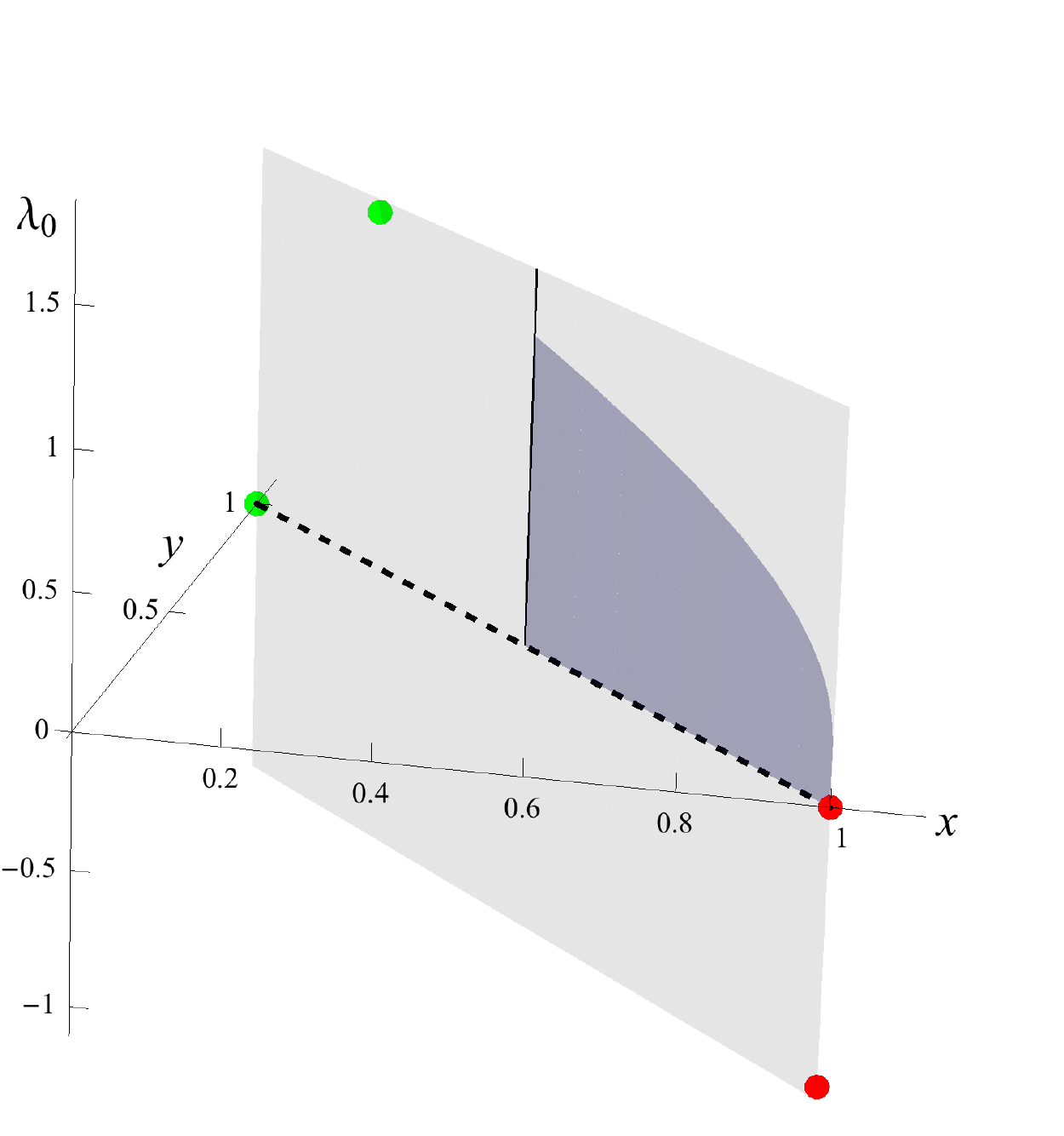}}
\end{minipage}
\begin{minipage}{.49\linewidth}
\centering
\subfloat[]{\includegraphics[scale=0.65]{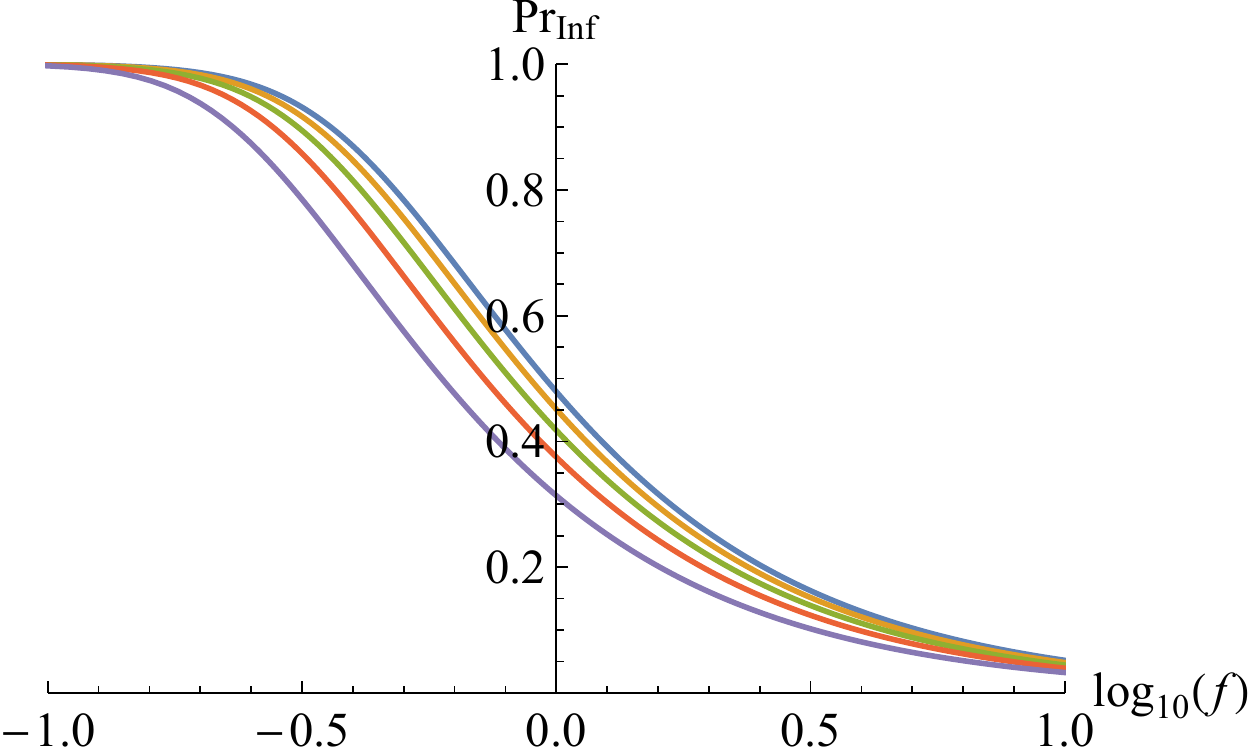}}
\end{minipage}
\begin{minipage}{.49\linewidth}
\centering
\subfloat[]{\includegraphics[scale=0.45]{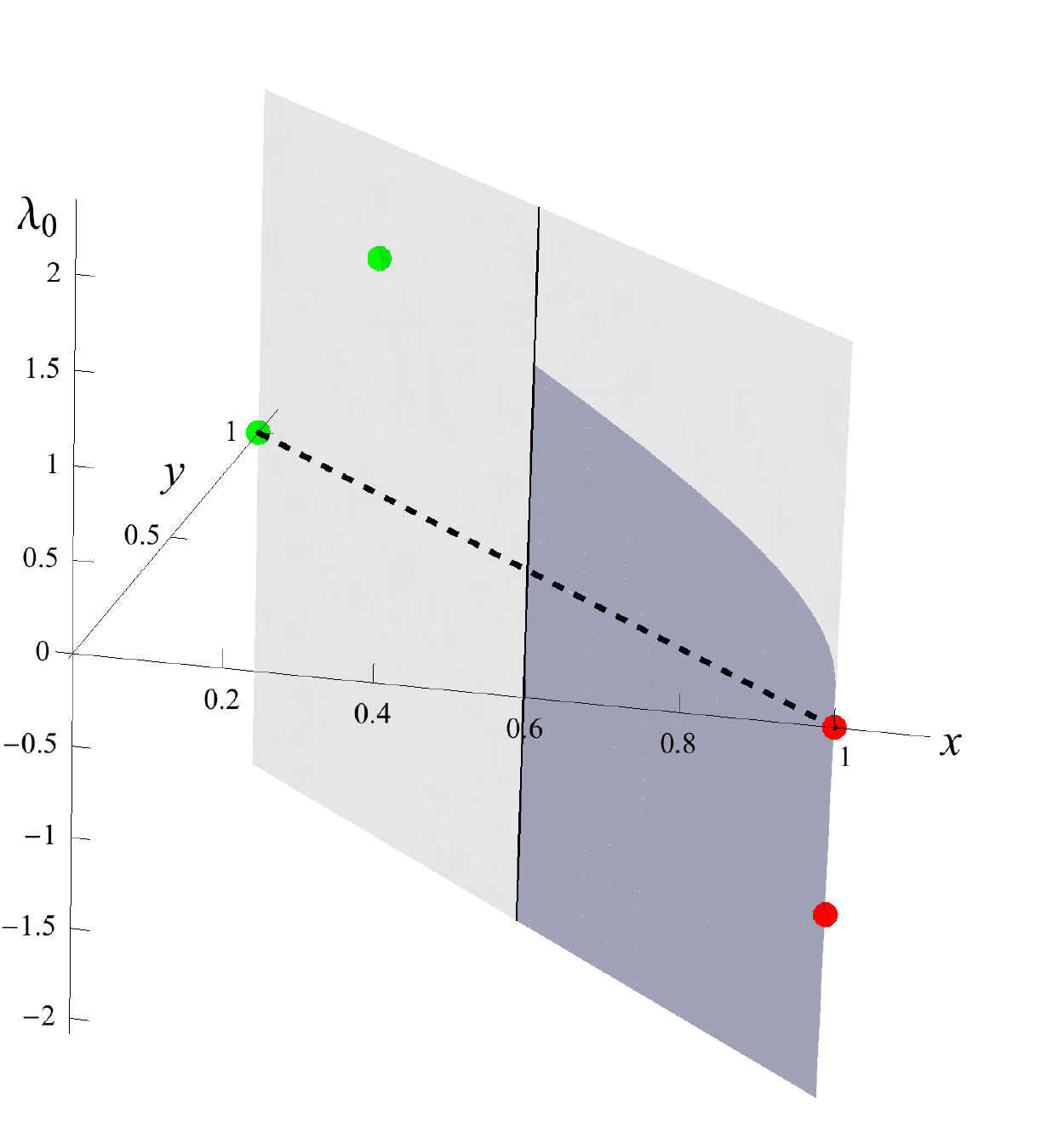}}
\end{minipage}
\begin{minipage}{.49\linewidth}
\centering
\subfloat[]{\includegraphics[scale=0.65]{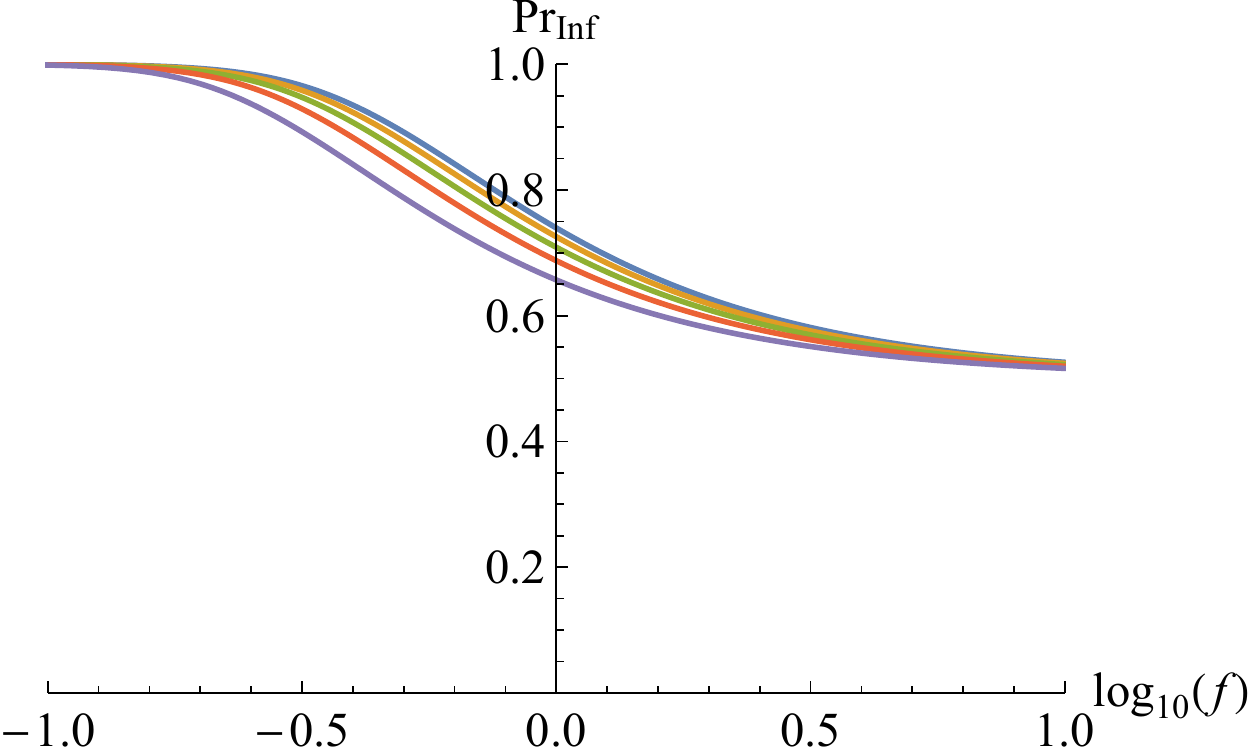}}
\end{minipage}
\caption{Basins of sufficient inflation, ${\cal R}$, and the probability of sufficient inflation, ${\rm Pr}_{\rm Inf}$, for first-order systems with $\lambda_{1}=2$ (case 1). In the plots on the left, fixed points {\bf FP1a} and {\bf FP1d} are shown in green, 
{\bf FP1b} and {\bf FP1c} are shown in red, and the purple regions denote ${\cal R}$ for (a) $\lambda_0 (\tau_i) \geq 0$, and (c) unrestricted $\lambda_0 (\tau_i)$. In both cases, we restrict $x (\tau_i) \in [0.5, 0.999]$. All points of phase space within the purple regions yield at least 60 efolds of inflation. The plots on the right show the corresponding behavior of ${\rm Pr}_{\rm Inf}$ for (b) $\lambda_0 (\tau_i) \geq 0$, and (d) unrestricted $\lambda_0 (\tau_i)$. In both plots, probabilities for various values of $x (\tau_i) = x_{\rm min}$ are shown, with $x_{\rm min} = 0.5, 0.6, 0.7, 0.8$, and $0.9$ (top to bottom). }
\label{FIG:EFTPS_lam1_2}
\end{figure*}

The basin of sufficient inflation, $\mathcal{R}$, for case 1 (with $\lambda_{1}=2$) and $\lambda_{0}(\tau_{i})\geq 0$ is presented in Fig.~\ref{FIG:EFTPS_lam1_2}a. This region includes all initial conditions with $x (\tau_i) \in [0.5, 0.999]$ for which the system flows towards the inflationary attractor {\bf FP1a} (lower green dot), as well as a small subset of initial conditions near the upper boundary of $\mathcal{R}$ whose subsequent flows are deflected upwards by the inflationary saddle point {\bf FP1d} (upper green dot) and inflate for at least 60 efolds. Systems that begin at 
$x (\tau_i) = 1$
and $\lambda_0 (\tau_i) \gtrsim 0$, just above the fixed point {\bf FP1b} at $(x,y,\lambda_0) = (1, 0, 0)$, shoot straight up (in the direction of increasing $\lambda_0$), leading to a suspected divergence in $\lambda_0 (\tau)$; such systems never inflate. 

To construct $P_{\tiny\textrm{KE}}(x,\lambda_{0})$, we set down a probability distribution that is uniform in the $x$-direction and Gaussian in the $\lambda_{0}$-direction, and treats these two directions independently. Since at first we restrict attention to $\lambda_{0}(\tau_{i})\geq 0$, we consider a half-Gaussian in the $\lambda_{0}$-direction. Thus we propose for $(x(\tau_{i}), \lambda_{0}(\tau_{i}))\in\mathcal{D}\equiv [x_{\textrm{min}},0.999]\times [0,\infty)$:
\begin{equation}\label{EQN:ProbKEextended}
P_{\tiny\textrm{KE}}(x,\lambda_{0}) \equiv \frac{2}{\Delta}  \frac{1}{\sqrt{2\pi (f\sigma)^2}}\exp\left[-\frac{1}{2(f\sigma)^2}\lambda_{0}^{\;2}\right],
\end{equation}
where $\Delta\equiv 0.999-x_{\textrm{min}}$ is the range of initial conditions considered in the $x$-direction. We vary $x_{\textrm{min}}=0.5,0.6,0.7,0.8$ or $0.9$, to produce 5 separate curves for the probability of flowing through sufficient amounts of inflation (each as a function of $f$).

The standard deviation of the Gaussian in the $\lambda_{0}$-direction, $f\sigma$, determines the scale over which the Gaussian has significant support. We parameterize the standard deviation with two terms. We set $\sigma$ equal to the $\lambda_0$-coordinate of the fixed point {\bf FP1d},
\begin{equation}
\sigma \equiv \frac{2\lambda_{1}}{3}\longrightarrow\frac{4}{3}\;\textrm{
(case 1)},
\label{EQN:sigmacase1}
\end{equation}
since {\bf FP1d} (upper green dot) plays a significant role in shaping the (inflationary) nature of trajectories that begin with kinetic-energy-dominated initial conditions; we therefore assume that this choice of $\sigma$ sets the scale for the region of phase space that is of dynamical interest. (One could select a different measure, such as the average distance between fixed points in the phase space, though this makes little numerical difference compared to our choice of $\sigma$.) We also include the multiplicative factor $f$, which we take to range between $f = 0.1$ and $f = 10$, with which we may explore how the resulting probability of flowing through sufficient inflation depends on the width of the Gaussian. (One may consider effects on the form of the probability distribution from averaging over finite-time intervals, as in Ref.~\cite{FordProbability}, though incorporating the factor $f$ suffices for our purposes.) For any choice of $f \sigma$, the probability distribution in Eq.~(\ref{EQN:ProbKEextended}) is properly normalized, with 
\begin{equation}
\int_{\mathcal{D}}\;dx\;d\lambda_{0}\;P_{\tiny\textrm{KE}}(x,\lambda_{0}) = 1.
\end{equation}

The final step is to integrate $P_{\rm KE} (x, \lambda_0)$ over the region ${\cal R}$ to find ${\rm Pr}_{\rm Inf}$ as a function of $f$. Results for ${\rm Pr}_{\rm Inf}$ for $\lambda_1 = 2$ and $\lambda (\tau_i) \geq 0$ are shown in Fig.~\ref{FIG:EFTPS_lam1_2}b. We find, as one might expect, that the highest probabilities occur for lower values of $x_{\textrm{min}}$. That is, for initial conditions such that the initial kinetic-energy density is less dominant, the probability of flowing through sufficient amounts of inflation is higher. Moreover, for any $x_{\rm min}$, the probability ${\rm Pr}_{\rm Inf}$ increases as $f$ decreases. This is because the width of the probability distribution over initial conditions becomes smaller as $f$ does, in which case a relatively greater amount of the support of the probability distribution comes from initial conditions that lead to trajectories that flow through sufficient inflation.

Next we relax the condition $\lambda (\tau_i) \geq 0$, and consider regions of phase space that include trajectories that expressly lie beyond those that are compatible with SSF realizations. As we found in Figs.~\ref{FIG:TrajOne_09}e,f, such scenarios include cases in which trajectories can traverse regions with $x < 0$, which violate each of the point-wise energy conditions identified in Eqs.~(\ref{EQN:nullenergy})--(\ref{EQN:strongenergy}), though (as noted above) such violations by an effective field theory need not signal pathologies \cite{creminelli_06}. For {\it every} case we investigated, with $x (\tau_i) \in [0.5, 0.999]$ and $\lambda (\tau_i) < 0$ --- as we varied $\lambda_0 (\tau_i)$ over 5 orders of magnitude --- the ensuing trajectory enters the regime with $x < 0$ en route to the inflationary attractor {\bf FP1a}, yielding at least 60 efolds of inflation. We therefore assume that, generically, first-order systems with $\lambda_1 = 2$ and $\lambda_0 (\tau_i) < 0$ yield sufficient amounts of inflation. The corresponding basin of attraction ${\cal R}$ is shown in Fig.~\ref{FIG:EFTPS_lam1_2}c.

To compute the probability of sufficient inflation for such cases, we again set down a probability distribution that is uniform in the $x$-direction and Gaussian in the $\lambda_{0}$-direction, though now we allow for all values of $\lambda_{0}(\tau_{i})$. Thus we propose, for $(x(\tau_{i}), \lambda_{0}(\tau_{i}))\in\mathcal{\hat{D}}\equiv [x_{\textrm{min}},0.999]\times (-\infty,\infty)$:
\begin{equation}\label{EQN:ProbKEextendedNEG}
\hat{P}_{\tiny\textrm{KE}}(x,\lambda_{0}) \equiv \frac{1}{\Delta} \frac{1}{\sqrt{2\pi (f\sigma)^2}}\exp\left[-\frac{1}{2(f\sigma)^2}\lambda_{0}^{\;2}\right],
\end{equation}
again with $\Delta\equiv 0.999-x_{\textrm{min}}$. We again select $x_{\textrm{min}}=0.5,0.6,0.7,0.8$ or $0.9$, and again use $\sigma = 2 \lambda_1 / 3 = 4/3$, based on the $\lambda_0$-coordinate of the fixed point {\bf FP1d}, with $f$ ranging between $f = 0.1$ and $10$. For any choice of $f \sigma$, we again find
\begin{equation}
\int_{\mathcal{\hat{D}}}\;dx\;d\lambda_{0}\;\hat{P}_{\tiny\textrm{KE}}(x,\lambda_{0}) = 1.
\end{equation}
In addition, we note that for any value of $x_{\textrm{min}}$, it is straightforward to show that the first-order probability for flowing through sufficient amounts of inflation can be written as
\begin{align}\label{EQN:1ProbInfNEG}
\textrm{Pr}_{\textrm{Inf}}&\equiv\int_{\mathcal{R}} dx\,d\lambda_{0}\;\hat{P}_{\tiny\textrm{KE}}(x,\lambda_{0})\nonumber\\
&=\frac{1}{2}+ \int_{\mathcal{R_{\rm U}}}dx\,d\lambda_{0}\;\hat{P}_{\tiny\textrm{KE}}(x,\lambda_{0}),
\end{align}
where $\mathcal{R}_{\textrm{U}}$ is the portion of the basin of sufficient inflation that lies in the `upper' part of $\mathcal{R}$, with $\lambda_{0}\geq 0$. For each $x_{\textrm{min}}$, the results of our numerical computation for the probability of flowing through sufficient amounts of inflation are presented in Fig.~\ref{FIG:EFTPS_lam1_2}d. Again we find that the highest probabilities occur for lower values of $x_{\textrm{min}}$, and that for any value of $x_{\rm min}$, ${\rm Pr}_{\rm Inf}$ increases with decreasing $f$.

\begin{figure*}
\begin{minipage}{.49\linewidth}
\centering
\subfloat[]{\includegraphics[scale=0.45]{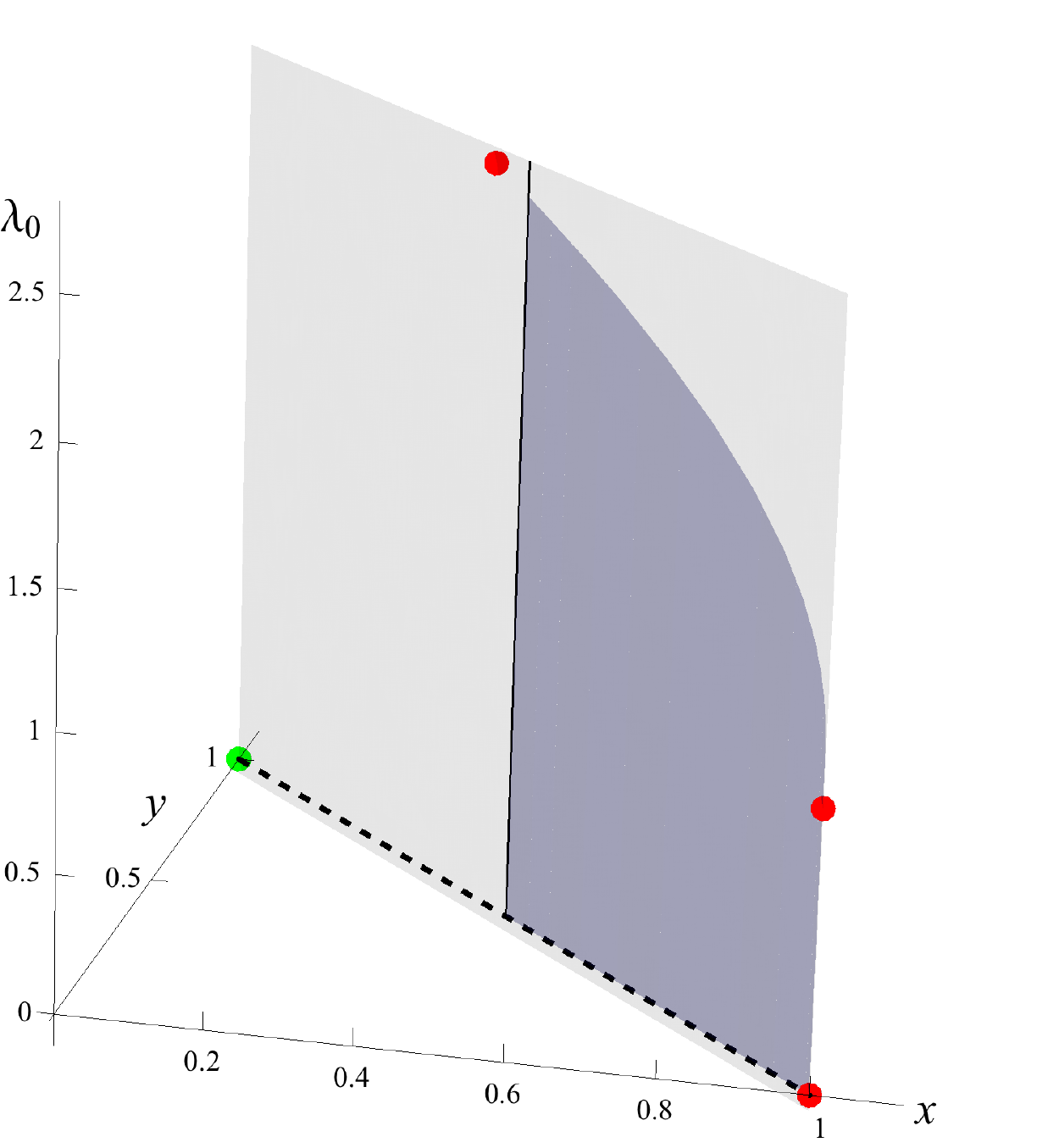}}
\end{minipage}
\begin{minipage}{.49\linewidth}
\centering
\subfloat[]{\includegraphics[scale=0.65]{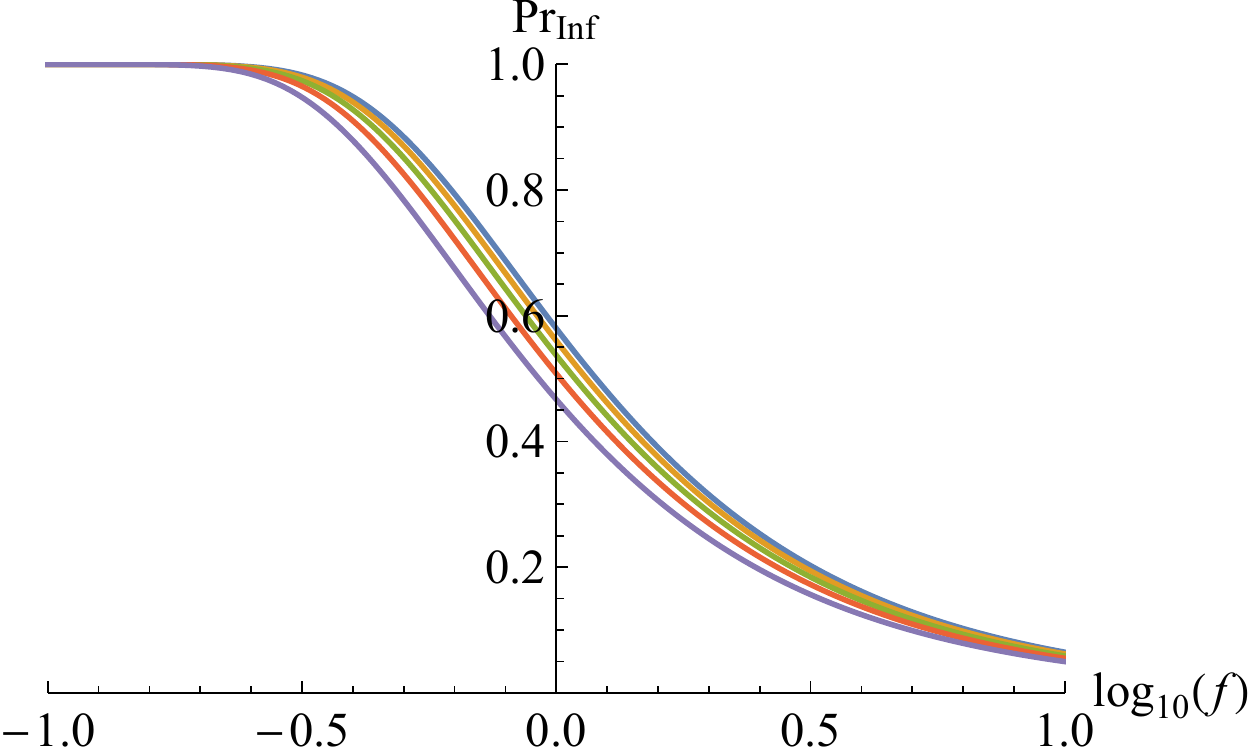}}
\end{minipage}
\begin{minipage}{.49\linewidth}
\centering
\subfloat[]{\includegraphics[scale=0.45]{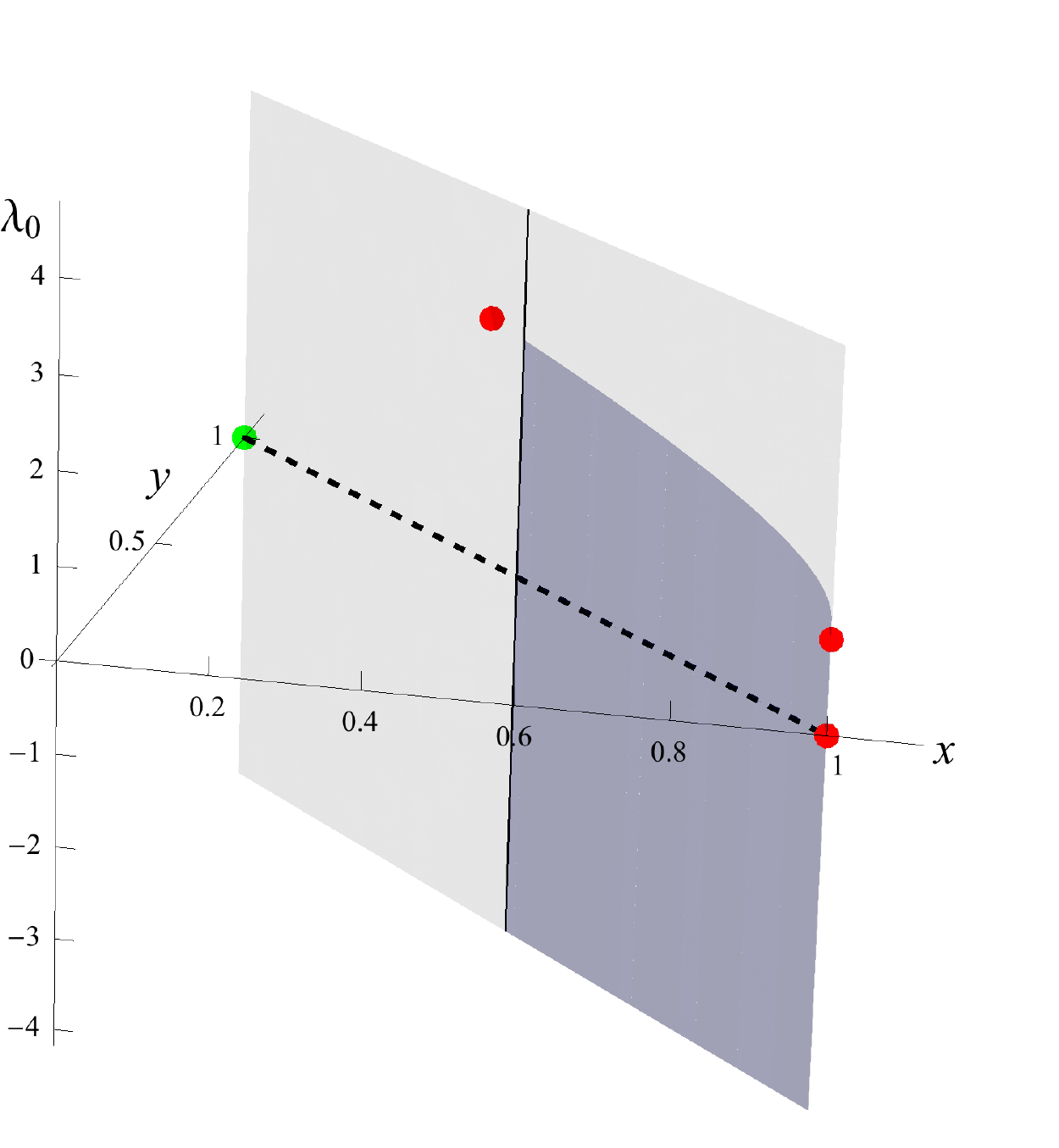}}
\end{minipage}
\begin{minipage}{.49\linewidth}
\centering
\subfloat[]{\includegraphics[scale=0.65]{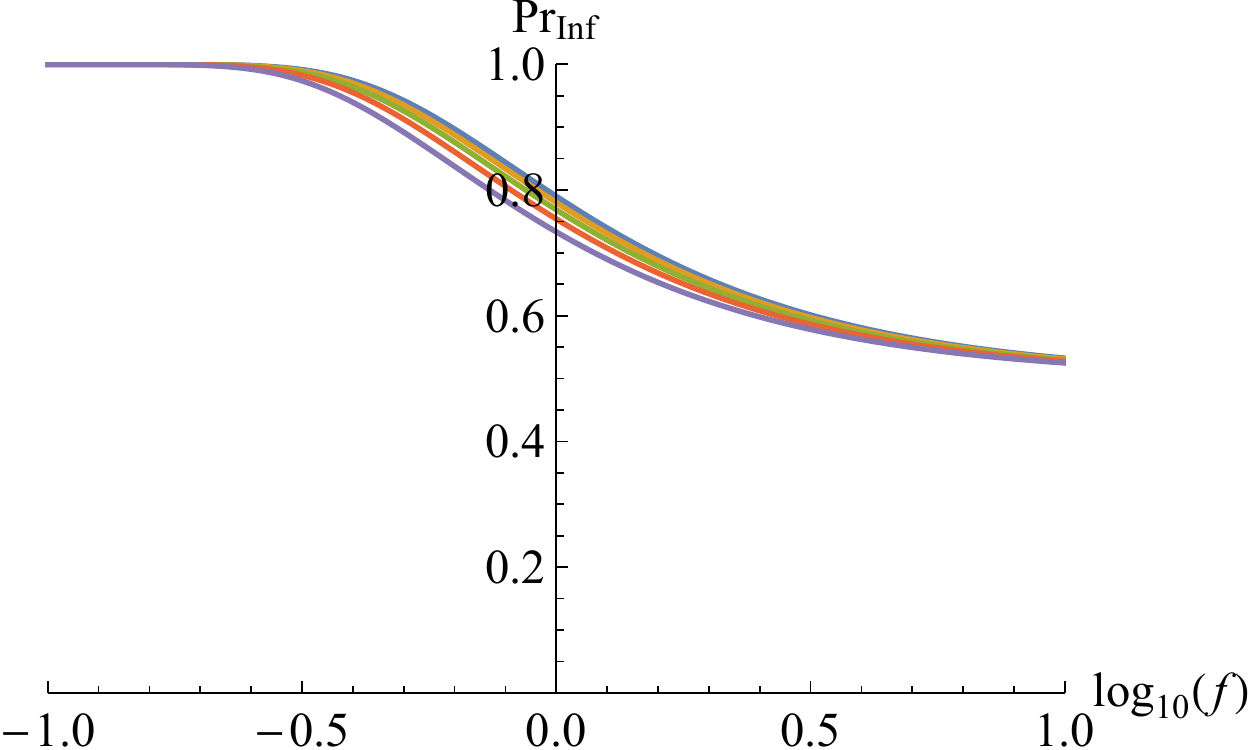}}
\end{minipage}
\caption{Basins of sufficient inflation, ${\cal R}$, and the probability of sufficient inflation, ${\rm Pr}_{\rm Inf}$, for first-order systems with $\lambda_{1}=4$ (case 2). In the plots on the left, fixed point {\bf FP1a} is shown in green, fixed points {\bf FP1b}, {\bf FP1c}, and {\bf FP1d} are shown in red, and the purple regions denote ${\cal R}$ for (a) $\lambda_0 (\tau_i) \geq 0$, and (c) unrestricted $\lambda_0 (\tau_i)$. In both cases, we restrict $x (\tau_i) \in [0.5, 0.999]$. All points of phase space within the purple regions yield at least 60 efolds of inflation. The plots on the right show the corresponding behavior of ${\rm Pr}_{\rm Inf}$ for (b) $\lambda_0 (\tau_i) \geq 0$, and (d) unrestricted $\lambda_0 (\tau_i)$. In both plots, probabilities for various values of $x (\tau_i) = x_{\rm min}$ are shown, with $x_{\rm min} = 0.5, 0.6, 0.7, 0.8$, and $0.9$ (top to bottom). }
\label{FIG:EFTPSandPROB_lam1_4}
\end{figure*}

We proceed similarly for case 2 ($\lambda_1 = 4$). The most important difference is that the fixed point {\bf FP1d} is no longer inflationary; only the point {\bf FP1a} remains an inflationary fixed point (in particular, an attractor). In this case, initial conditions whose subsequent flows are deflected upwards by the noninflationary saddle point {\bf FP1d} do not inflate. As before, we first consider the case $\lambda_0 (\tau_i) \geq 0$, which is compatible with SSF realizations, and examine initial conditions $x (\tau_i) \in [0.5, 0.999]$. We again use the probability distribution $P_{\rm KE} (x,\lambda_0)$ of Eq.~(\ref{EQN:ProbKEextended}) with $\sigma = 2 \lambda_1 / 3 = 8/3$. In Figs.~\ref{FIG:EFTPSandPROB_lam1_4}a,b we show the basin of sufficient inflation, ${\cal R}$, and the corresponding probability to flow through sufficient inflation, ${\rm Pr}_{\rm Inf}$, as we vary the width of the Gaussian, $f$. We may also relax the restriction on $\lambda_0 (\tau_i)$ and include negative initial values (which are not compatible with SSF realizations). As in case 1, we find that $\lambda_0 (\tau_i) < 0$ generically yields trajectories that flow through at least 60 efolds of inflation, and hence the basin of sufficient inflation extends uniformly below $\lambda_0  = 0$. When we use the probability distribution $\hat{P}_{\rm KE} (x, \lambda_0)$ of Eq.~(\ref{EQN:ProbKEextendedNEG}) in this case, we again find a corresponding increase in ${\rm Pr}_{\rm Inf}$, as shown in Figs.~\ref{FIG:EFTPSandPROB_lam1_4}c,d. As in case 1, we find highest probabilities for lower values of $x_{\rm min}$ and smaller $f$.

The results in Figs.~\ref{FIG:EFTPS_lam1_2} and \ref{FIG:EFTPSandPROB_lam1_4} for ${\rm Pr}_{\rm Inf}$ are essentially unchanged if we adopt a box-like probability distribution of the form $P_{\rm KE} (x, \lambda_0) = 1/ (\Delta f \sigma)$ (for $\lambda_0 \geq 0$) and $\hat{P}_{\rm KE} (x, \lambda_0) = 1/ (2 \Delta  f \sigma )$ (for unrestricted $\lambda_0$), for $\vert \lambda_0 \vert \leq f \sigma$, with $P_{\rm KE} (x, \lambda_0) = \hat{P}_{\rm KE} (x,\lambda_0) = 0$ for $\vert \lambda_0 \vert > f\sigma$. Here $\Delta = 0.999 - x_{\rm min}$ and $\sigma = 2 \lambda_1 / 3$, as above. In both cases, the Gaussian distributions of Eqs.~(\ref{EQN:ProbKEextended}) and (\ref{EQN:ProbKEextendedNEG}) yield modestly more conservative results for ${\rm Pr}_{\rm Inf}$ than the box-like probability distributions.  

\section{A general potential for dynamical trajectories\label{SEC:GenPotential}}

To place trajectories like those shown in Fig.~\ref{FIG:TrajZero}a (for a zeroth-order system) and Fig.~\ref{FIG:TrajOne_09}a (for a first-order system) in a more familiar context, it is helpful to construct SSF realizations of such dynamical systems. In this section we explore such realizations for zeroth- and first-order systems, and demonstrate that a single functional form for the effective potential, $V (\phi)$, is compatible with such dynamical trajectories through phase space.

We may generate an SSF realization of an EFT dynamical system at arbitrary order by solving Eqs.~(\ref{EQN:xFtimeM})--(\ref{EQN:lMminusonetimeM}), together with Eq.~(\ref{EQN:Ep}), for $x(t)$, $y(t)$, and $H(t)$, in terms of cosmic time $t$. By employing the mapping provided by Eqs.~(\ref{EQN:xSF}) and~(\ref{EQN:ySF}), we may then derive the time evolution of SSF quantities of interest. In what follows, we will be particularly interested in $\phi(t)$, $\dot{\phi}(t)$, and $V(t)$. For clarity, we will first collect some relevant results.

From Eq.~(\ref{EQN:xSF}), we may write
\begin{equation}
    \phi (t) = \phi (t_i) \pm \int_{t_i}^t dt' \sqrt{ 6 M_{\rm pl}^2 H^2 (t') x (t') } ,
    \label{EQN:phiSSF}
\end{equation}
where $\phi (t_i)$, an integration constant, is the initial value of the field at time $t_i$. Likewise, from Eq.~(\ref{EQN:ySF}) we have%
\begin{align}
V(t) = 3 \Mpl^{2} H^2(t)y(t) \label{EQN:VSSF}.
\end{align}
One can then construct an explicit functional form for $V(\phi)$ from Eqs.~(\ref{EQN:phiSSF}) and (\ref{EQN:VSSF}). 

In the cases in which we analyze SSF realizations of trajectories that correspond to (hyperbolic inflationary) fixed points, we will be able to construct $V(\phi)$ analytically. For more general flows in the phase space --- especially for flows that start with kinetic-energy-density dominated initial conditions and which subsequently flow into inflationary states --- we will do so parametrically, and then fit a functional form to the parametrically determined $V(\phi (t))$. 

Remarkably, we find that a single functional form is sufficient to fit a wide variety of such flows in the EFT phase space for zeroth- and first-order systems. (This trend continues for second-order systems, which we explore in Appendix~\ref{APP:secondorder}.) This functional form is given by
\begin{equation}\label{EQN:FFfit}
V(\phi) = V_{0}\exp\left[-\alpha\left(\frac{\phi}{\Mpl}\right)^{\beta}\right],
\end{equation}
where $\alpha$ and $\beta$ are positive constants. Potentials of this form have recently been explored, in a different context, in Refs.~\cite{mukhanov_13, geng_15, geng_17}.

The potential of Eq.~(\ref{EQN:FFfit}) has two interesting limits. For $\alpha = 0$, the potential reduces to $V (\phi) \rightarrow V_0$, corresponding to evolution in a pure de Sitter background. For $\beta = 1$, the potential reduces to the familiar form for power-law inflation \cite{abbott+wise_84,lucchin_85,SalopekBond90}, which is typically written as
\begin{equation}\label{EQN:VPLI}
V(\phi)=V_{0}\exp\left(-\sqrt{\frac{2}{p}}\frac{\phi}{\Mpl}\right) ,
\end{equation}
with $\alpha = \sqrt{2 / p}$ (and $p > 1$). As we demonstrate in the following subsections (for $M = 0, 1$) and in the appendices (for $M \geq 2$), at each order $M$ the effective phase space includes at most two inflationary hyperbolic fixed points: one corresponding to evolution in a pure de Sitter state, and the other corresponding to evolution with the power-law potential of Eq.~(\ref{EQN:VPLI}). The more general form for $V (\phi)$ in Eq.~(\ref{EQN:FFfit}) that we infer for SSF-compatible trajectories through an $M$th-order phase space (at least up through $M = 2$) incorporates the behavior at these two fixed points.

For evolution with the 
exponential potential of Eq.~(\ref{EQN:VPLI}) in a spatially flat background, the Friedmann equations yield the particular solutions
\begin{align}
\frac{\phi(t)}{\Mpl}&=\sqrt{2 p}\ln\left(\sqrt{\frac{V_{0}}{p(3p-1)}}\frac{t}{\Mpl}\right), \label{EQN:phiPLI}
\\
a(t)&=a(t_{i})\left(\frac{t}{t_{i}}\right)^{p},\label{EQN:aPLI}
\end{align}
for some initial time $t_{i}$. We may compare these results with the behavior we infer for various zeroth- and first-order systems evolving at the appropriate fixed point.

\subsection{Zeroth-Order Systems}\label{SEC:ZerothOrderSystems}

For the zeroth-order system, we are interested in two types of trajectories: those that correspond to fixed point {\bf FP0b}, with $0 < \lambda_0 < 2$ (see Table~\ref{TAB:ZO}); and those that begin with kinetic-energy-dominated initial conditions, $x (t_i) > y (t_i)$, but which satisfy $x, y \geq 0$ throughout the ensuing evolution, so as to remain compatible with SSF realizations.

We first consider evolution of the system at fixed point {\bf FP0b}, which (as we will see) reduces to the power-law inflation scenario of Eqs.~(\ref{EQN:VPLI})--(\ref{EQN:aPLI}). We fix $0 < \lambda_0 < 2$ and set $x (t_i) = \lambda_0 / 6$, $y (t_i) = 1 - x (t_i)$, and follow the system for times $t \in [t_i, t_f]$. Because {\bf FP0b} is a fixed point, $x (t)$ and $y(t)$ remain at these initial values. Then we may solve for the corresponding SSF quantities from Eqs.~(\ref{EQN:phiSSF})--(\ref{EQN:VSSF}). We select $\dot{\phi} (t_i) > 0$ and find
\begin{align}
\phi (t) &= \phi(t_{i})+\int_{t_{i}}^{t}dt'\sqrt{\lambda_{0}\Mpl^2 H^2(t')},\label{EQN:phiSSFFP2}\\
V(t)&= \left( \frac{6-\lambda_{0}}{2}\right) \Mpl^2 H^{2}(t).\label{EQN:VSSFFP2}
\end{align}
We may find an analytic expression for $H(t)$ as well. In particular, from Eq.~(\ref{EQN:Ep}), we have
\begin{equation}\label{EQN:HzeroSSF}
-\frac{\dot H}{H^2}=3 x = \frac{\lambda_{0}}{2} .
\end{equation}
The general solution to this differential equation is easily found:
\begin{equation}\label{EQN:HSS}
H(t)=\frac{2}{\lambda_{0}t-C},
\end{equation}
where $C$ is a constant of integration. In particular, evaluating Eq.~(\ref{EQN:HSS}) at $t=t_{i}$ yields $C=\lambda_{0}t_{i}-2/H(t_{i})$, so that Eq.~(\ref{EQN:HSS}) becomes
\begin{equation}
H(t)=\frac{2}{\lambda_{0}}\left(t-t_{i}+\frac{2}{\lambda_{0}H(t_{i})}\right)^{-1}.\label{EQN:HSSFFP2}
\end{equation}

To find an expression for $V(\phi)$ we substitute Eq.~(\ref{EQN:HSSFFP2}) for $H(t)$ into Eq.~(\ref{EQN:VSSFFP2}) for $V(t)$ to find
\begin{equation}
V(t)= \left( \frac{6-\lambda_{0}}{2} \right) \Mpl^2\left(\frac{2}{\lambda_{0}}\right)^{2}\left(t-t_{i}+\frac{2}{\lambda_{0}H(t_{i})}\right)^{-2}.\label{EQN:VintPLI}
\end{equation}
We may likewise substitute our expression for $H (t)$ into Eq.~(\ref{EQN:phiSSFFP2}) for $\phi (t)$ to find
\begin{equation}
\phi(t)=\phi(t_{i})+\Mpl\frac{2}{\sqrt{\lambda_{0}}}\ln\left[\frac{\lambda_{0}H(t_{i})}{2}\left(t-t_{i}+\frac{2}{\lambda_{0}H(t_{i})}\right)\right].\label{EQN:phiintPLI}
\end{equation}
Straightforward algebra then yields
\begin{equation}
V(\phi) = V_0 \exp\left(-\sqrt{\lambda_{0}}\frac{\phi}{\Mpl}\right) ,
\label{EQN:VphiPLIrequired}
\end{equation}
where we have defined
\begin{equation}
 V_{0}\equiv \left( \frac{6-\lambda_{0}}{2} \right)\Mpl^2 H^2(t_{i})\exp\left[\sqrt{\lambda_{0}} \,\frac{\phi(t_{i})}{\Mpl}\right] . \label{EQN:VzeroPLIrequired}
\end{equation}
Eq.~(\ref{EQN:VphiPLIrequired}) for $V (\phi)$ agrees with the potential for power-law inflation, Eq.~(\ref{EQN:VPLI}), upon setting
\begin{equation}
p = \frac{2}{\lambda_{0}}.\label{EQN:EquivAGAIN}
\end{equation}
Using Eqs.~(\ref{EQN:VzeroPLIrequired})--(\ref{EQN:EquivAGAIN}), we may rewrite Eq.~(\ref{EQN:phiintPLI}) as
\begin{equation}
    \frac{\phi (u) }{M_{\rm pl} } = \sqrt{2 p} \, \ln \left( \sqrt{ \frac{V_0}{p (3p - 1)}} \, \frac{u}{M_{\rm pl}} \right) ,
    \label{EQN:phiPLImatch}
\end{equation}
in terms of
\begin{equation}
    u (t) \equiv t - t_i + \frac{ 2}{\lambda_0 H (t_i) } .
    \label{EQN:udef}
\end{equation}
Eq.~(\ref{EQN:phiPLImatch}) for 
$\phi (u)$
matches Eq.~(\ref{EQN:phiPLI}) for power-law inflation. Similar manipulations, using Eq.~(\ref{EQN:HSSFFP2}) and $H = \dot{a} / a$, yield
\begin{equation}
    a (u) = a (u_i) \left( \frac{u}{u_i} \right)^p ,
    \label{EQN:aPLIfinal}
\end{equation}
where $u_i \equiv u (t_i)$. This solution reproduces Eq.~(\ref{EQN:aPLI}), and is indeed inflationary (with $p > 1$), given $p = 2 / \lambda_0$ and $\lambda_0 < 2$. We thus find for our first case of interest that zeroth-order systems that begin at fixed point {\bf FP0b} evolve exactly like models of power-law inflation, with $p = 2 / \lambda_0$.

\begin{figure*}[htp!]
\begin{minipage}{.3\linewidth}
\subfloat[]{\includegraphics[scale=0.3]{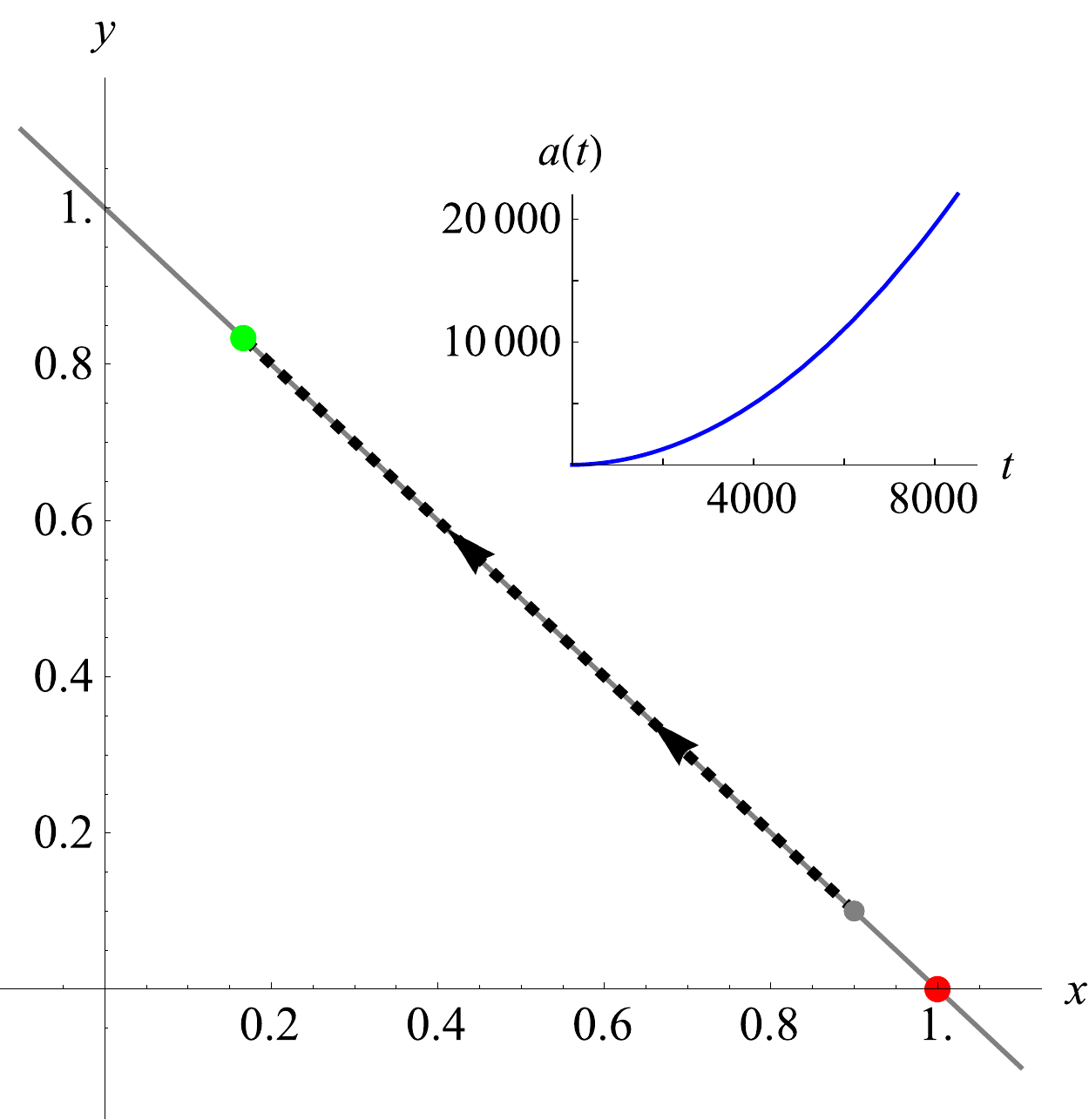}}
\end{minipage}
\begin{minipage}{.3\linewidth}
\subfloat[]{\includegraphics[scale=0.4]{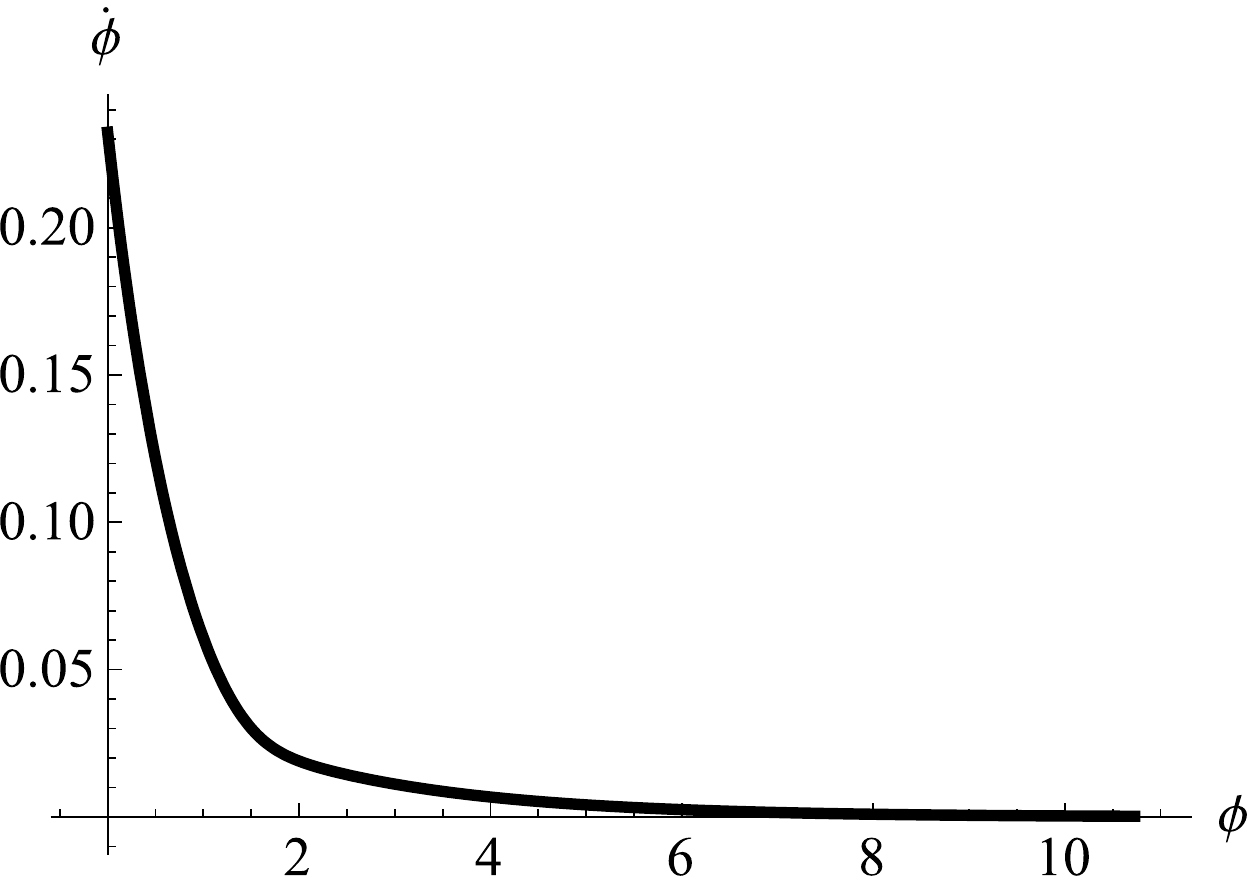}}
\end{minipage}
\hspace{0.5cm}
\begin{minipage}{.3\linewidth}
\subfloat[]{\includegraphics[scale=0.4]{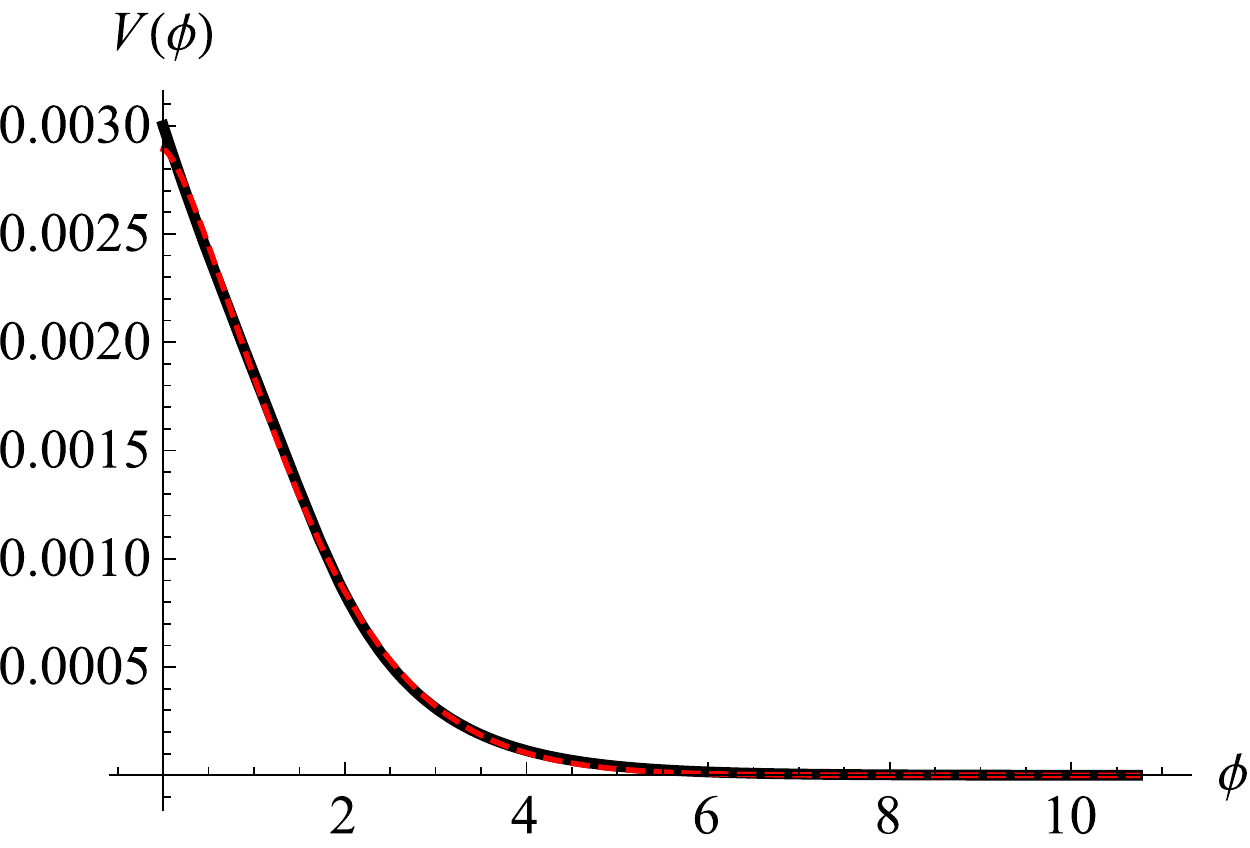}}
\end{minipage}\par\medskip
\begin{minipage}{.3\linewidth}
\subfloat[]{\includegraphics[scale=0.3]{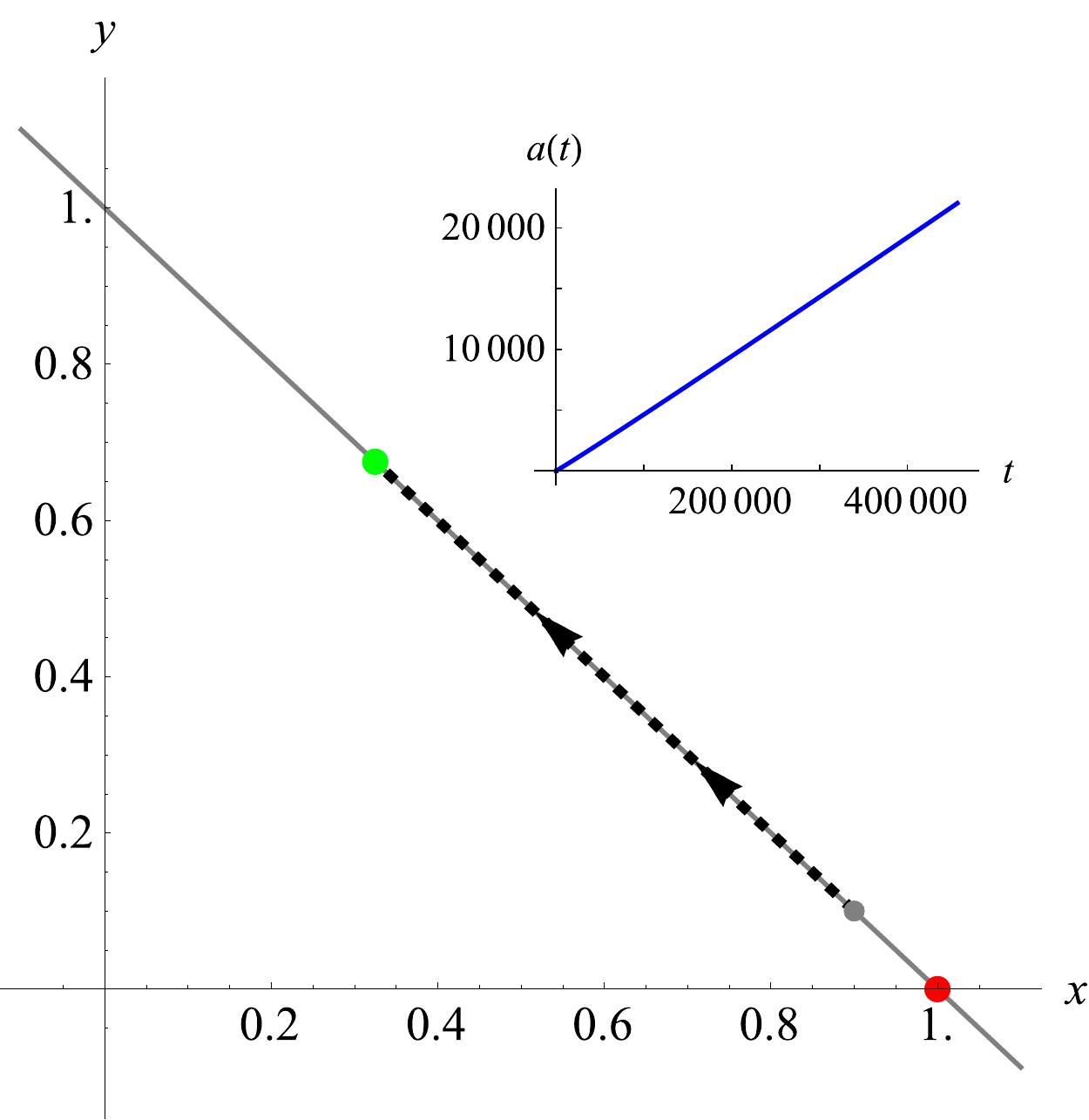}}
\end{minipage}
\begin{minipage}{.3\linewidth}
\subfloat[]{\includegraphics[scale=0.4]{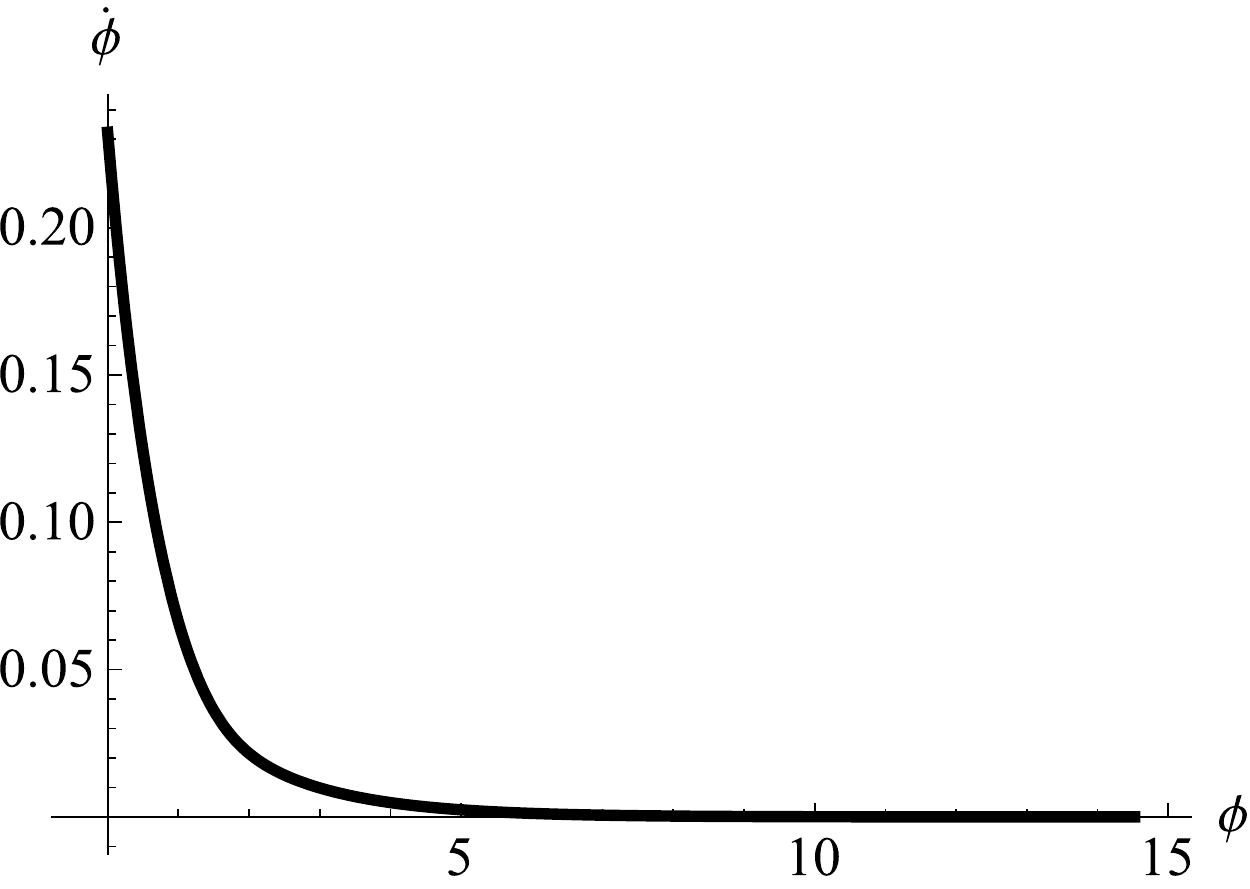}}
\end{minipage}
\hspace{0.5cm}
\begin{minipage}{.3\linewidth}
\subfloat[]{\includegraphics[scale=0.4]{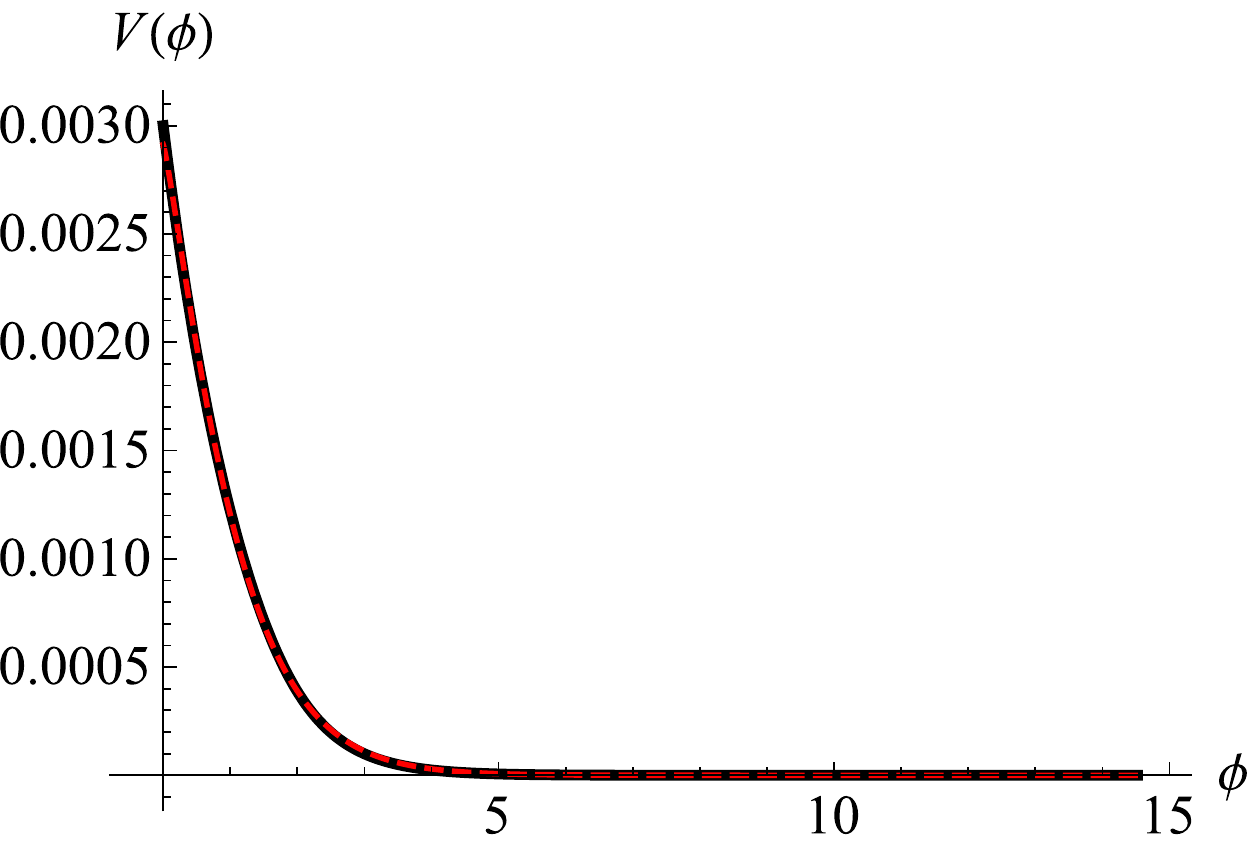}}
\end{minipage}
\caption{Zeroth-order EFT phase space and SSF realizations, for $\lambda_0 = 1$ (top row) and $\lambda_0 = 1.95$ (bottom row). In both cases, we set $(x (\tau_i), y (\tau_i)) = (0.9, 0.1)$ and $H (\tau_i) = 0.1$ (in units of $M_{\rm pl}$). In each row, the first plot displays the system's trajectory through the EFT phase space, with $a (t)$ displayed in the inset; the second plot displays $\dot{\phi}$ vs.~$\phi$; and the third plot displays $V (\phi)$ as obtained parametrically from the EFT dynamical system (black) and as a fit to the form of $V (\phi)$ in Eq.~(\ref{EQN:FFfit}) (red dashed). Parameters for each fit are given in Table~\ref{TAB:SyZero}. In plots b, c, e, and f, time evolution flows from left to right. Each plot is evolved forward for a total of 10 efolds.
}
\label{FIG:SyZero}
\end{figure*}

Next we consider zeroth-order systems that do not begin at a fixed point, but whose initial conditions satisfy $x (t_i) > y (t_i)$ and whose ensuing trajectories satisfy $x (t) , y (t) \geq 0$. Given the form of {\bf FP0b} in Table~\ref{TAB:ZO}, we consider two cases: $\lambda_0 = 1$ and $\lambda_0 = 1.95$. For each of these values, {\bf FP0b} serves as an inflationary fixed point, though for $\lambda_0 = 1.95$, {\bf FP0b} lies near the edge of the inflationary region ($x < 1/3)$. For both $\lambda_0 = 1$ and $\lambda_0 = 1.95$, we may follow the evolution of the system through phase space, and fit $V (\phi (t))$ from the behavior of $\phi (t)$, as shown in Fig.~\ref{FIG:SyZero}. In Fig.~\ref{FIG:SyZero} we show results for the case $x (\tau_i) = 0.9$; the corresponding plots for 
$x (\tau_i) = 0.6$ to $x (\tau_i) = 0.8$ appear quite similar, albeit with slightly different inferred best-fit values for the parameters $V_0$, $\alpha$, and $\beta$ that appear in Eq.~(\ref{EQN:FFfit}). In Table~\ref{TAB:SyZero} we present best-fit values for $V_0$, $\alpha$, and $\beta$ for both $\lambda_0 = 1$ and $\lambda_0 = 1.95$, as we vary $x (\tau_i)$ between $0.6$ and $0.9$.\footnote{The best-fit values for $V_0$, $\alpha$, and $\beta$ that are inferred for a given trajectory through the EFT phase space depend on the portion of the trajectory that is considered. In particular, one may find modest differences in the inferred values if one fits the system's trajectory beginning at initial time $t_i$, or if one only fits some portion of the trajectory after the system has begun to inflate. Likewise, one finds modest shifts in the best-fit values depending on the duration of a given trajectory that is considered. 
Unless otherwise specified, throughout our analysis we present best-fit values for $V_0$, $\alpha$, and $\beta$ based on fits that begin at $t_i$ and persist for $10$ efolds of expansion (not necessarily inflation). 
}

%
\begin{table}[htpb!]
\begin{center}
\begin{tabular}{ c | c|c|c|c}
 $\lambda_0$ & $x(\tau_{i})$ & $V_{0}$ & $\alpha$ & $\beta$ \\ \Xhline{2pt}
 $1$ &$0.9$ & $2.89 \times 10^{-3}$ & 0.452 & 1.44 \\
 $1$ & $0.8$ & $6.09 \times 10^{-3}$ & 0.608 & 1.28 \\
 $1$ & $0.7$ & $9.02 \times 10^{-3}$ & 0.665 & 1.26 \\
 $1$ &$0.6$ & $1.20 \times 10^{-2}$ & 0.729 & 1.21 \\
 \hline
 $1.95$ &  $0.9$ & $2.91 \times 10^{-3}$ & 0.884 & 1.19 \\
 $1.95$ & $0.8$ & $5.87 \times 10^{-3}$ & 0.994 & 1.17 \\
 $1.95$ & $0.7$ & $8.90 \times 10^{-3}$ & 1.10 & 1.14 \\
 $1.95$ & $0.6$ & $1.19 \times 10^{-2}$ & 1.18 & 1.11 \\
\end{tabular}
\caption{Best-fit values for the parameters $V_0$, $\alpha$, and $\beta$ for $V (\phi)$ in Eq.~(\ref{EQN:FFfit}) as obtained parametrically for the zeroth-order EFT dynamical system, with $H (\tau_i) = 0.1$ (in units of $M_{\rm pl}$). The values for $x (\tau_i) = 0.9$ correspond to the dashed, red curves in Figs.~\ref{FIG:SyZero}c,f for $\lambda_0 = 1$ and $\lambda_0 = 1.95$, respectively.}
\label{TAB:SyZero}
\end{center}
\end{table}

Thus we see that for zeroth-order systems, trajectories that begin from kinetic-energy-dominated initial conditions can flow into inflationary states, along the constraint surface $x + y = 1$, towards {\bf FP0b}. Systems that begin at {\bf FP0b} evolve with an effective potential corresponding to power-law inflation, Eq.~(\ref{EQN:VPLI}), whereas trajectories that begin with more general initial conditions evolve with an effective potential that may be parameterized as in Eq.~(\ref{EQN:FFfit}).

\subsection{First-Order Systems}\label{SEC:FirstOrderSystems}

For first-order systems, we again consider two types of trajectories: those that begin (and hence remain) at inflationary fixed points, and those that begin with $x (t_i) > y (t_i)$ and which retain $x (t) , y (t) \geq 0$ throughout their subsequent evolution, so as to be compatible with SSF realizations.

As shown in Table~\ref{TAB:FO}, for first-order systems there exist at most four hyperbolic fixed points, at most two of which can be inflationary ({\bf FP1a} and {\bf FP1d}), and only one of which ({\bf FP1a}) corresponds to an inflationary attractor for $\lambda_1 > 0$. We therefore begin by analyzing SSF realizations that evolve at {\bf FP1a} and {\bf FP1d} before considering more general trajectories.

Simplest to analyze is evolution at fixed point {\bf FP1a}, which corresponds to $(x, y , \lambda_0) = (0, 1, 0)$. 
From Eqs.~(\ref{EQN:xSF}) and (\ref{EQN:Ep}) we note that $x = 0$
corresponds to $\dot{\phi} = \dot{H} = 0$, and hence $\phi (t) = \phi (t_i) = {\rm constant}$ and $H (t) = H (t_i) = {\rm constant}$. With $\dot{H} = 0$, such evolution corresponds to an unending de Sitter phase.

To explore the trajectory that corresponds to {\bf FP1d}, we fix $0<\lambda_{1}<3$, set $x (t_i) = \lambda_1 / 9$, $y (t_i) = 1 - x (t_i)$, and follow the system for times $t \in [ t_i, t_f]$. From Eqs.~(\ref{EQN:phiSSF}) and (\ref{EQN:VSSF}), assuming $\dot{\phi}(t_i)>0$, we have
\begin{align}
\phi (t) &= \phi(t_{i})+\int_{t_{i}}^{t}dt'\sqrt{\frac{2\lambda_{1}}{3}\Mpl^2 H^2(t')},\label{EQN:phiSSFFP1d}\\
V(t)&= \left( \frac{9-\lambda_{1}}{3} \right) \Mpl^2 H^{2}(t) ,\label{EQN:VSSFFP1d}
\end{align}
and, from Eq.~(\ref{EQN:Ep}), 
\begin{equation}\label{EQN:HxONE}
-\frac{\dot H}{H^2}=3 x = \frac{\lambda_{1}}{3} .
\end{equation}
Comparing Eqs.~(\ref{EQN:phiSSFFP1d})--(\ref{EQN:HxONE}) with Eqs.~(\ref{EQN:phiSSFFP2})--(\ref{EQN:HzeroSSF}), we see that evolution of a first-order system at {\bf FP1d} is identical to that of a zeroth-order system at {\bf FP0b}, under the substitution
\begin{equation}\label{EQN:FZident}
\frac{\lambda_{1}}{9}\to\frac{\lambda_{0}}{6}.
\end{equation}
We immediately find
\begin{equation}
H(t)=\frac{3}{\lambda_{1}}\left(t-t_{i}+\frac{3}{\lambda_{1}H(t_{i})}\right)^{-1}  \label{EQN:HSSFFP1d}
\end{equation}
and
\begin{equation}
V(\phi)= V_{0}\exp\left(-\sqrt{\frac{2\lambda_{1}}{3}}\frac{\phi}{\Mpl}\right),\label{EQN:VphiPLIrequiredONE}
\end{equation}
where
\begin{equation}
 V_{0}\equiv \left( \frac{9-\lambda_{1}}{3} \right) \Mpl^2 H^2(t_{i})\exp\left[\sqrt{\frac{2\lambda_{1}}{3}} \, \frac{\phi(t_{i})}{\Mpl}\right].\label{EQN:VzeroPLIrequiredONE}
\end{equation}
This agrees with Eq.~(\ref{EQN:VPLI}) for the potential for power-law inflation provided that 
\begin{equation}
p\equiv\frac{3}{\lambda_{1}}.\label{EQN:EquivAGAINONE}
\end{equation}
Again we define a new time coordinate $u(t)$ by 
\begin{equation}\label{EQN:utimeONE}
u(t)\equiv t-t_{i}+\frac{3}{\lambda_{1}H(t_{i})} ,
\end{equation}
in terms of which the evolution of the scalar field may be written
\begin{equation}
\frac{\phi(u)}{\Mpl}=\sqrt{2p}\ln\left(
\sqrt{\frac{V_{0}}{p{(3p-1)}}}
\frac{u}{\Mpl}\right),
\label{EQN:phiPLIfinalONE}
\end{equation}
in agreement with Eq.~(\ref{EQN:phiPLI}). Finally, the scale factor is given by
\begin{equation}
a(u)=a(u_{i})\left(\frac{u}{u_{i}}\right)^{p},
\label{EQN:aPLIfinalONE}
\end{equation}
again using $u_i = u (t_i)$, in agreement with Eq.~(\ref{EQN:aPLI}). Hence we find that such a solution is inflationary ($p>1$), given $p = 3/\lambda_{1}$ and $0 < \lambda_{1}<3$.

We now turn to an exploration of SSF realizations at first order, where, again, initial conditions are chosen such that the initial kinetic-energy density is dominant. We consider scenarios that flow into inflation, and in particular, into {\bf FP1a} (the pure de Sitter attractor). As we will see, one can fit the same general functional form, Eq.~(\ref{EQN:FFfit}), to a variety of initially kinetic-energy dominated trajectories that flow into inflation for the first order, just as we had found in the zeroth-order case. 

\begin{figure*}[htp!]
\begin{minipage}{.3\linewidth}
\subfloat[]{\includegraphics[scale=0.4]{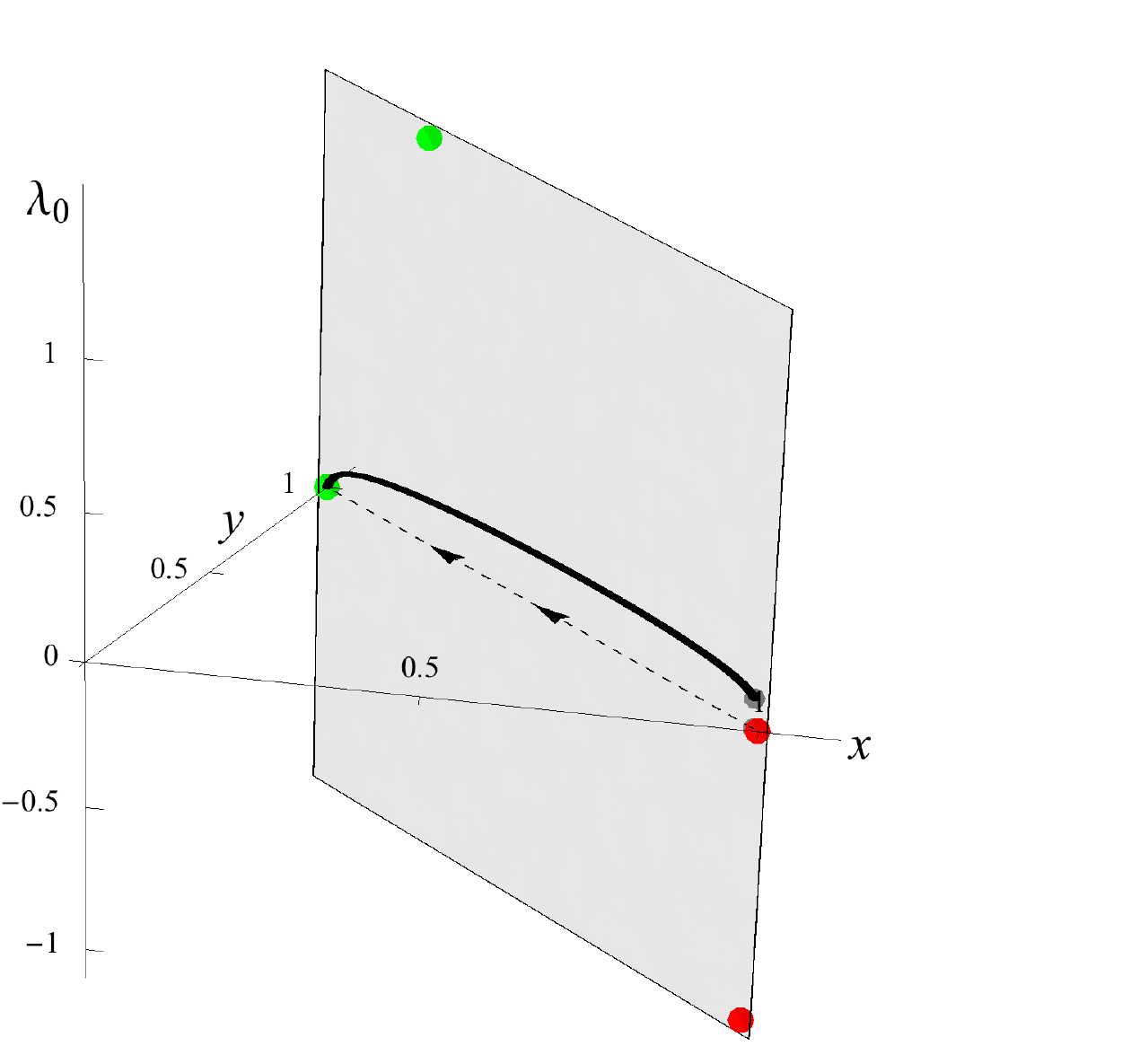}}
\end{minipage}
\begin{minipage}{.3\linewidth}
\subfloat[]{\includegraphics[scale=0.4]{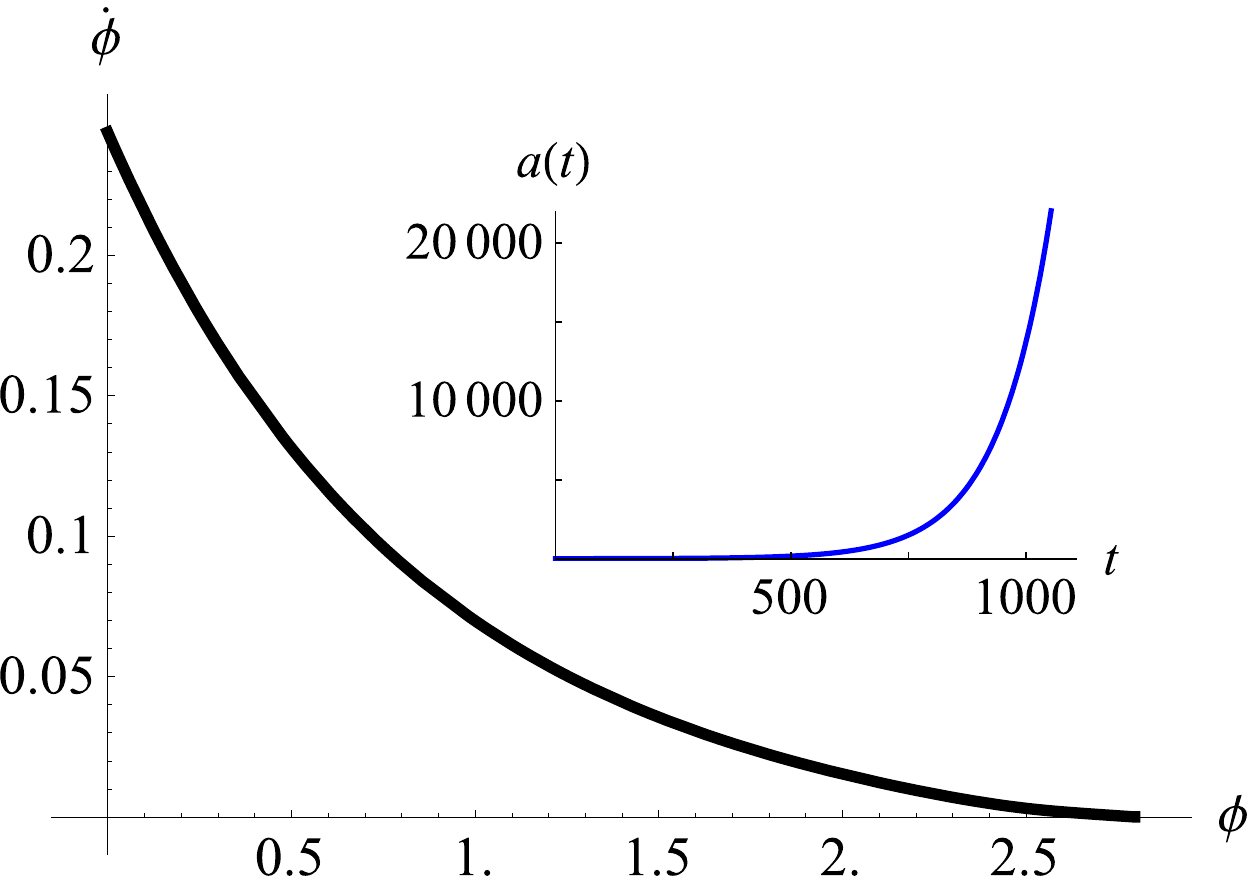}}
\end{minipage}
\hspace{0.5cm}
\begin{minipage}{.3\linewidth}
\subfloat[]{\includegraphics[scale=0.4]{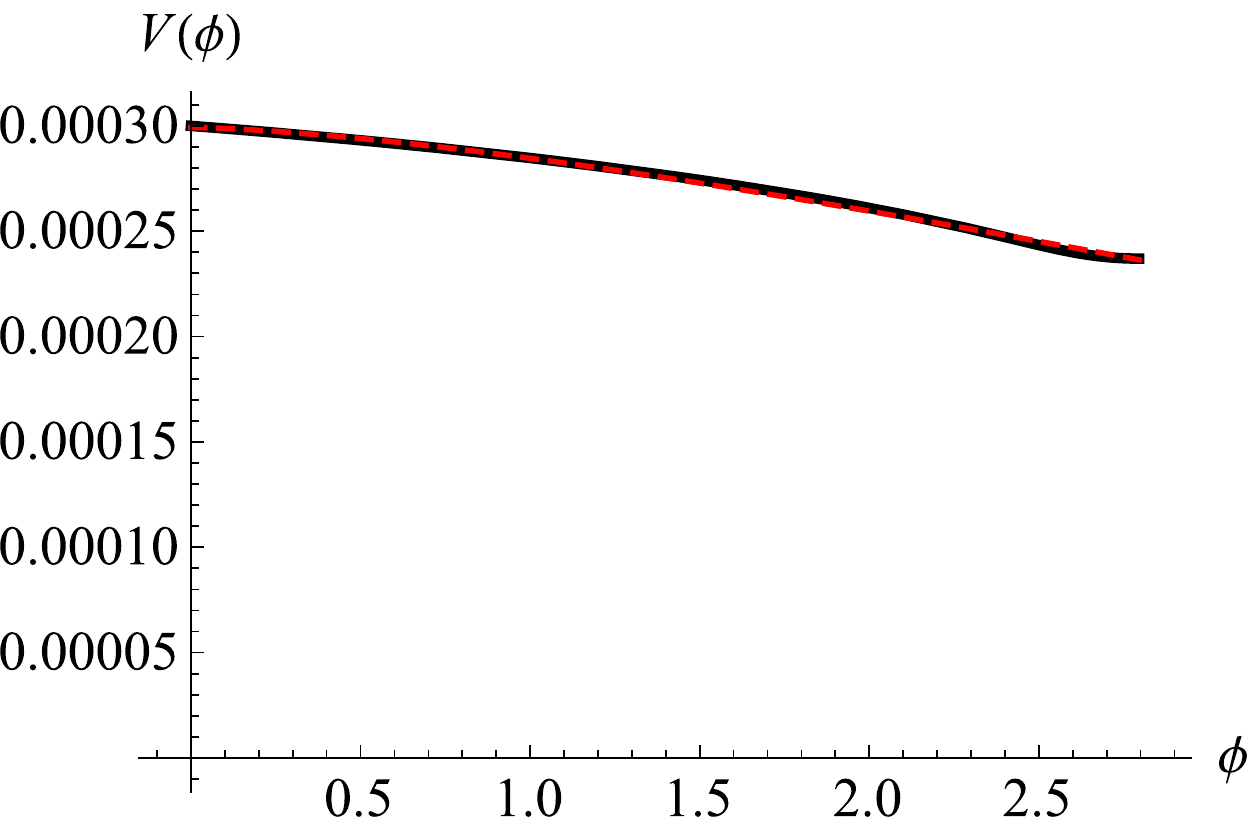}}
\end{minipage}\par\medskip
\begin{minipage}{.3\linewidth}
\subfloat[]{\includegraphics[scale=0.4]{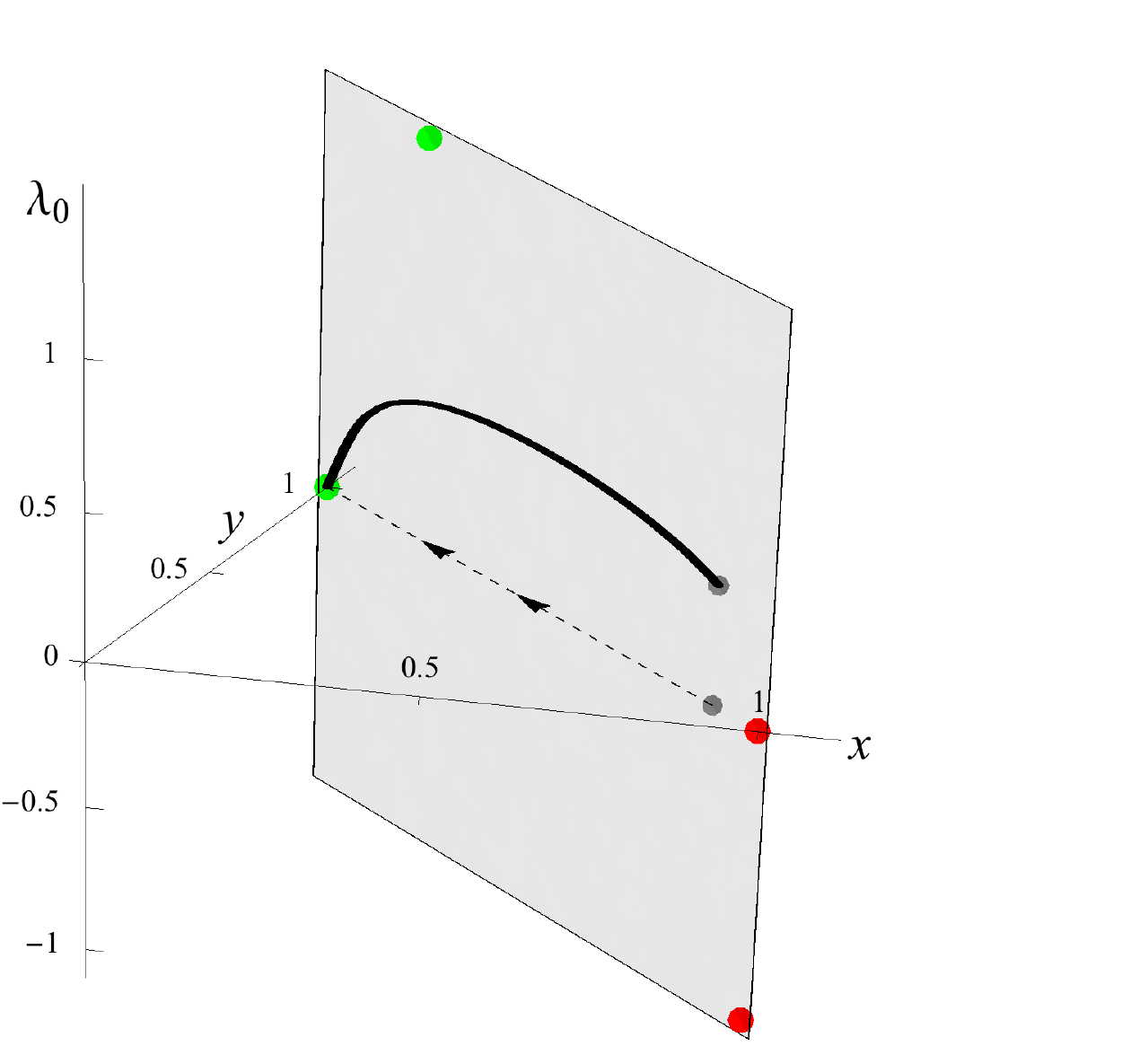}}
\end{minipage}
\begin{minipage}{.3\linewidth}
\subfloat[]{\includegraphics[scale=0.4]{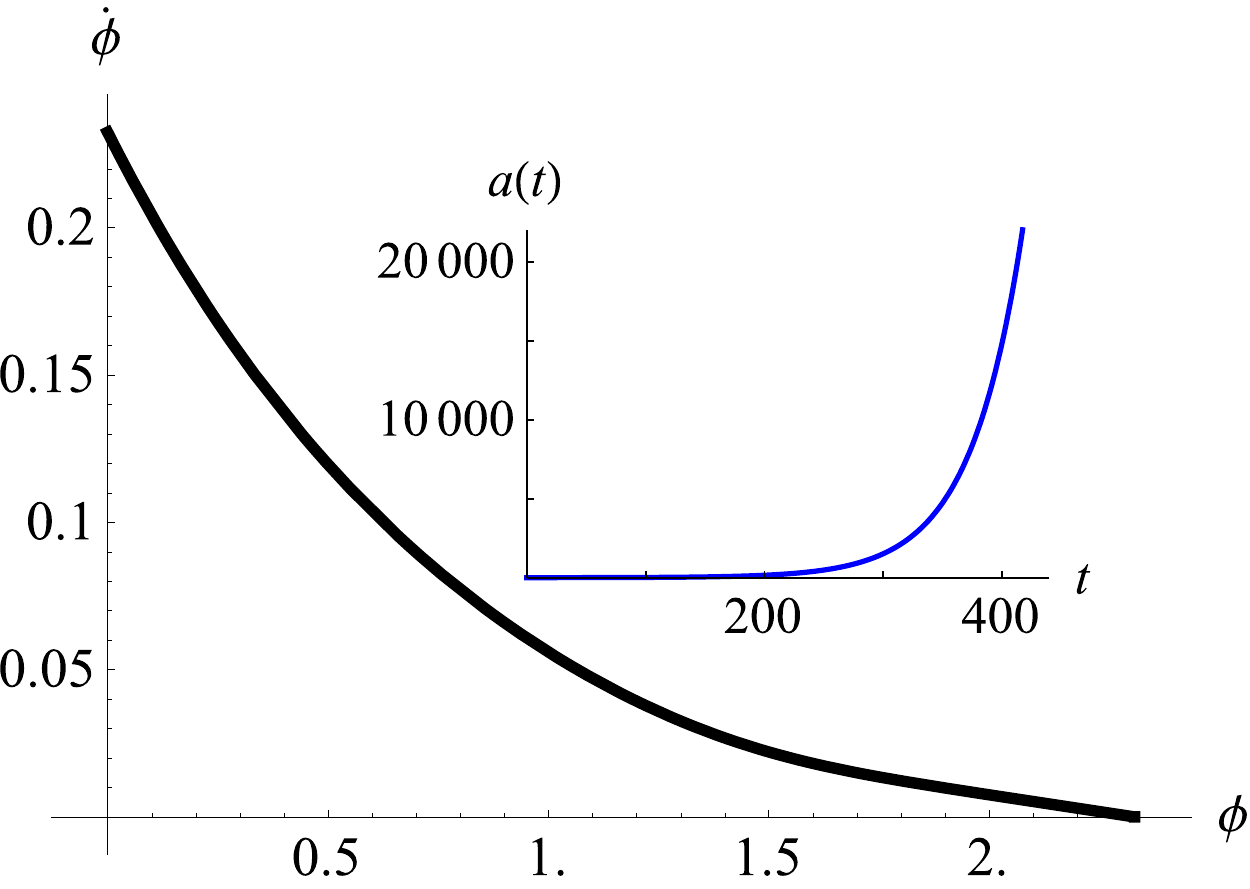}}
\end{minipage}
\hspace{0.5cm}
\begin{minipage}{.3\linewidth}
\subfloat[]{\includegraphics[scale=0.4]{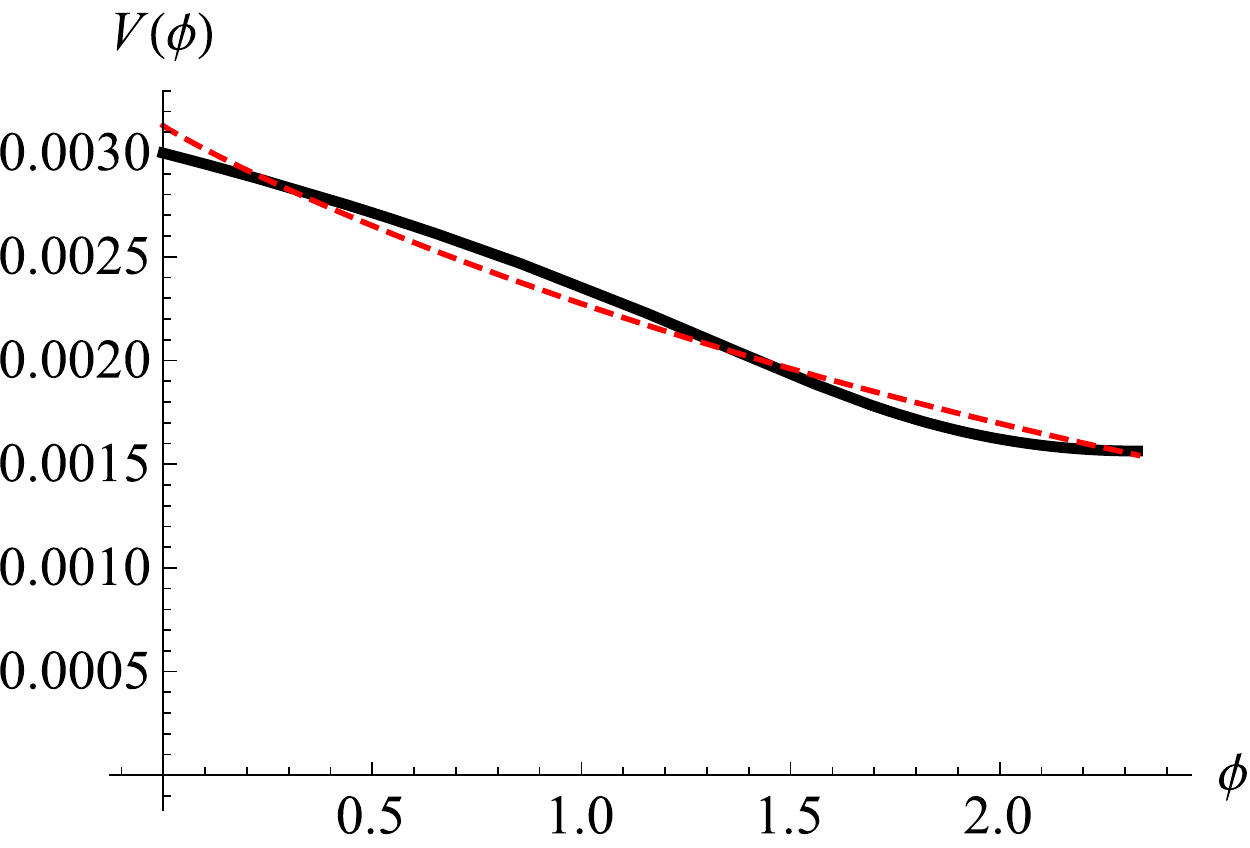}}
\end{minipage}\par\medskip
\begin{minipage}{.3\linewidth}
\subfloat[]{\includegraphics[scale=0.4]{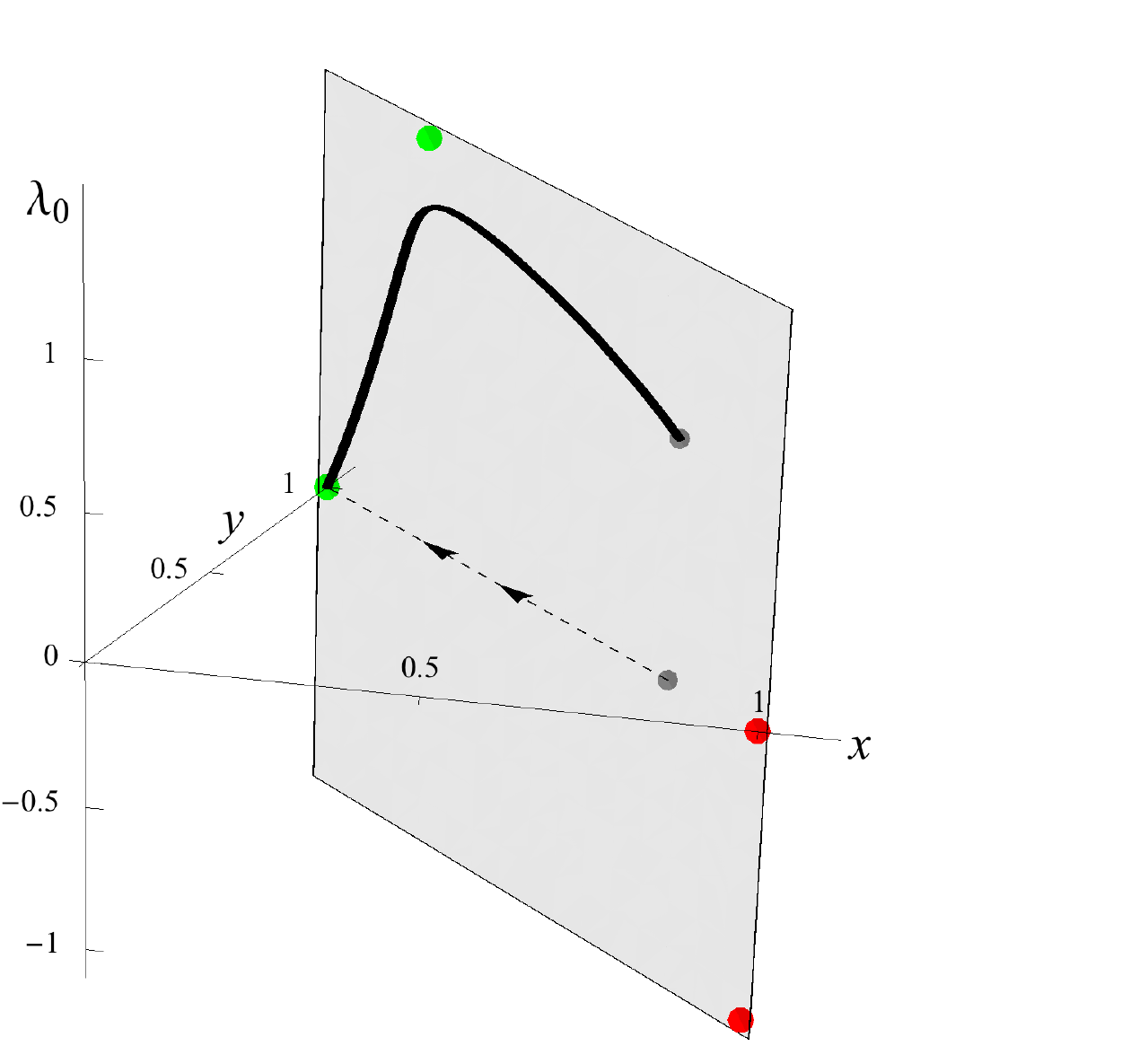}}
\end{minipage}
\begin{minipage}{.3\linewidth}
\subfloat[]{\includegraphics[scale=0.4]{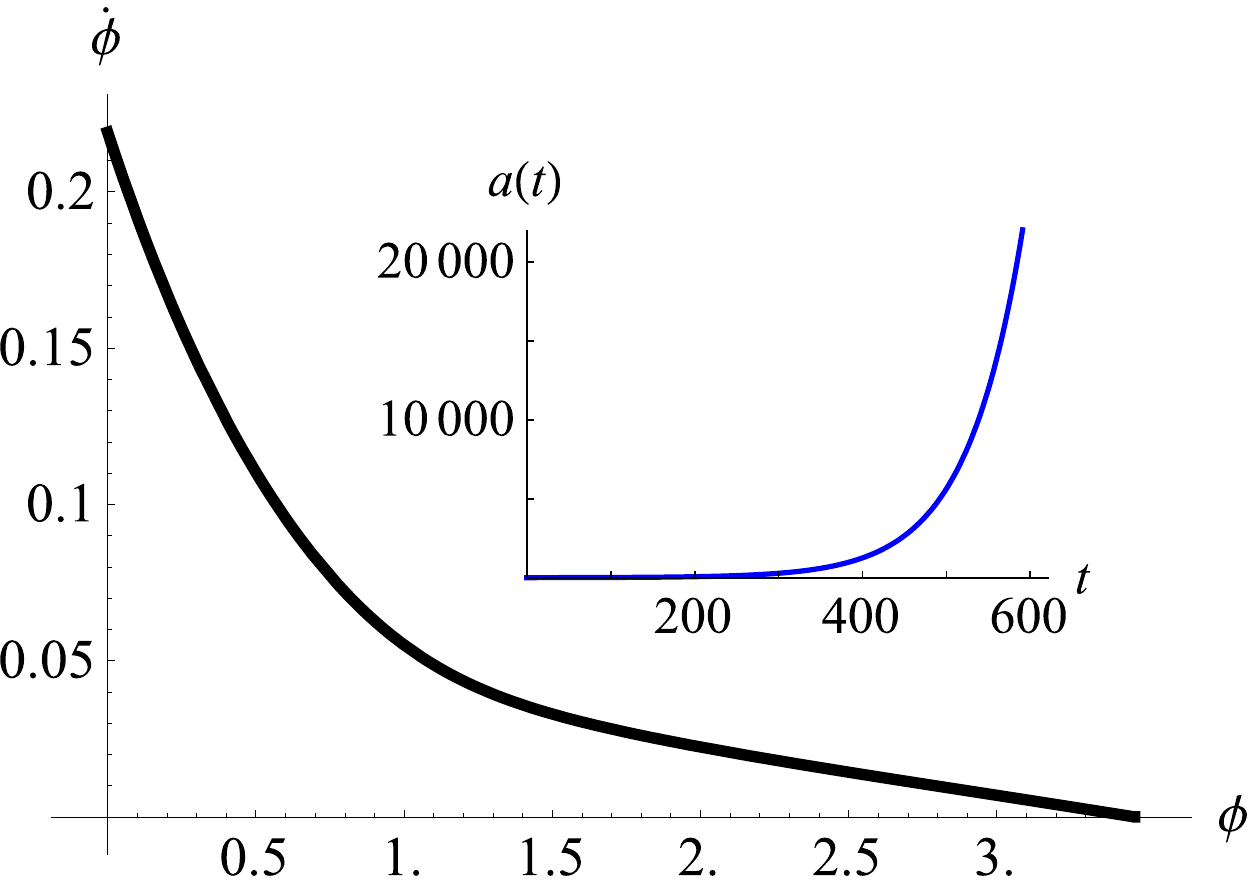}}
\end{minipage}
\hspace{0.5cm}
\begin{minipage}{.3\linewidth}
\subfloat[]{\includegraphics[scale=0.4]{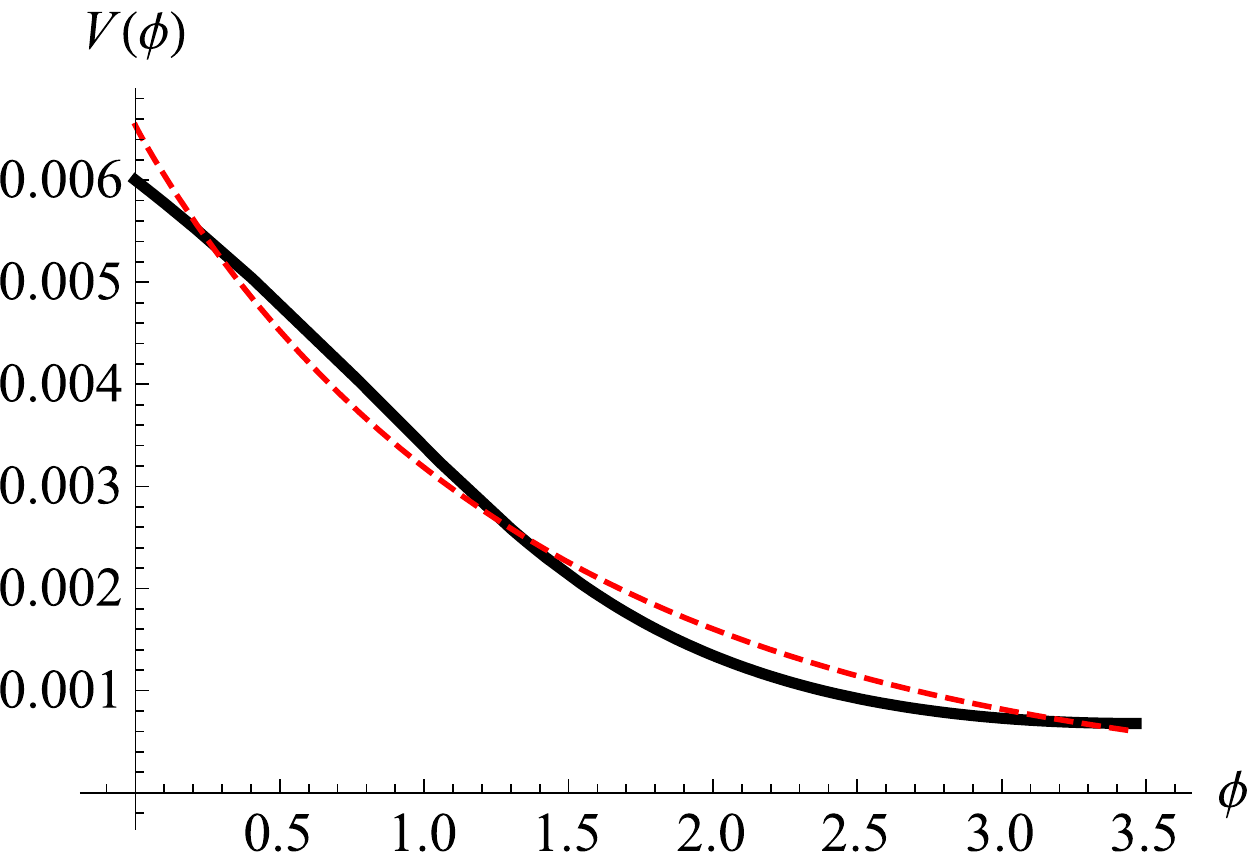}}
\end{minipage}
\caption{First-order EFT phase space and SSF realizations for $\lambda_{1}=2$ and $H(\tau_{i})=0.1$ (in units of $M_{\rm pl}$), for $(x (\tau_i), \lambda_0 (\tau_i)) = (0.99, 0.1)$ (top row), $(0.9, 0.4)$ (middle row), and ($0.8, 0.8$) (bottom row). In each row, the first plot displays the system's trajectory through the EFT phase space; the second plot displays $\dot{\phi}$ vs.~$\phi$, with $a (t)$ displayed in the inset; and the third plot displays $V (\phi)$ as obtained parametrically from the EFT dynamical system (black) and as a fit to the form of $V (\phi)$ in Eq.~(\ref{EQN:FFfit}) (red dashed). Parameters for each fit are given in Table~\ref{TAB:SyOneCase1abc}. For each EFT phase space, the trajectory starts out at the gray dot and flows towards the green (inflationary) attractor; red dots indicate non-inflationary fixed points. The `shadow' of this trajectory, as projected onto the $x$-$y$ plane, is shown as a dashed black line, with arrows indicating the direction of time evolution of the shadow. In the latter two plots in each row, time evolution flows from left to right. Each plot is evolved forward for a total of 10 efolds.
}
\label{FIG:SyOneCase1abc}
\end{figure*}

\begin{figure*}[htp!]
\begin{minipage}{.3\linewidth}
\subfloat[]{\includegraphics[scale=0.4]{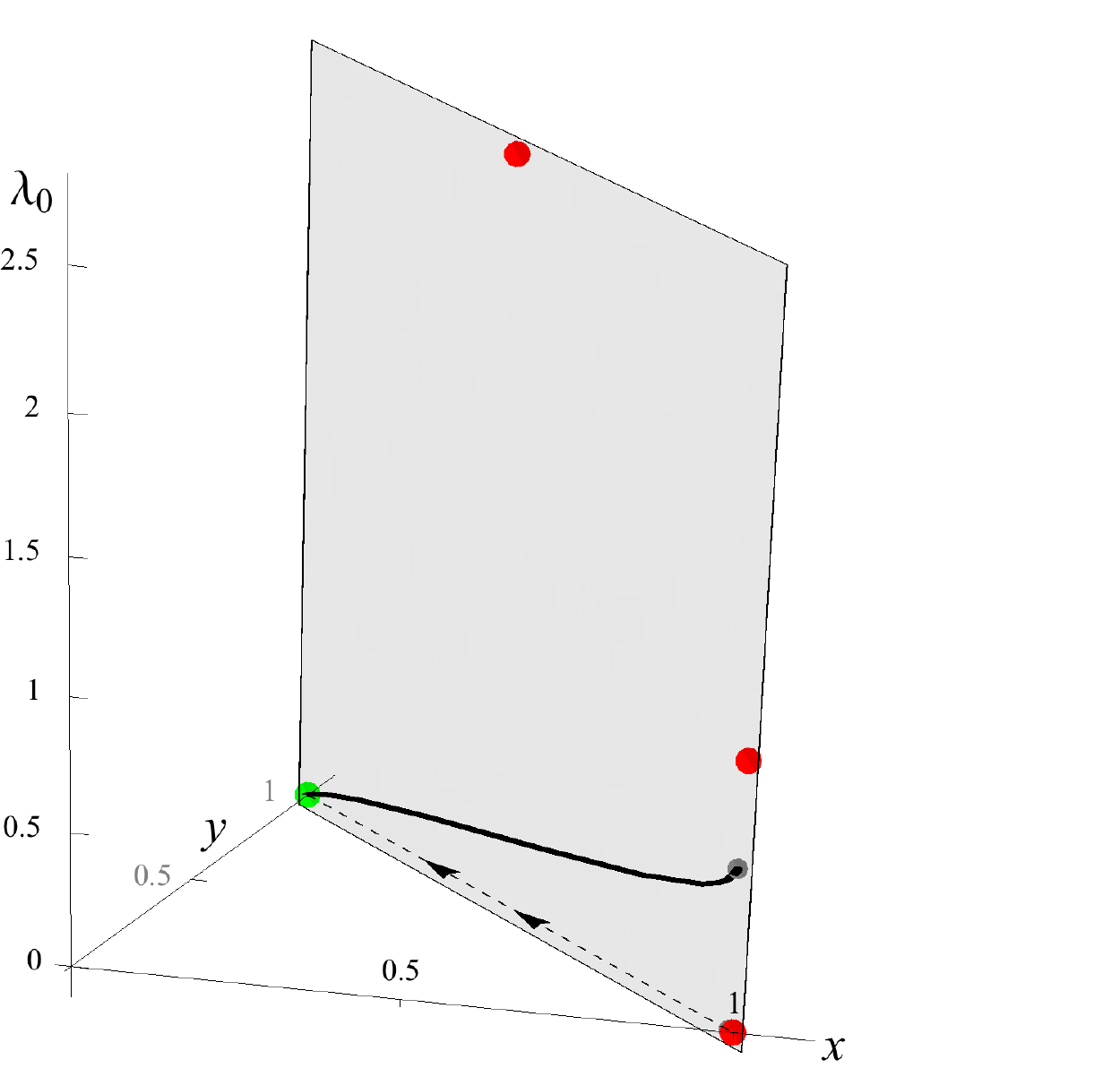}}
\end{minipage}
\begin{minipage}{.3\linewidth}
\subfloat[]{\includegraphics[scale=0.4]{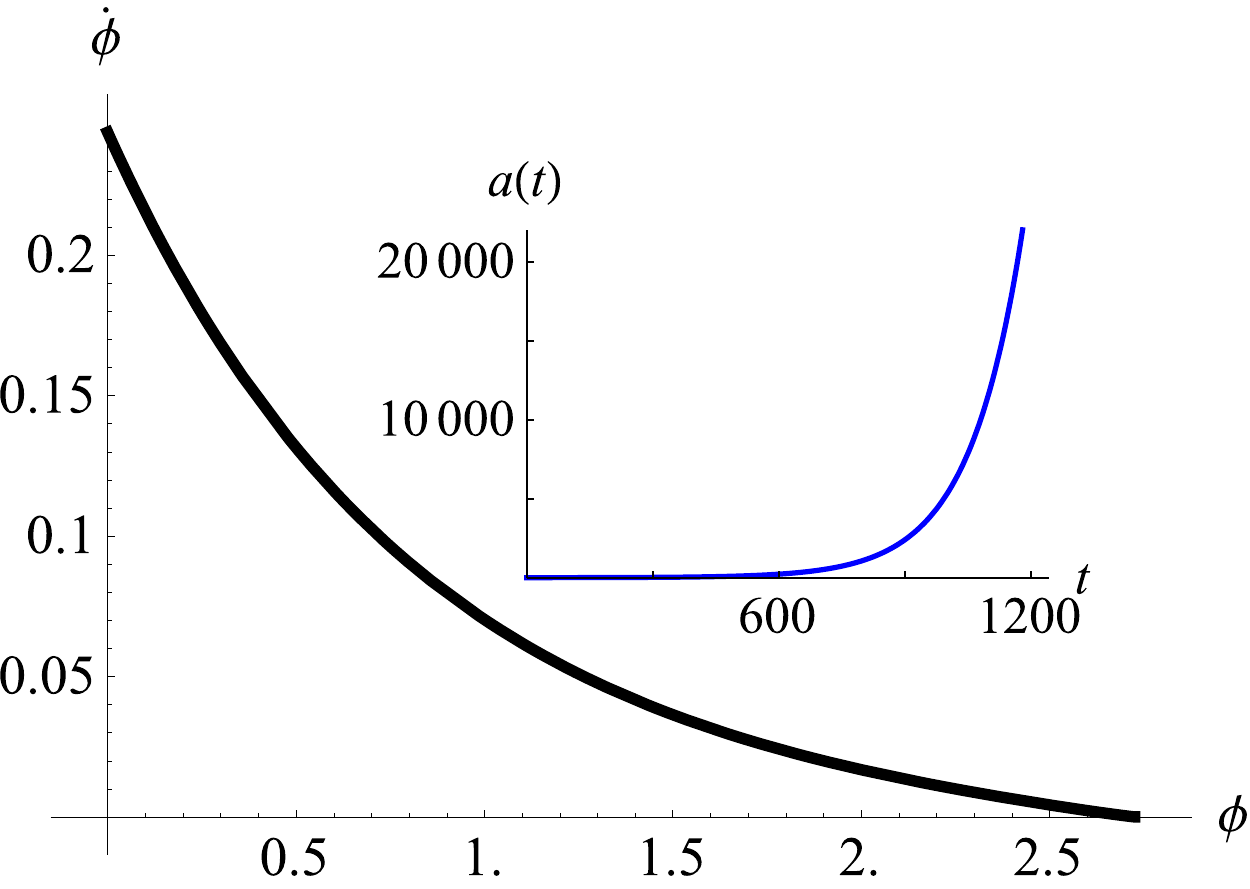}}
\end{minipage}
\hspace{0.5cm}
\begin{minipage}{.3\linewidth}
\subfloat[]{\includegraphics[scale=0.4]{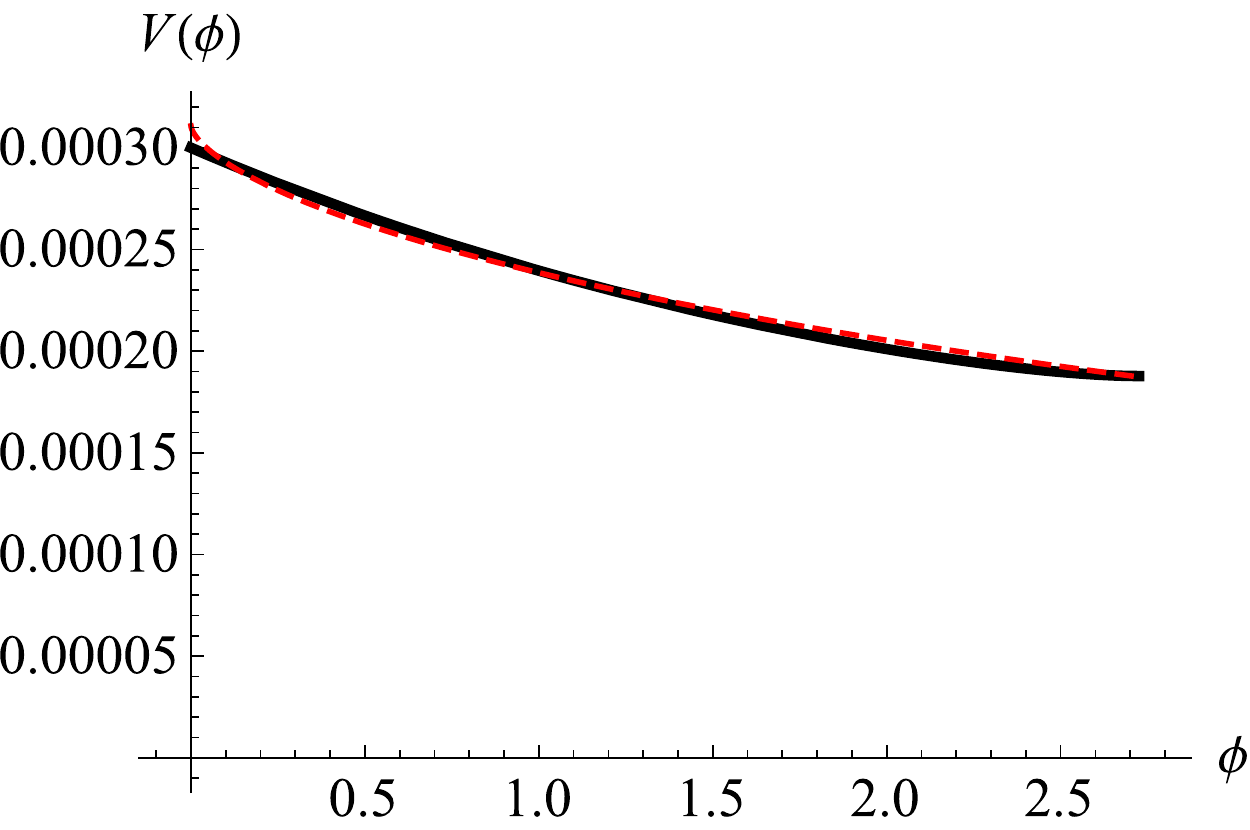}}
\end{minipage}\par\medskip
\begin{minipage}{.3\linewidth}
\subfloat[]{\includegraphics[scale=0.4]{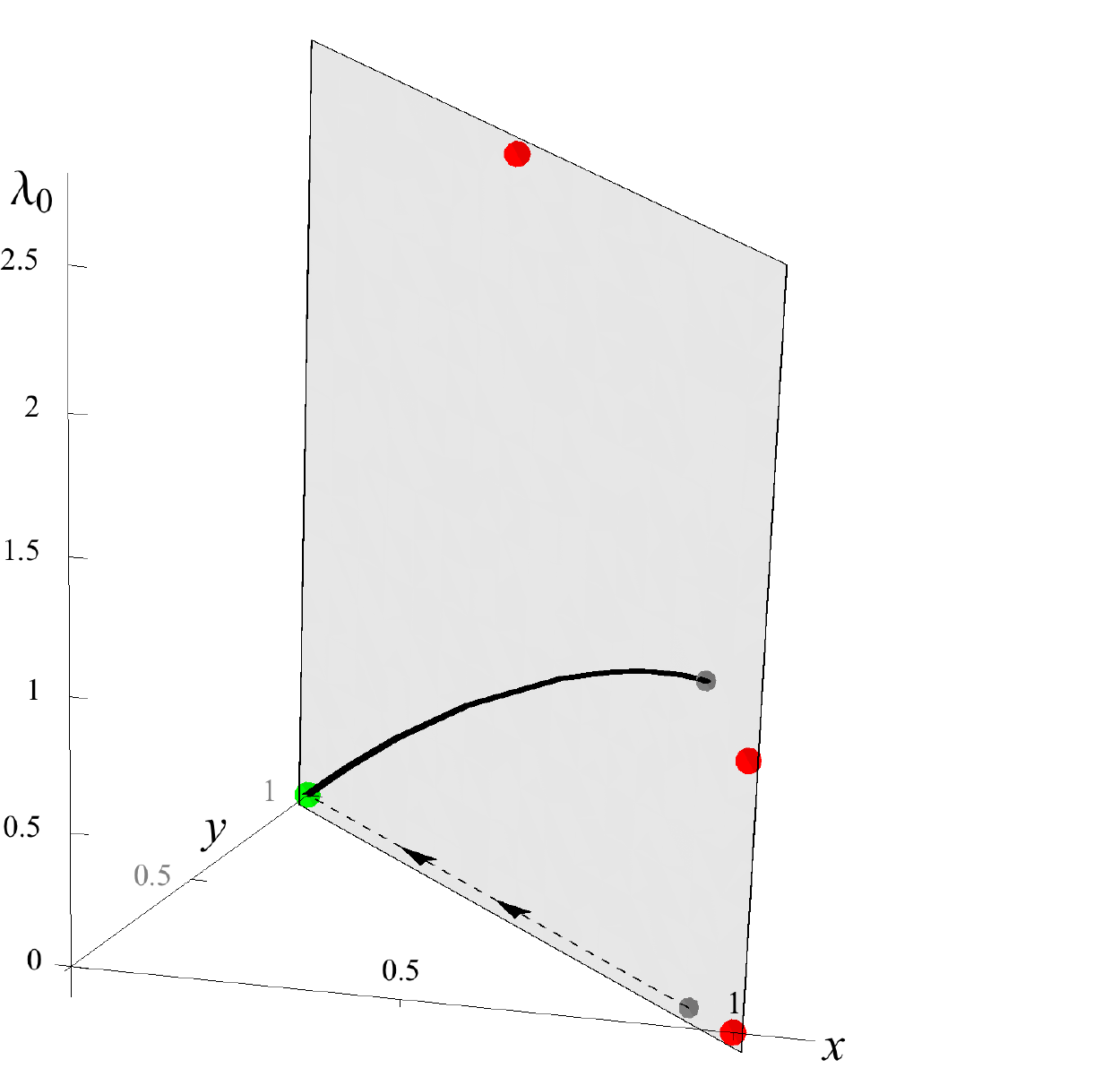}}
\end{minipage}
\begin{minipage}{.3\linewidth}
\subfloat[]{\includegraphics[scale=0.4]{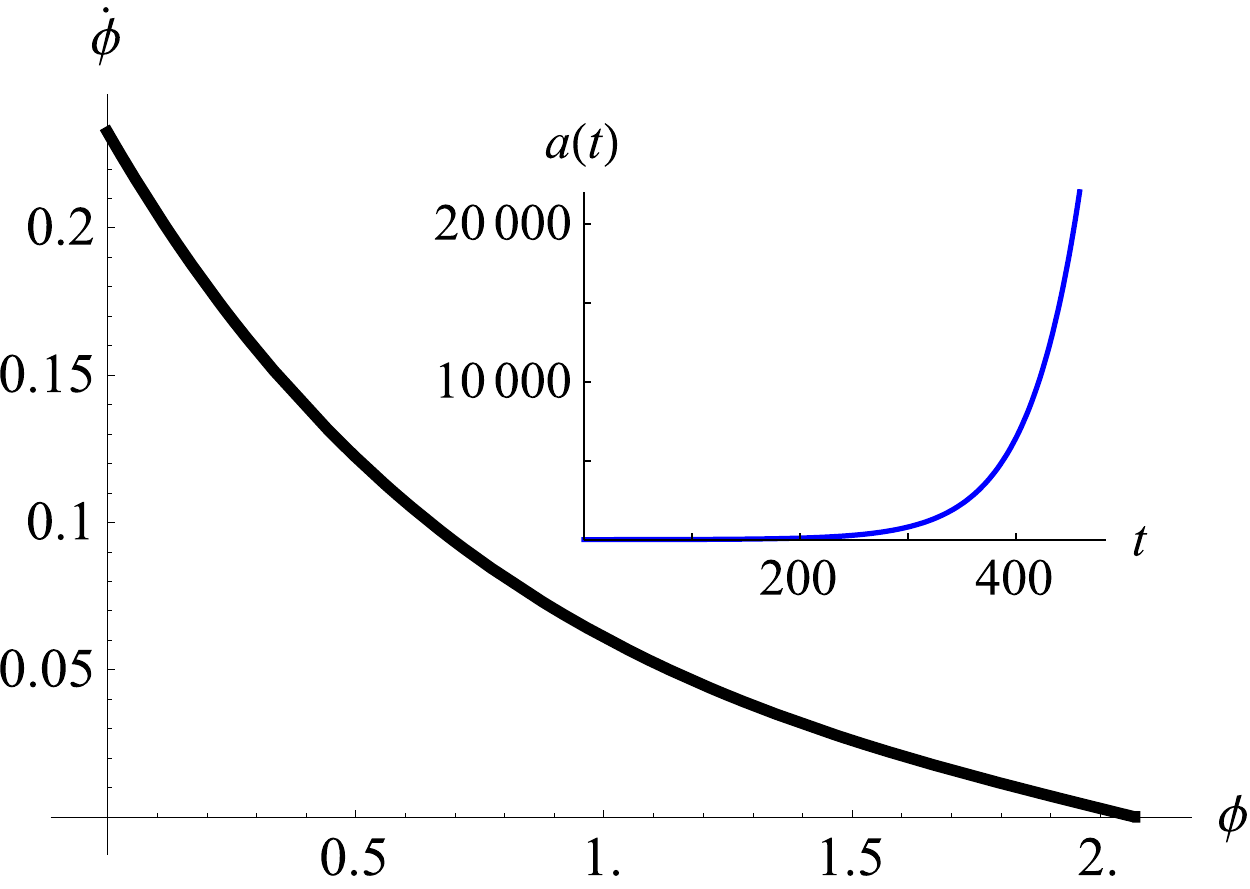}}
\end{minipage}
\hspace{0.5cm}
\begin{minipage}{.3\linewidth}
\subfloat[]{\includegraphics[scale=0.4]{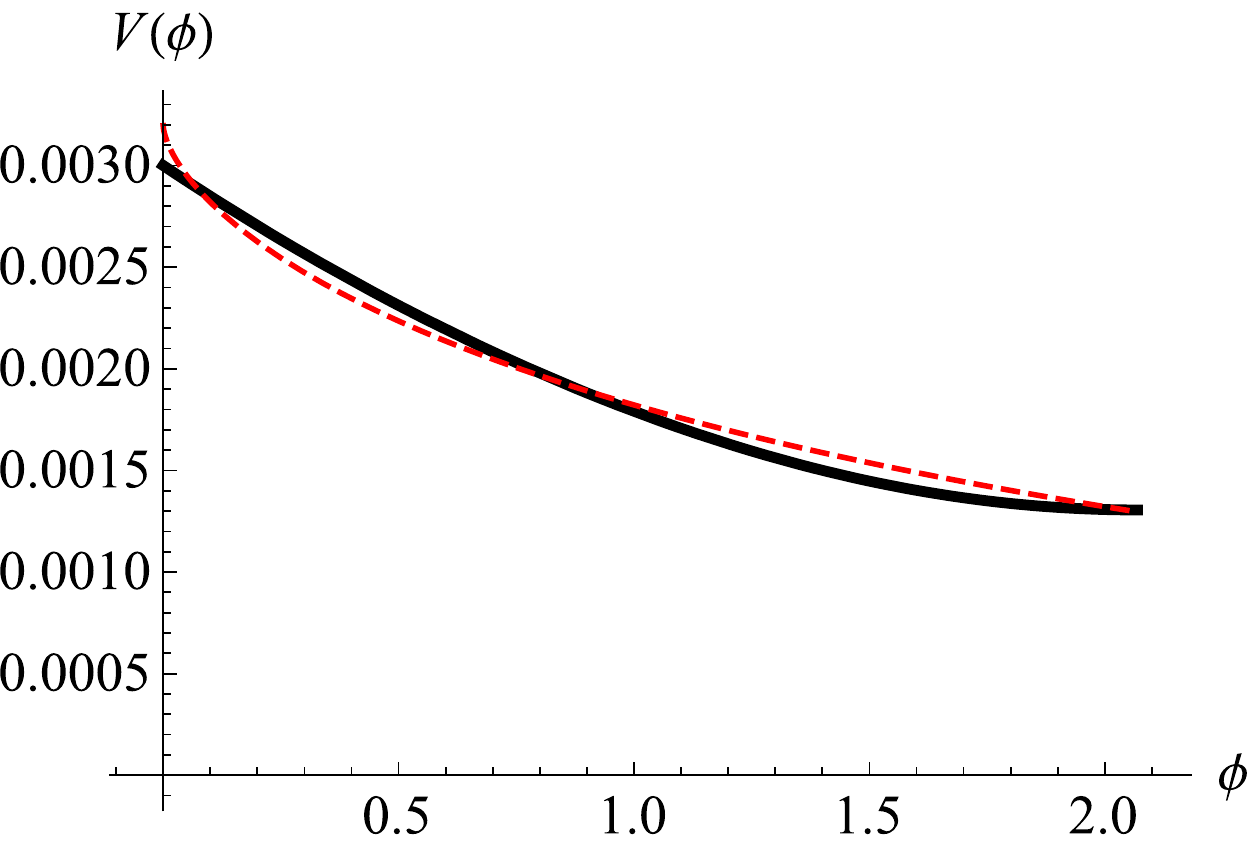}}
\end{minipage}\par\medskip
\begin{minipage}{.3\linewidth}
\subfloat[]{\includegraphics[scale=0.4]{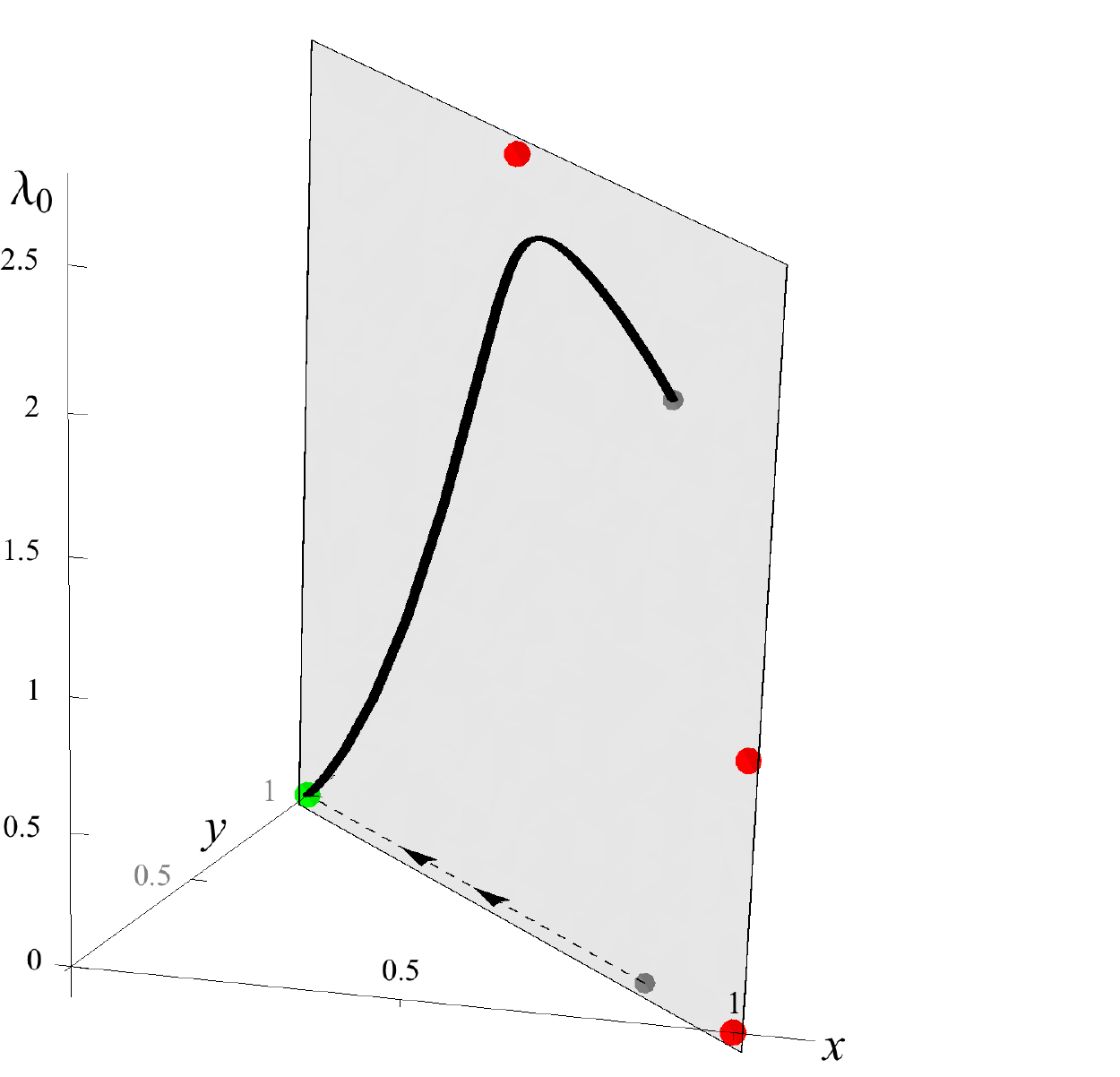}}
\end{minipage}
\begin{minipage}{.3\linewidth}
\subfloat[]{\includegraphics[scale=0.4]{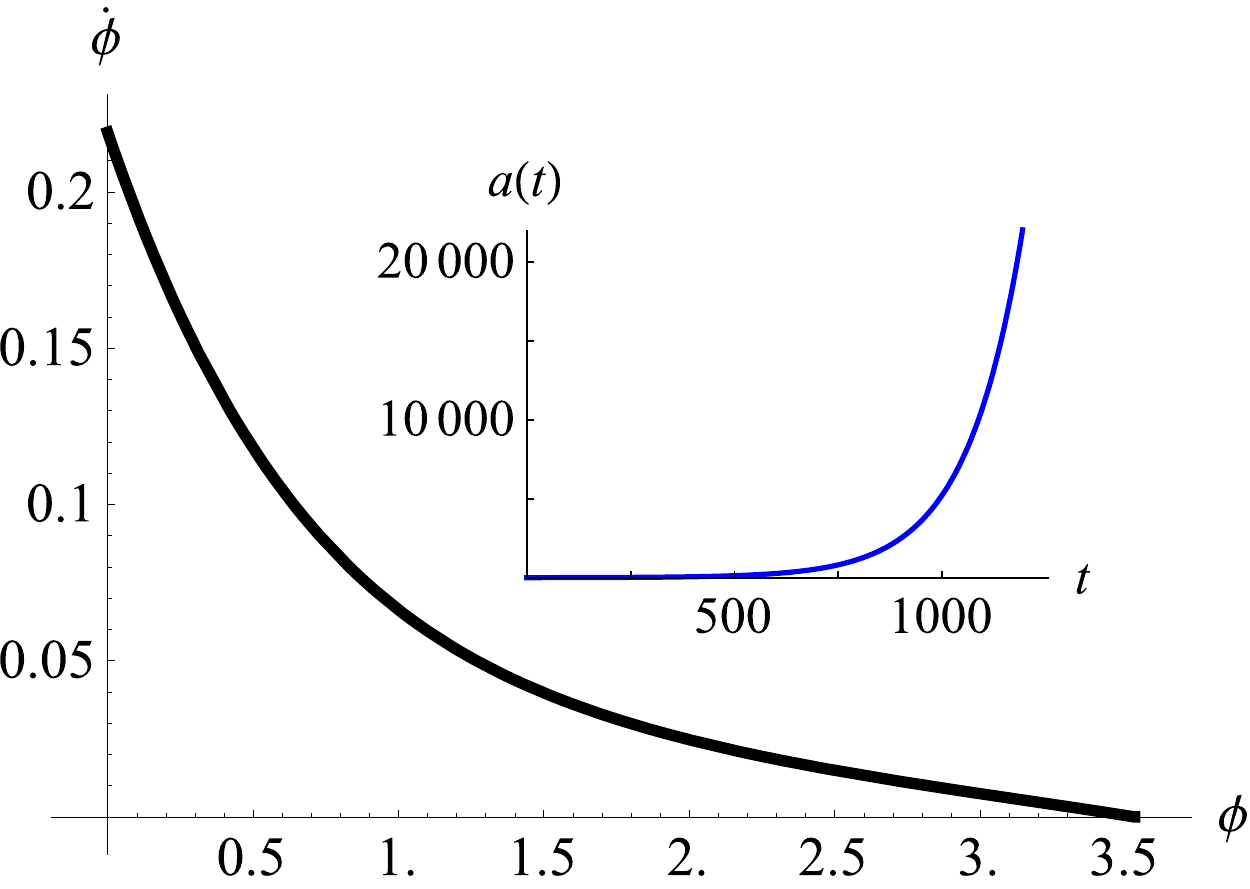}}
\end{minipage}
\hspace{0.5cm}
\begin{minipage}{.3\linewidth}
\subfloat[]{\includegraphics[scale=0.4]{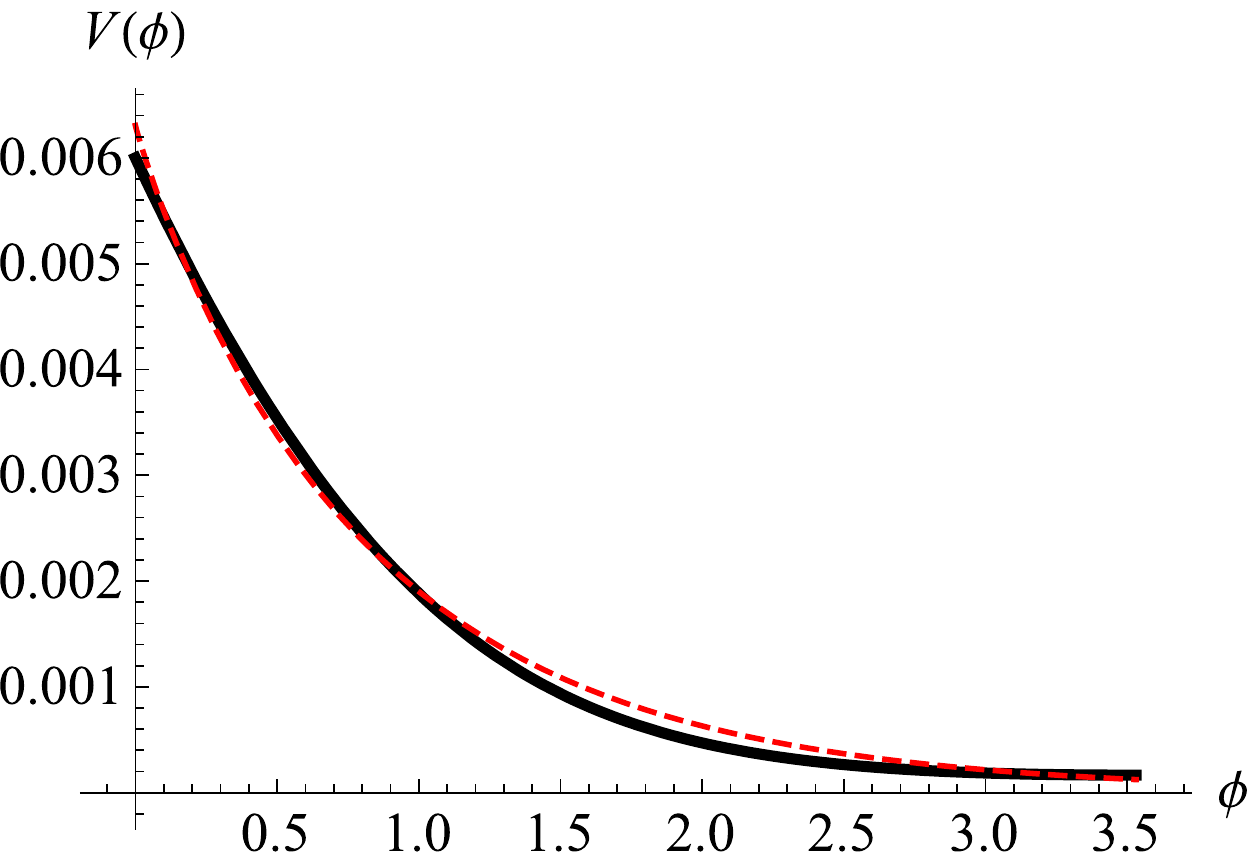}}
\end{minipage}
\caption{First-order EFT phase space and SSF realizations for $\lambda_{1}=4$ and $H(\tau_{i})=0.1$ (in units of $M_{\rm pl}$), for $(x (\tau_i), \lambda_0 (\tau_i)) = (0.99, 0.6)$ (top row), $(0.9, 1.2)$ (middle row), and ($0.8, 2.1$) (bottom row). In each row, the first plot displays the system's trajectory through the EFT phase space; the second plot displays $\dot{\phi}$ vs.~$\phi$, with $a (t)$ displayed in the inset; and the third plot displays $V (\phi)$ as obtained parametrically from the EFT dynamical system (black) and as a fit to the form of $V (\phi)$ in Eq.~(\ref{EQN:FFfit}) (red dashed). Parameters for each fit are given in Table~\ref{TAB:SyOneCase2abc}. For each EFT phase space, the trajectory starts out at the gray dot and flows towards the green (inflationary) attractor; red dots indicate non-inflationary fixed points. The `shadow' of this trajectory, as projected onto the $x$-$y$ plane, is shown as a dashed black line, with arrows indicating the direction of time evolution of the shadow. In the latter two plots in each row, time evolution flows from left to right. Each plot is evolved forward for a total of 10 efolds.
}
\label{FIG:SyOneCase2abc}
\end{figure*}

In what follows, we present results for $\lambda_{1}=2$ and $\lambda_{1}=4$. These values were chosen so that {\bf FP1d} is an inflationary fixed point (when $\lambda_{1}=2$), or a non-inflationary fixed point (when $\lambda_{1}=4$). Given the richer range of behaviors that are possible within the expanded phase space for first-order systems compared to zeroth-order ones, we consider a wider range of initial conditions $x (\tau_i)$ and $\lambda_0 (\tau_i)$ than we did for the zeroth-order system. In particular, for each $x (\tau_i) = 0.99, 0.9$, and $0.8$, we explore values $\lambda (\tau_i) > 0$ such that the ensuing trajectories are not deflected upward by {\bf FP1d} (as in Figs.~\ref{FIG:TrajOne_09}c,d). We likewise neglect cases with $\lambda_0 (\tau_i) < 0$, which are incompatible with SSF realizations. Best-fit values for $V_0$, $\alpha$, and $\beta$, with which we parameterize the SSF effective potential as in Eq.~(\ref{EQN:FFfit}), are shown in Table~\ref{TAB:SyOneCase1abc} (for $\lambda_1 = 2$) and Table~\ref{TAB:SyOneCase2abc} (for $\lambda_1 = 4$), as we vary $x (\tau_i)$ and $\lambda_0 (\tau_i)$. Corresponding trajectories are shown in Fig.~\ref{FIG:SyOneCase1abc} (for $\lambda_1 = 2$) and Fig.~\ref{FIG:SyOneCase2abc} (for 
$\lambda_1 = 4$).

\begin{table}[htp!]
\begin{center}
\begin{tabular}{ c | c | c | c | c}
  $x(\tau_{i})$ &$\lambda_{0}(\tau_{i})$& $V_{0}$ & $\alpha$ & $\beta$ \\ \Xhline{2pt}
 $0.99$& 0.1 & $2.99 \times 10^{-4}$ & 0.0502 & 1.51\\
 $0.99$& 0.2 & $3.03 \times 10^{-4}$ & 0.127 & 1.35\\
 $0.99$& 0.3 & $3.07 \times 10^{-4}$ & 0.215 & 1.36\\
 \Xhline{0.5pt}
 $0.9$& 0.2 & $3.10 \times 10^{-3}$ & 0.134 & 1.24\\
 $0.9$& 0.4 & $3.13 \times 10^{-3}$ & 0.320 & 0.938\\
 $0.9$& 0.6 & $3.22 \times 10^{-3}$ & 0.529 & 1.00\\ 
 \Xhline{0.5pt}
 $0.8$& 0.1 & $6.01 \times 10^{-3}$ & 0.0660 & 1.13\\
 $0.8$& 0.5 & $6.37 \times 10^{-3}$ & 0.409 & 0.806\\
 $0.8$& 0.8 & $6.54 \times 10^{-3}$ & 0.719 & 0.966 
\end{tabular}
\caption{Best-fit parameters for $V(\phi)$ of Eq.~(\ref{EQN:FFfit}) for a first-order system with $\lambda_1 = 2$ and $H (\tau_i) = 0.1$, in units of $M_{\rm pl}$. Parameters in the first, fifth, and ninth rows describe the dashed, red curves in Figs.~\ref{FIG:SyOneCase1abc}c, f, and i, respectively.}\label{TAB:SyOneCase1abc}
\end{center}
\end{table}

\begin{table}[h]
\begin{center}
\begin{tabular}{ c | c | c | c | c}
$x(\tau_{i})$ &$\lambda_{0}(\tau_{i})$& $V_{0}$ & $\alpha$ & $\beta$ \\ \Xhline{2pt}
$0.99$& 0.6 & $3.11 \times 10^{-4}$ & 0.264 & 0.648\\
$0.99$& 1.0 & $3.16 \times 10^{-4}$ & 0.476 & 0.747\\
$0.99$& 1.4 & $2.98 \times 10^{-4}$ & 0.650 & 1.15\\
\Xhline{0.5pt}
$0.9$& 0.6 & $3.09 \times 10^{-3}$ & 0.247 & 0.610\\
$0.9$& 1.2 & $3.20 \times 10^{-3}$ & 0.562 & 0.652\\
$0.9$& 1.8 & $3.19 \times 10^{-3}$ & 0.990 & 0.887\\
\Xhline{0.5pt}
$0.8$& 0.6 & $6.16 \times 10^{-3}$ & 0.241 & 0.582\\
$0.8$& 1.2 & $6.38 \times 10^{-3}$ & 0.537 & 0.593\\
$0.8$& 2.1 & $6.31 \times 10^{-3}$ & 1.20 & 0.942
\end{tabular}
\caption{Best-fit parameters for $V (\phi)$ of Eq.~(\ref{EQN:FFfit}) for a first-order system with $\lambda_1 = 4$ and $H (\tau_i) = 0.1$, in units of $M_{\rm pl}$. Parameters in the first, fifth, and ninth rows describe the dashed, red curves in Figs.~\ref{FIG:SyOneCase2abc}c, f, and i, respectively.}\label{TAB:SyOneCase2abc}
\end{center}
\end{table}

We see that for first-order systems, for trajectories that begin dominated by kinetic energy (with a select, but dynamically interesting set of values of $\lambda_{1}$), the effective scalar potential $V(\phi)$ again takes the simple form of Eq.~(\ref{EQN:FFfit}), and represents a variant of the potential for power-law inflation, Eq.~(\ref{EQN:VPLI}). One can show that the special role being played by power-law inflation continues at all higher orders as well. In Appendix~\ref{APP:secondorder}, we explore rudiments of the second-order system, and subsequently derive, in Appendix~\ref{APP:Mthorder}, analytical results for the general $M$th-order case. 

\section{Observational Constraints}
\label{SEC:observables}

For trajectories through the EFT phase space that are compatible with SSF realizations, we may consider whether they are compatible with observations. To relate the form of the effective potential $V (\phi)$ in Eq.~(\ref{EQN:FFfit}) to observables, such as the primordial spectral index ($n_s$) and the tensor-to-scalar ratio ($r$), we compute the usual slow-roll parameters \cite{bassett_05,liddle+lyth_09}
\begin{equation}
\begin{split}    
\epsilon(\phi) &= \frac{\Mpl^2}{2}\left(\frac{V_{,\phi}}{V}\right)^{2}
= \frac{1}{2} \alpha^2 \beta^2 \left( \frac{ \phi}{M_{\rm pl}} \right)^{2\beta - 2} , \\
\eta(\phi) &= \Mpl^2 \left( \frac{V_{,\phi \phi}}{V} \right) \\
&= \alpha \beta \left( \frac{ \phi}{M_{\rm pl}} \right)^{\beta - 2}  \left[ 1 + \beta \left( \alpha \left( \frac{ \phi}{M_{\rm pl} } \right)^\beta - 1 \right) \right],
\label{EQN:etaSlowRoll}
\end{split}
\end{equation}
where $V_{, \phi}  \equiv  \partial V / \partial \phi$, $V_{,\phi\phi}  \equiv \partial^2 V / \partial \phi^2$, and so on. To lowest order in the slow-roll parameters, the primordial observables are given by $n_s = 1 - 6 \epsilon (\phi_*) + 2 \eta (\phi_*)$ and $r = 16 \epsilon (\phi_*)$ \cite{bassett_05,liddle+lyth_09}, which yields
%
\begin{equation} \label{EQN:rstar}
\begin{split}
n_{s}&= 1 - 2 \alpha \beta (\beta - 1) \left( \frac{ \phi_*}{M_{\rm pl}} \right)^{\beta - 2} - \alpha^2 \beta^2 \left( \frac{ \phi_*}{M_{\rm pl}} \right)^{2 \beta - 2} , \\
r&= 8 \alpha^2 \beta^2 \left( \frac{ \phi_*}{M_{\rm pl}} \right)^{2 \beta - 2}.
\end{split}
\end{equation}
%
(See also Ref.~\cite{geng_15}.) Here $\phi_{*}$ indicates that parameters are to be evaluated at the time during inflation when cosmologically relevant perturbations of comoving wavenumber $k$ first crossed the Hubble radius, $k_{*}=a (t_*) H (t_*)$. Up to modest uncertainties from the reheating epoch, this time is typically assumed to occur $N_{*}=50$ to $60$ efolds before the end of inflation \cite{amin+al_15}. 

For models with $V (\phi)$ as in Eq.~(\ref{EQN:FFfit}), we consider trajectories in which the field  begins at $\phi = 0$ and rolls to larger and larger field values; the potential does not have a global minimum. Within the slow-roll regime we may estimate the time when inflation ends, $t_{\rm end}$, from the condition $\epsilon (t_{\rm end}) = 1$. From Eqs.~(\ref{EQN:FFfit}) and (\ref{EQN:etaSlowRoll}), this yields
\begin{equation}
    \frac{ \phi_{\rm end}  }{M_{\rm pl} } = \left( \frac{ 2}{\alpha^2 \beta^2 } \right)^{1/(2\beta - 2)} ,
    \label{EQN:phiend}
\end{equation}
where $\phi_{\rm end} \equiv \phi (t_{\rm end})$. We may likewise estimate \cite{liddle+lyth_09}
\begin{equation}\label{EQN:Nstar}
\begin{split}
N_{*} &\simeq \frac{1}{\Mpl}\int_{\phi_{*}}^{\phi_{\rm end}}\frac{d\phi}{\sqrt{2\epsilon(\phi)}}\\
&=\frac{1}{ \alpha \beta (2 - \beta) } \left[ \left( \frac{\phi_{\rm end} }{M_{\rm pl} } \right)^{2 - \beta} - \left( \frac{ \phi_*}{M_{\rm pl}} \right)^{2 - \beta} \right],
\end{split}
\end{equation}
which yields
\begin{equation}
    \frac{ \phi_*}{M_{\rm pl}} \simeq \left[ \left( \frac{ \phi_{\rm end} }{M_{\rm pl}} \right)^{2 - \beta} - \alpha \beta (2 - \beta) N_* \right]^{1/(2 - \beta)}.
    \label{phistar}
\end{equation}
From Eqs.~(\ref{EQN:etaSlowRoll}) and (\ref{EQN:phiend}), we see that for $\beta = 1$ and $\alpha^2 < 2$, inflation never ends: $\epsilon \rightarrow \alpha^2 / 2$, independent of $\phi$, and hence there is no finite value of $\phi$ such that $\epsilon (\phi_{\rm end}) = 1$. For $1 < \beta < 2$, inflation will end, but, for $\beta \rightarrow 1$, only after the field has undergone a very large excursion, to values $\phi_{\rm end} \gg M_{\rm pl}$. For example, for a typical value of $\alpha \sim {\cal O} (0.1)$ and $1.1 \leq \beta \leq 1.5$, we find $\phi_*, \phi_{\rm end} \sim {\cal O} (10^1 - 10^{11}) \, M_{\rm pl}$, corresponding to very long durations of inflation, with $N_{\rm tot} \sim {\cal O} (10^2 - 10^{11})$ efolds. (We may estimate $N_{\rm tot}$ from Eqs.~(\ref{EQN:phiend}) and (\ref{EQN:Nstar}), substituting $\phi_* \rightarrow \phi (t_i) \sim 0$.)

Within the context of our EFT framework, we do not take such exponentially large field excursions at face value. In particular, there is no reason to expect that our (classical) analysis of the field dynamics should continue to hold at arbitrarily large field values, $\phi \gg M_{\rm pl}$. Rather, our goal is to analyze the flow into inflation, and to consider features of such dynamical systems for values of the field in the vicinity of $M_{\rm pl}$. Hence we consider predictions for observables for values of $\phi_*$ within the range $0.1 \leq (\phi_* / M_{\rm pl} ) \leq 10$.

\begin{figure}
    \centering
    \includegraphics[width=0.49\textwidth]{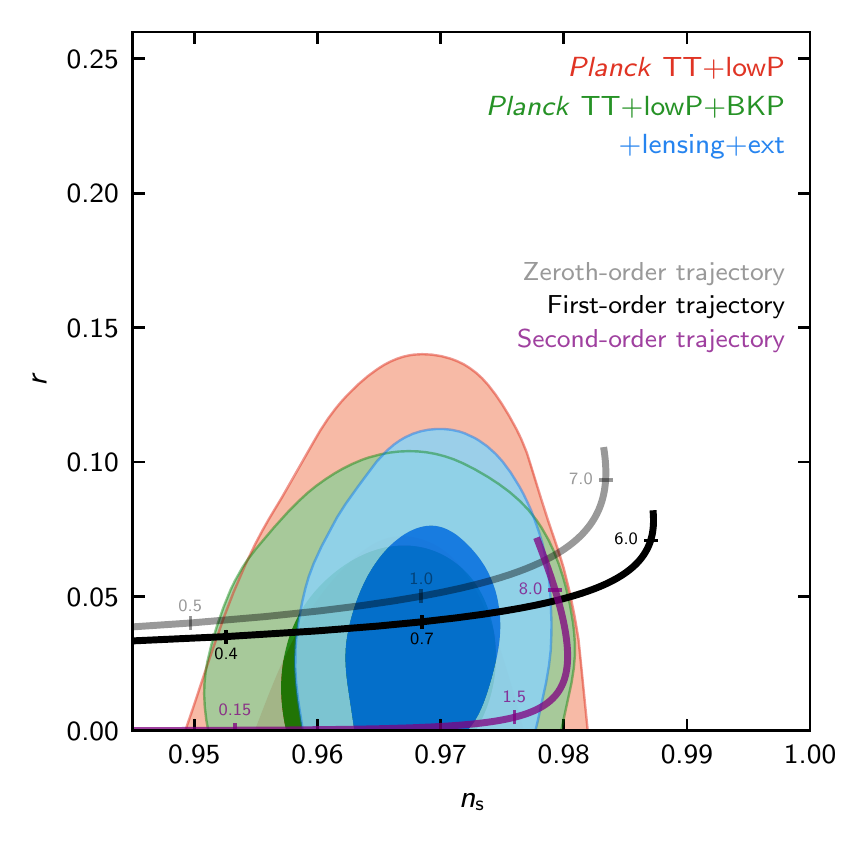}
    \caption{Predictions for $n_s$ and $r$ based on best-fit values for the parameters $\alpha$ and $\beta$ for various trajectories through the EFT phase space. For the particular trajectories shown, the zeroth-order trajectory corresponds to $(\alpha, \beta) =  (0.0682, 1.16)$ [with $\lambda_{0}=0.01$ and $x(\tau_{i})=0.9$]; the first-order trajectory to $(\alpha, \beta) =  (0.0660, 1.13)$ (from Table~\ref{TAB:SyOneCase1abc}); and the second-order trajectory to $(\alpha, \beta) = (0.0111, 1.70)$ (from Table~\ref{TAB:SyTwoCase1a(i)}). In each case, numbers along a given curve indicate values of $(\phi_* / M_{\rm pl})$.}
    \label{FIG:Planckfit}
\end{figure}

The \emph{Planck} collaboration has measured \citep{Planck_2015_ConstraintsOnInflation}
%
\begin{equation}
\begin{split}
n_{s}&=0.968\pm 0.006,
\\
r&<0.11.\label{EQN:rPLANCK}
\end{split}
\end{equation}
%
The value of $n_s$ is quoted for pivot-scale $k_{*}=0.05\;\textrm{Mpc}^{-1}$ (at the 68\% confidence level), whereas $r$ is quoted for $k_{*}=0.002\;\textrm{Mpc}^{-1}$ (at the 95\% confidence level). (Our discussion in this section would change little if we adopted the updated constraint $r < 0.09$ at $k_{*} = 0.05 \; {\rm Mpc}^{-1}$ \cite{PlanckBICEP}; we use the constraint in Eq.~(\ref{EQN:rPLANCK}) because the underlying data from the {\it Planck} mission are more readily available.) As shown in Fig.~\ref{FIG:Planckfit}, there exist trajectories for zeroth-order, first-order, and second-order systems that are readily compatible with the observational constraints of Eq.~(\ref{EQN:rPLANCK}), for $\phi_*$ within the range $0.1 \leq (\phi_* / M_{\rm pl} ) \leq 10$.

We next explore the range of parameters $(\alpha, \beta)$ that yield predictions for $n_s$ and $r$ which remain consistent with Eq.~(\ref{EQN:rPLANCK}), for $0.1 \leq (\phi_* / M_{\rm pl}) \leq 10$. From Eq.~(\ref{EQN:rstar}), we immediately see that the case of power-law inflation, with $\beta = 1$, is incompatible with the observational constraints of Eq.~(\ref{EQN:rPLANCK}). In particular, for $\beta = 1$, $n_s$ and $r$ reduce to constants that depend only on $\alpha$:
\begin{equation}
n_{s}=1-\alpha^2,\;\; r=8\alpha^2.
\label{nsrbeta1}
\end{equation}
The bound on $r$ in Eq.~(\ref{EQN:rPLANCK}) constrains $\alpha<0.12$, which in turn yields $n_{s}>0.986$, fully $3 \sigma$ away from the central value in Eq.~(\ref{EQN:rPLANCK}). Or, working the other way, the $2\sigma$ bounds on $n_{s}$ require $0.14 \leq \alpha \leq 0.21$, which yields $r \geq 0.16$. 

The situation is similar for the range $0 < \beta < 1$. In that case, we may find values of $(\alpha, \beta)$ that yield predictions for $n_s$ within the $2\sigma$ bound of the {\it Planck} value.
However, none of these values is also consistent with the constraint $r < 0.11$, across the entire range $\alpha > 0$ and $0.1 \leq (\phi_* / M_{\rm pl} ) \leq 10$. Hence trajectories for the dynamical system's evolution through the EFT phase space that yield $0 < \beta < 1$ are inconsistent with the observational constraints, at least under the assumption that perturbations on cosmologically relevant scales cross outside the Hubble radius for some $\phi_*$ within the range $0.1 \leq (\phi_* / M_{\rm pl}) \leq 10$.

\begin{figure*}[!]
    \centering
    \includegraphics[width=0.49\textwidth]{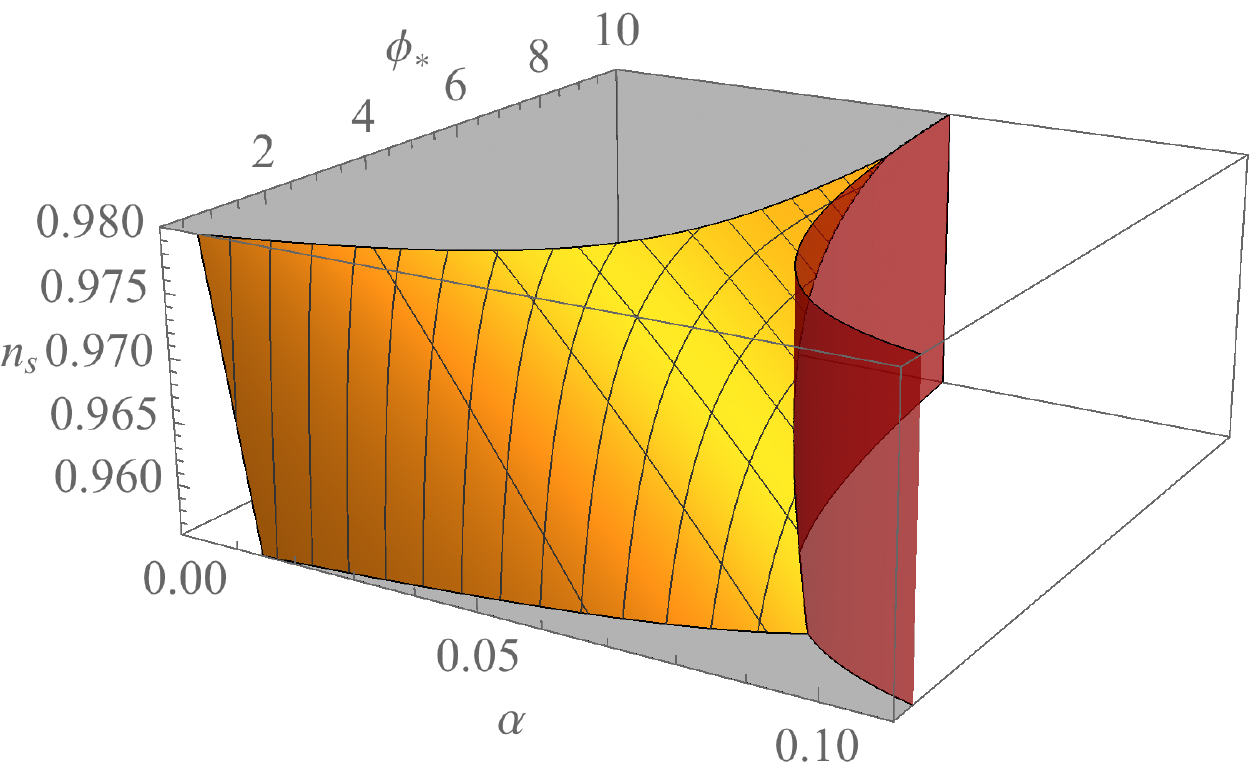} 
    \includegraphics[width=0.49\textwidth]{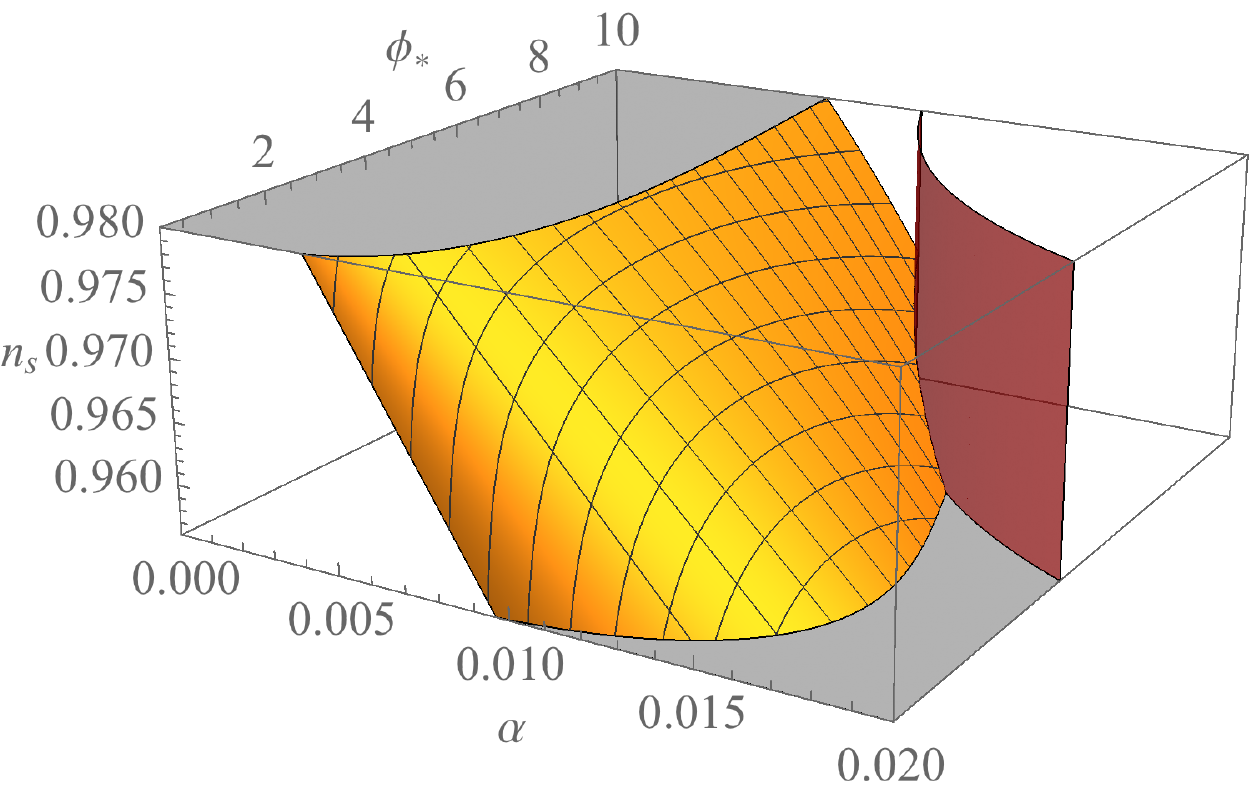}
    \caption{Predictions for $n_s$ (gold surface) as a function of $\alpha$ and $\phi_*$ (in units of $M_{\rm pl}$), consistent with the bound $r < 0.11$ (effected by the red surface), for $\beta = 1.2$ (left) and $\beta = 1.8$ (right).}
    \label{FIG:nsrmediumvA}
\end{figure*}

For the range $1 < \beta \leq 2$, we do find values of $(\alpha, \beta)$ that are consistent with the observational constraints of Eq.~(\ref{EQN:rPLANCK}) for $0.1 \leq (\phi_* / M_{\rm pl}) \leq 10$, examples of which are shown in Fig.~\ref{FIG:nsrmediumvA} for the cases $\beta = 1.2$ and $\beta = 1.8$. 
We first consider the constraints on $n_s$. 

For a given value of $n_s$, we may use the expression for $n_s$ in Eq.~(\ref{EQN:rstar}) to solve for $\alpha \rightarrow \alpha^{(n_s)} (\beta, \phi_*, n_s)$. Straightforward algebra yields
\begin{equation}
\alpha^{(n_s)} (\beta, \phi_*, n_s ) = - \frac{ B}{2} + \frac{1}{2} \left[ B^2 + 4 C \right]^{1/2} ,
\label{EQN:alphans}
\end{equation}
with
\begin{equation}
\begin{split}
    B (\beta, \phi_*) &\equiv 2 \left( \frac{ \beta - 1}{\beta} \right) \left( \frac{ \phi_* }{M_{\rm pl} } \right)^{-\beta} , \\
    C (\beta, \phi_*, n_s) &\equiv \left(\frac{  1 - n_s  }{\beta^2}\right) \left(\frac{ \phi_* }{M_{\rm pl}} \right)^{2 - 2\beta} .
    \end{split}
    \label{EQN:BCdef}
\end{equation}
We may then set $\partial \alpha^{(n_s)}  / \partial \phi_* = 0$ and solve for $\bar{\phi}_* (\beta, n_s)$, the value of $\phi_*$ at which $\alpha^{(n_s)}$ is an extremum. We find
\begin{equation}
    \frac{\bar{\phi}_* (\beta, n_s)}{M_{\rm pl} } = \sqrt{ \frac{ \beta (2 - \beta) }{1 - n_s} } ,
    \label{EQN:phistarns}
\end{equation}
which remains well-behaved for the range we are considering, with $1 < \beta \leq 2$ and $(1 - n_s) > 0$. For a given value of $\beta$, the maximum value of $\alpha$ that will keep $n_s$ within the $2\sigma$ bound of the {\it Planck} value in Eq.~(\ref{EQN:rPLANCK}) will occur for the maximum value of $(1 - n_s)$, which is to say, for $n_s^{\rm min} = 0.956$. This yields 
\begin{equation}
    \alpha_{\rm max}^{(n_s)} (\beta) = \alpha^{(n_s)} \left( \beta, \bar{\Phi}_* (\beta), n_s^{\rm min} \right)\;\textrm{for}\; 1 < \beta \leq 2,
    \label{EQN:alphamaxns}
\end{equation}
where $\alpha^{(n_s)} (\beta,\phi_*, n_s)$ is given by Eqs.~(\ref{EQN:alphans})--(\ref{EQN:BCdef}), and $\bar{\Phi}_* (\beta) \equiv \bar{\phi}_* (\beta, n_s^{\rm min})$.\footnote{Note that Eq.~(\ref{EQN:alphamaxns}) does not quite hold everywhere in the range $1 < \beta \leq 2$, as for $1.9998<\beta\leq 2$, $\bar{\Phi}_* (\beta)< \phi_{\textrm{min}}\equiv 0.1\Mpl$ (which is the minimum value of $\phi_{*}$ we have elected to consider). Given that the empirical constraints of Eq.~(\ref{EQN:rPLANCK}) are only known to 2 - 3 significant figures, however, we consider the value $\beta = 1.9998$ to be indistinguishable (for all practical purposes) from $\beta = 2$. Hence we may work with $\alpha_{\rm max}^{(n_s)} (\beta)$ as given in Eq.~(\ref{EQN:alphamaxns}). }


Now for certain values of $\beta$ within the range $1 < \beta \leq 2$, the value of $\alpha_{\rm max}^{(n_s)} (\beta)$ in Eq.~(\ref{EQN:alphamaxns}) yields a value of $r (\alpha_{\rm max}^{(n_s)}, \beta) \geq r_{\rm max} = 0.11$, violating the observational bound on the tensor-to-scalar ratio. The ratio $r$ rises monotonically with $\phi_*$. We label $\phi_*^{(r)}$ the value that saturates the bound $r \rightarrow r_{\rm max}$. From Eq.~(\ref{EQN:rstar}), we find
\begin{equation}
    \frac{ \phi_*^{(r)} (\alpha, \beta)}{M_{\rm pl}} = \left[ \frac{ r_{\rm max}}{8 \alpha^2 \beta^2 } \right]^{1/(2\beta - 2)} .
    \label{EQN:phistar_r}
\end{equation}
For a given value of $\beta$, values of $\alpha_{\rm max}^{(n_s)} (\beta)$ such that $\bar{\Phi}_* \geq \phi_*^{(r)}$ yield $r \geq r_{\rm max}$.

For cases in which $\bar{\Phi}_* \geq \phi_*^{(r)}$, we calculate an alternate form for $\alpha_{\rm max} (\beta)$ that remains consistent with the observational constraints on both $n_s$ and $r$. In particular, we substitute $\phi_*^{(r)}$ from Eq.~(\ref{EQN:phistar_r}) into Eq.~(\ref{EQN:rstar}) for $n_s$. After some straightforward algebra, we find
\begin{equation}
    \alpha_{\rm max}^{(r1)} (\beta) = \frac{ \left[ 1 - n_s^{\rm min} - (r_{\rm max} / 8) \right]^{\beta -1} }{ \beta (2 \beta - 2)^{\beta - 1}  (r_{\rm max} / 8)^{ (\beta - 2) / 2} } .
    \label{EQN:alphamax}
\end{equation}
We can find the cross-over $\beta$-value as follows. Note from Eq.~(\ref{EQN:rstar}) that the ratio $r$ rises monotonically with $\alpha$. We let $\alpha^{(r0)}$ correspond to the $\alpha$-value that saturates the bound $r\to r_{\rm max}$. Then we find
\begin{equation}
\alpha^{(r0)}(\beta, \phi_{*}) \equiv \left[\frac{r_{\rm max}}{8 \beta^2}\left(\frac{\phi_*}{\Mpl}\right)^{-2(\beta-1)}\right]^{1/2}.
\end{equation}
The cross-over $\beta$-value occurs for $\beta_{1}$ such that
\begin{equation}
\alpha^{(r0)}(\beta_{1}, \bar{\Phi}_*(\beta_1))=\alpha_{\rm max}^{(n_s)}(\beta_{1}), 
\end{equation}
which yields $\beta_{1}= 1.524$.

\begin{figure}
    \centering
    \includegraphics[width=0.49\textwidth]{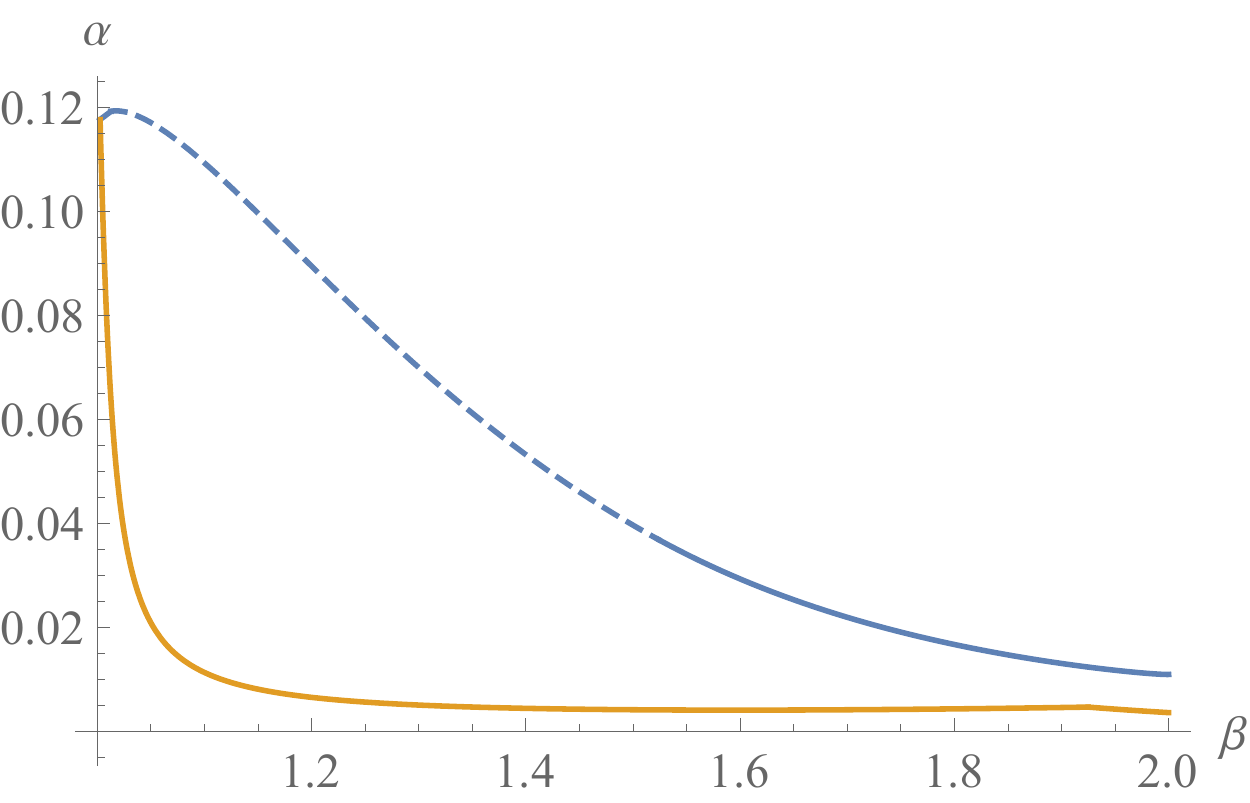}
    \caption{The values of $\alpha_{\rm max} (\beta)$ (blue) and $\alpha_{\rm min} (\beta)$ (gold), from Eqs.~(\ref{EQN:alphamaxfull}) and (\ref{EQN:alphamin}), respectively, for a given value of $\beta$ in the range $1 < \beta \leq 2$ that will yield predictions for $n_s$ and $r$ consistent with the constraints of Eq.~(\ref{EQN:rPLANCK}), for some values of $\phi_*$ within the range $0.1 \leq (\phi_* / M_{\rm pl} \leq 10$. The dashed line signifies a constraint derived from the inequality on the tensor-to-scalar ratio $r$, so that admissable values of $\alpha$ are those that lie strictly beneath the dashed line.}
    \label{FIG:alpharange}
\end{figure}

There is one further regime of interest. Solving the equation
\begin{equation}\label{EQN:betacrit}
\alpha^{(r0)}(\beta, \phi_{\rm min}) = \alpha^{(n_s)} \left( \beta, \phi_{\rm min}, n_s^{\rm max} \right),
\end{equation}
for $\beta$ (with $\phi_{\textrm{min}}\equiv 0.1\Mpl$), we obtain a critical value of $\beta$, $\beta_{\rm crit}\equiv 
1.003$,
below which there are no values of $(\alpha,\beta)$ that are consistent with the constraints on $n_s$ and $r$. Above (and including) $\beta_{\rm crit}$, the maximum value of $\alpha$ is given by $\alpha^{(r0)}(\beta, \phi_{\rm min})$. This continues to hold for values of $\beta$ up to and including $\beta_{0}\equiv 
1.013$, which can be found by solving $\alpha^{(r0)}(\beta, \phi_{\rm min})=\alpha^{(n_s)} \left( \beta, \phi_{\rm min}, n_s^{\rm min} \right)$ for $\beta$. When the maximum allowable value of $\alpha$ is affected by the $r$-constraint, we therefore find
\begin{align}\label{EQN:alphamaxrFINAL}
  \alpha_{\rm max}^{(r)} (\beta) =\begin{cases}
                 \alpha^{(r0)}(\beta, \phi_{\rm min})&\;\textrm{for}\; \beta_{\rm crit} \leq \beta \leq \beta_{0}\\
              \alpha_{\rm max}^{(r1)} (\beta)&\;\textrm{for}\; \beta_{0} < \beta \leq \beta_{1} ,
                          \end{cases}
\end{align}
with $(\beta_{\rm crit}, \beta_0, \beta_1) = (1.003, 1.013, 1.524)$.
Combining Eqs.~(\ref{EQN:alphamaxns}) and (\ref{EQN:alphamaxrFINAL}), for a given value $\beta_{\rm crit} \leq \beta \leq 2$, we find the maximum value of $\alpha$ that will remain consistent with the observational constraints on both $n_s$ and $r$:
\begin{align}
    \alpha_{\rm max} (\beta) = \begin{cases}
                  \alpha_{\rm max}^{(r)} (\beta)&\;\textrm{for}\; \beta_{\rm crit } \leq \beta \leq \beta_{1}\\
              \alpha_{\rm max}^{(n_s)} (\beta)&\;\textrm{for}\; \beta_{1} < \beta \leq 2,
                            \end{cases}
    \label{EQN:alphamaxfull}
\end{align}
A plot of $\alpha_{\rm max} (\beta)$ is shown in Fig.~\ref{FIG:alpharange} (in blue).

We may find $\alpha_{\rm min} (\beta)$ similarly. The minimum allowable value of $\alpha$ will correspond to the minimum value of $(1 - n_s)$, and hence to $n_s \rightarrow n_s^{\rm max} = 0.980$. The function $\alpha^{(n_s)} (\beta, \phi_*, n_s^{\rm max})$ has a nontrivial dependence on $\phi_*$. For most of the range $1 < \beta \leq 2$, $\alpha^{(n_s)} (\beta, \phi_*, n_s^{\rm max})$ will be minimized for $\phi_* \rightarrow \phi_{\rm min}$. Only near the upper end of the range $\beta \rightarrow 2$ will the minimum of $\alpha^{(n_s)} (\beta, \phi_*, n_s^{\rm max})$ occur at $\phi_* = \phi_{\rm max} = 10 \, M_{\rm pl}$, the maximum value of $\phi_*$ under consideration. The two values become equal, with $\alpha^{(n_s)} (\beta_2, \phi_{\rm min} , n_s^{\rm max}) = \alpha^{(n_s)} (\beta_2, \phi_{\rm max} , n_s^{\rm max})$, at $\beta_2 = 
1.925$. Hence across the full range $\beta_{\rm crit} \leq \beta \leq 2$, we have
\begin{align}
    \alpha_{\rm min} (\beta) = \begin{cases}
                  \alpha^{(n_s)} (\beta, \phi_{\rm min} , n_s^{\rm max} )&\;\textrm{for}\; \beta_{\rm crit } \leq \beta \leq \beta_{2}\\
              \alpha^{(n_s)} (\beta, \phi_{\rm max}, n_s^{\rm max} )&\;\textrm{for}\; \beta_{2} \leq \beta \leq 2.
                            \end{cases}
    \label{EQN:alphamin}
\end{align}
Note that we do not need to make any additional adjustments to our expression for $\alpha_{\rm min} (\beta)$ in Eq.~(\ref{EQN:alphamin}) in order to accommodate observational constraints on $r$. In addition, from Eq.~(\ref{EQN:betacrit}), we see that $\alpha_{\rm max}(\beta_{\rm crit})=\alpha_{\rm min}(\beta_{\rm crit})$. A plot of $\alpha_{\rm min}(\beta)$ is presented in Fig.~\ref{FIG:alpharange} (in gold).

To summarize: for any value of $\beta$ within the range $\beta_{\rm crit} \leq \beta \leq 2$, the best-fit parameters ($\alpha, \beta$) for a given trajectory will be consistent with the observational constraints of Eq.~(\ref{EQN:rPLANCK}) for some values $\phi_*$ within the range $0.1 \leq (\phi_* / M_{\rm pl} ) \leq 10$ and $\alpha_{\rm min} (\beta) \leq \alpha \leq \alpha_{\rm max} (\beta)$.

Finally, an analysis similar to the one we have carried out for $1 < \beta \leq 2$ is possible for $\beta > 2$. In that case, in accord with Ref.~\citep{geng_15}, we find that broad ranges of $\alpha$ remain consistent with the {\it Planck} constraints.

We therefore find that there exist trajectories through the effective phase space for systems at various orders $M$ that are compatible with SSF realizations and that remain consistent with observational constraints. In particular, there exist non-trivial windows within which the inferred values for $\alpha$ and $\beta$ of the effective potential $V (\phi)$ in Eq.~(\ref{EQN:FFfit}) yield predictions for the primordial spectral index for scalar curvature perturbations, $n_s$, and for the ratio of tensor-to-scalar perturbations, $r$, consistent with the latest observations, under reasonable assumptions about the cross-out scale $\phi_* / M_{\rm pl}$. We defer to future research the question of how representative such observationally consistent values of $\alpha$ and $\beta$ are, for a given order $M$, among inflationary trajectories through the effective phase space.

\section{Discussion\label{SEC:Discussion}}

By combining techniques from effective field theory (EFT) approaches to inflation with dynamical-systems analyses, we have developed a framework within which one may assess how generic (or otherwise) the flow into early-universe inflation may be. Our approach applies to all single-clock scenarios, including, but not limited to, single-scalar-field (SSF) realizations. Rather than specify a functional form $V (\phi)$ for the effective potential, we study the dynamics of systems under various assumptions about the behavior of the $M$th time derivative of a potential-like quantity in the effective action, $L (t)$. 

When we fix the $M$th time derivative of $L(t)$ --- thereby reducing the dynamical system to $M$th order --- we find that there exist at most two hyperbolic inflationary fixed points within the effective phase space. One of these fixed points corresponds to evolution of the system in a pure de Sitter state, while the other corresponds to evolution in a quasi-de Sitter state akin to that of power-law inflation. For zeroth-order and first-order systems (corresponding to $M = 0$ and $1$, respectively), we find significant probability for systems to flow into inflation, and for inflation to persist for at least 60 efolds, even for initial conditions such that kinetic energy dominates potential energy at early times. For first-order systems, we also identify trajectories through the effective phase space that do not correspond to any SSF realization. Including such trajectories further increases the probability that dynamical systems will flow into inflation. 

We further find that all trajectories through the effective phase space that are compatible with SSF realizations (at least up to and including order $M = 2$) may be characterized by a single functional form for the (inferred) effective potential, $V (\phi)$: a generalization of the familiar potential for power-law inflation. The specific form of $V (\phi)$ that we infer, $V (\phi) = V_0 \exp [- \alpha (\phi / M_{\rm pl} )^\beta]$, includes the two fixed points as special cases: $\alpha = 0$ for a de Sitter phase, $\beta = 1$ for power-law inflation.

Given the functional form for $V (\phi)$ for $M$th-order systems that are compatible with SSF realizations, we identify ranges for the (inferred) parameters of the potential that are compatible with observational constraints, including the measured value of the primordial spectral index ($n_s$) and the upper bound on the tensor-to-scalar ratio ($r$). For zeroth-order, first-order, and second-order systems, we find examples of trajectories through the effective phase space that yield a sufficient amount of inflation and can also remain compatible with observations.

Our aim in this work has been to establish a formalism for assessing the flow into inflation without needing to specify a particular form for $V (\phi)$, thereby complementing recent numerical \cite{clough+al_16,east+al_16} and semi-analytic \cite{remmen_13,remmen+carroll_14,MarshBarrow} approaches. Hence we have restricted attention to the simple case in which the background spacetime is (already) homogeneous, isotropic, and spatially flat. An obvious next step is to expand the analysis presented here to background spacetimes that have nonvanishing spatial curvature, initial anisotropy, and/or initial inhomogeneities. In the presence of inhomogeneities, we would no longer expect dynamical trajectories to remain on the constraint surface $x + y = 1$, given additional contributions from fluctuations to the effective energy density. Such extensions remain the subject of further research.

\appendix

\section{Second-order system\label{APP:secondorder}}

In this first appendix, we collect some results of interest for the second-order system. In particular, we discuss fixed points at second order, as well as certain SSF realizations. 

To obtain the second-order system, we set $\lambda_{2}$ = constant. Under this assumption, the equations governing the dynamics, Eqs.~(\ref{EQN:xFtimeM})--(\ref{EQN:ConstraintM}), take the form
\begin{subequations}
\begin{align}
\frac{dx}{d\ln a} &=\lambda_{0} y -3 x +3 x^2-3 x y,\label{EQN:xFtimeTWO}\\
\frac{dy}{d\ln a} &=-\lambda_{0} y +3 y +3 x y-3 y^2,\label{EQN:yFtimeTWO}\\
\frac{d\lambda_{0}}{d\ln a} &= \left[-\lambda_{1}+\lambda_{0}+\frac{3}{2}(1+x-y)\right]\lambda_{0},\label{EQN:lzerotimeTWO}\\
\frac{d\lambda_{1}}{d\ln a} &= \left[-\lambda_{2}+\lambda_{1}+\frac{3}{2}(1+x-y)\right]\lambda_{1},\label{EQN:lonetimeTWO}\\
1&=x+y, \label{EQN:ConstraintTWO}
\end{align}
\end{subequations}
where, as in Eq.~(\ref{EQN:Ep}), the slow-roll parameter is given by $\epsilon=3x$. 

\begin{figure*}[htp!]
\begin{minipage}{.32\linewidth}
\subfloat[]{\includegraphics[scale=0.25]{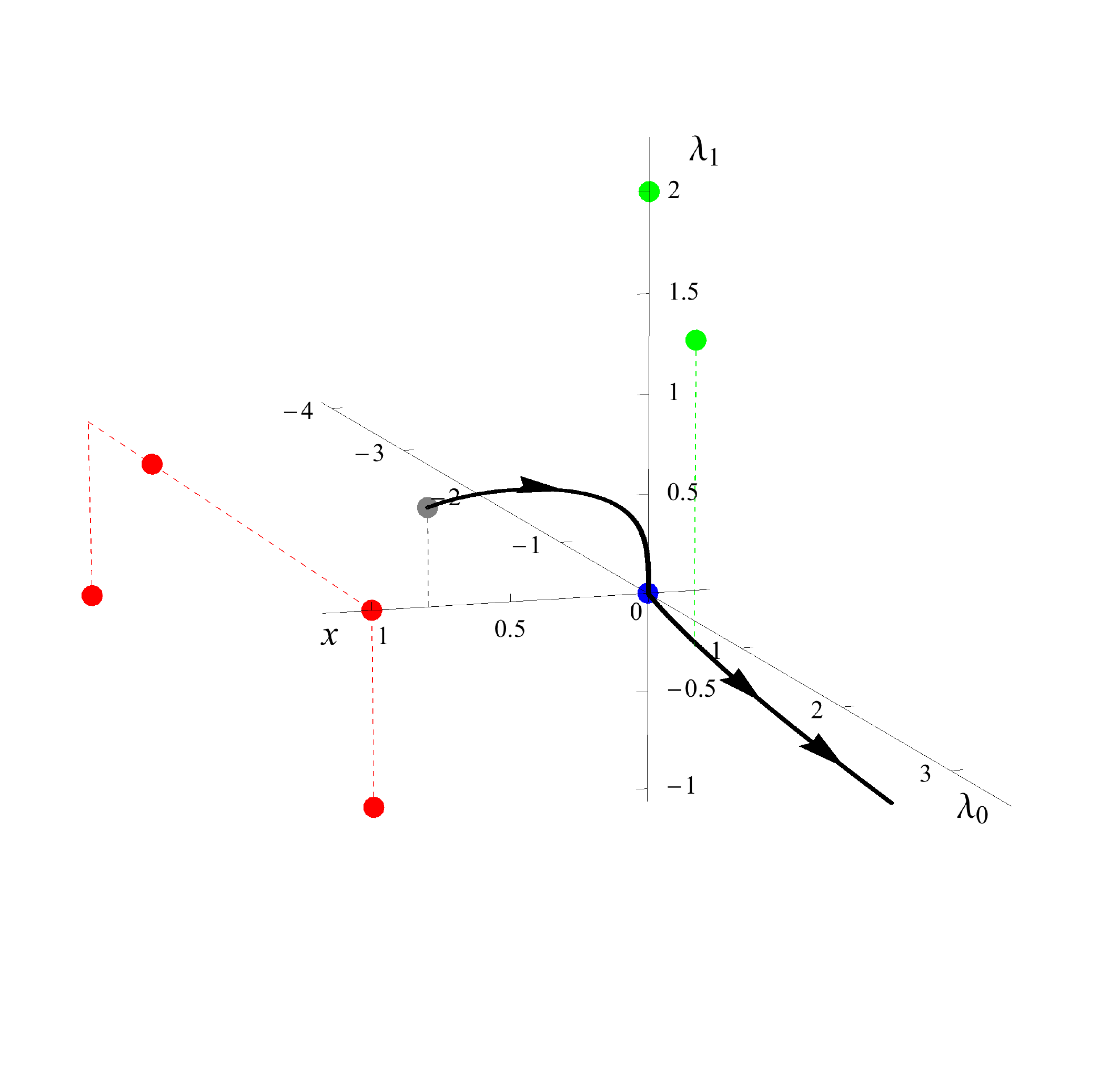}}
\end{minipage}
\begin{minipage}{.32\linewidth}
\subfloat[]{\includegraphics[scale=0.4]{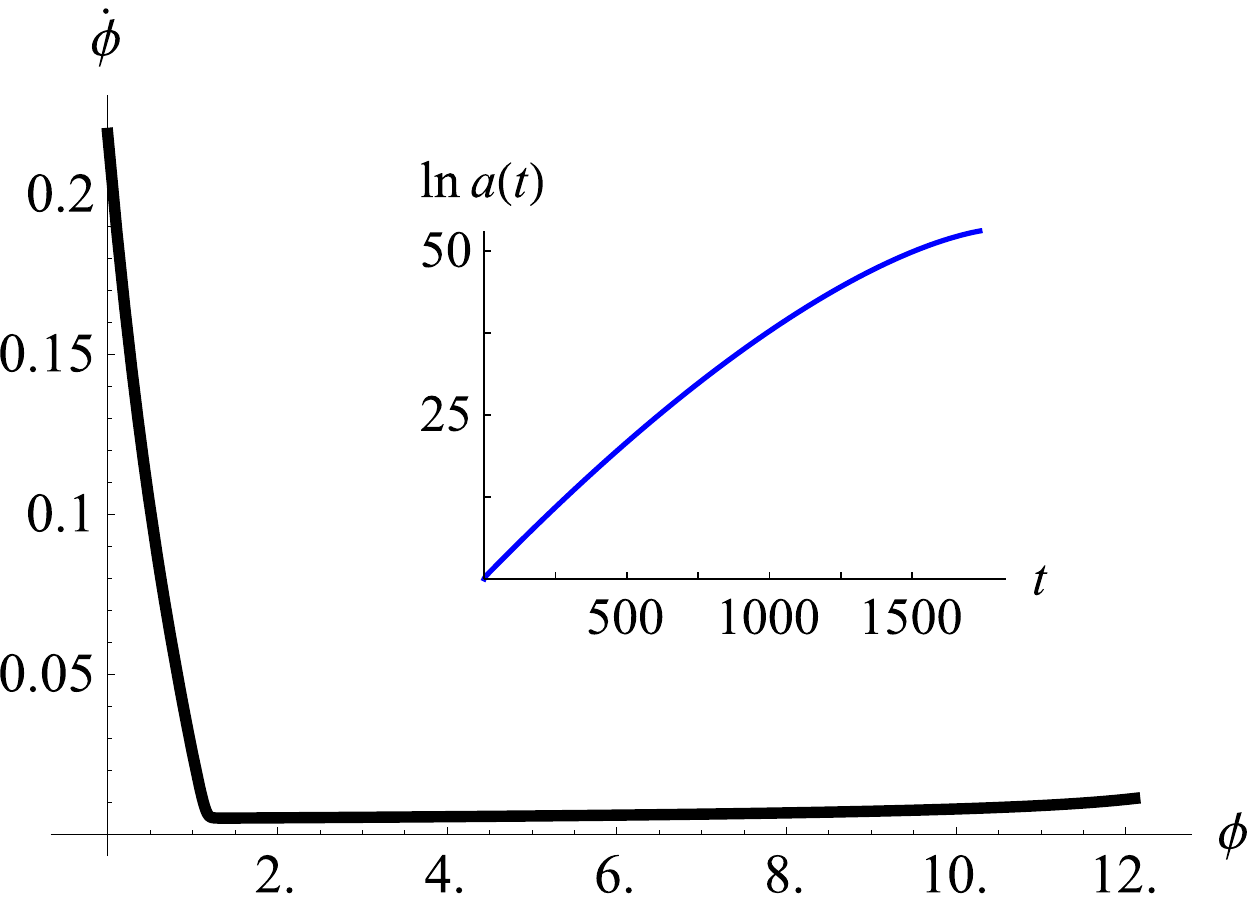}}
\end{minipage}
\begin{minipage}{.32\linewidth}
\subfloat[]{\includegraphics[scale=0.4]{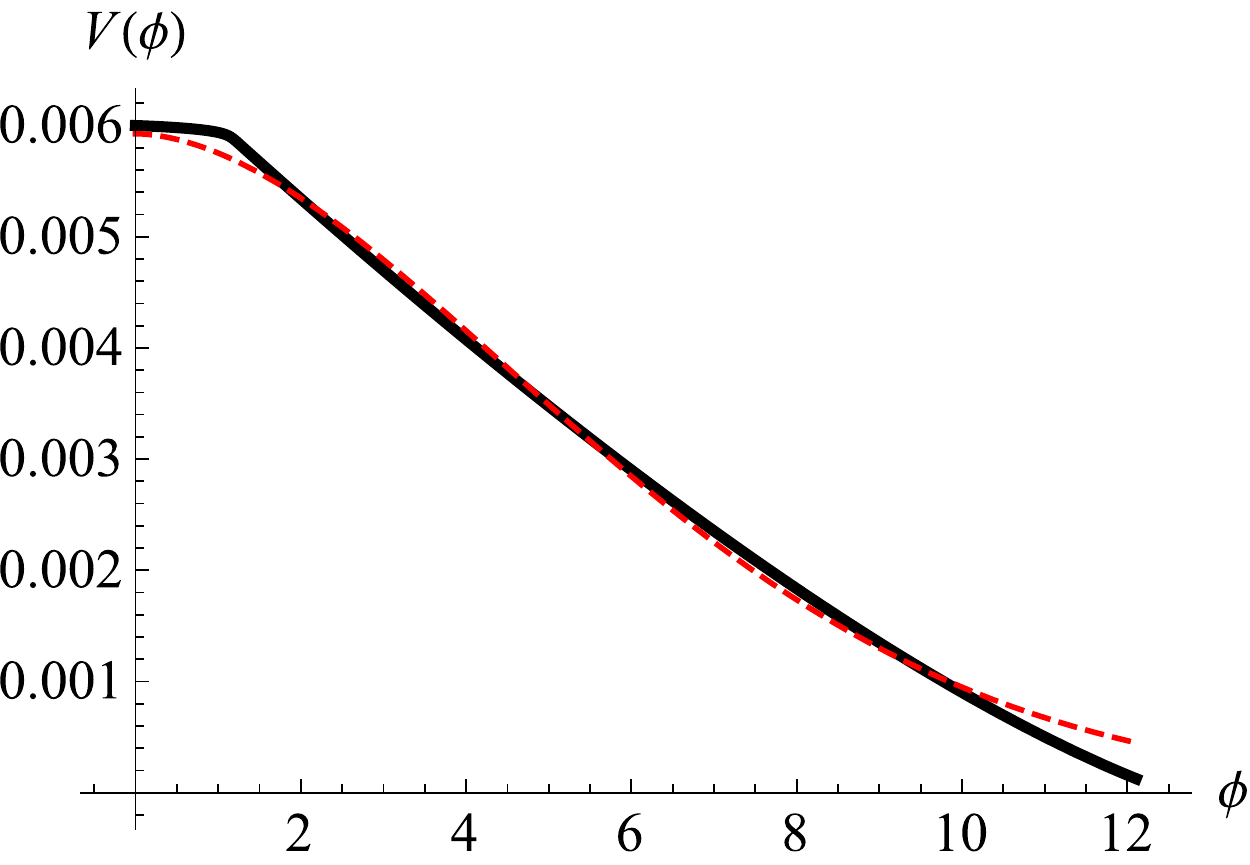}}
\end{minipage}\par\medskip
\begin{minipage}{.32\linewidth}
\subfloat[]{\includegraphics[scale=0.25]{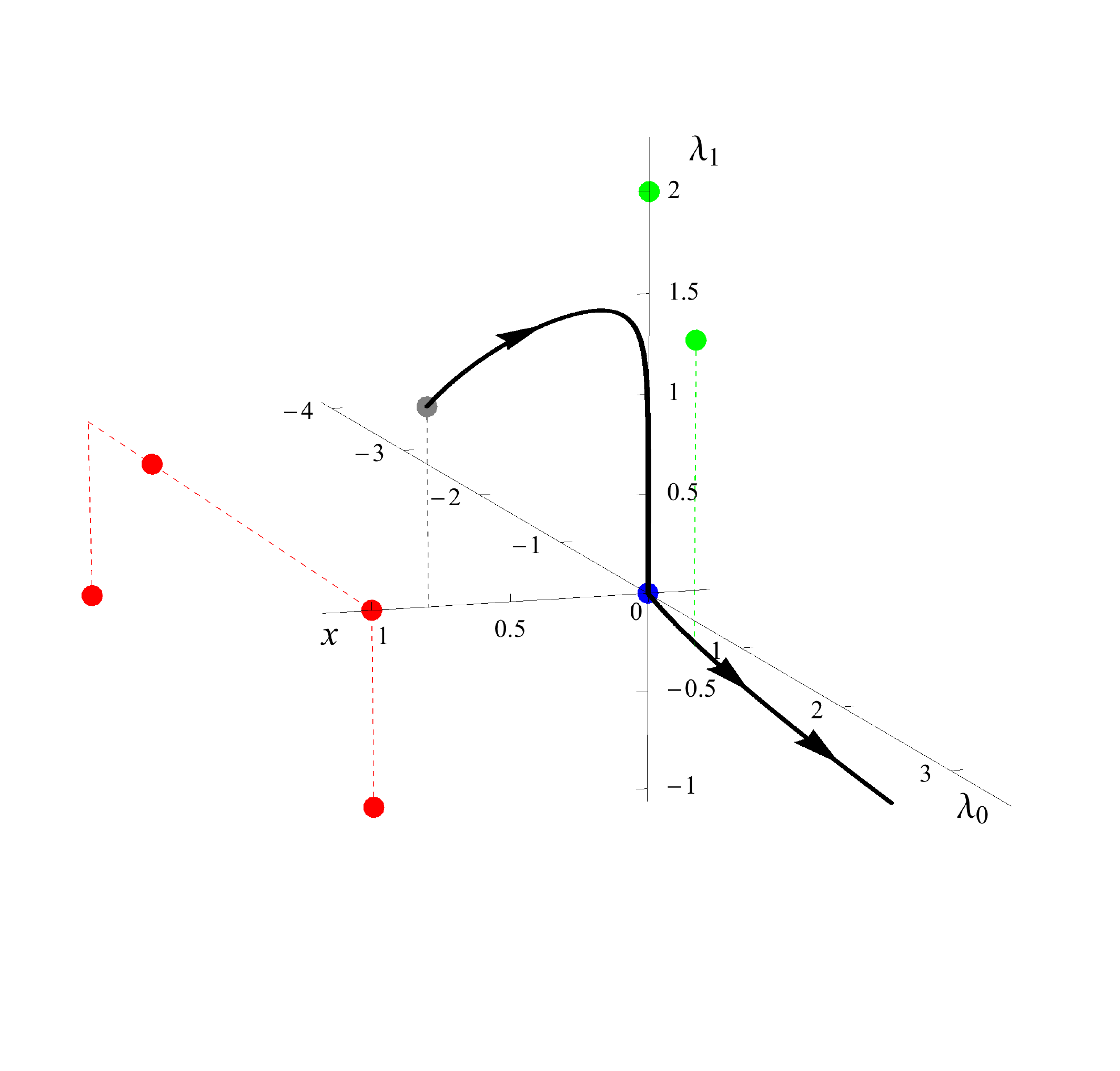}}
\end{minipage}
\begin{minipage}{.32\linewidth}
\subfloat[]{\includegraphics[scale=0.4]{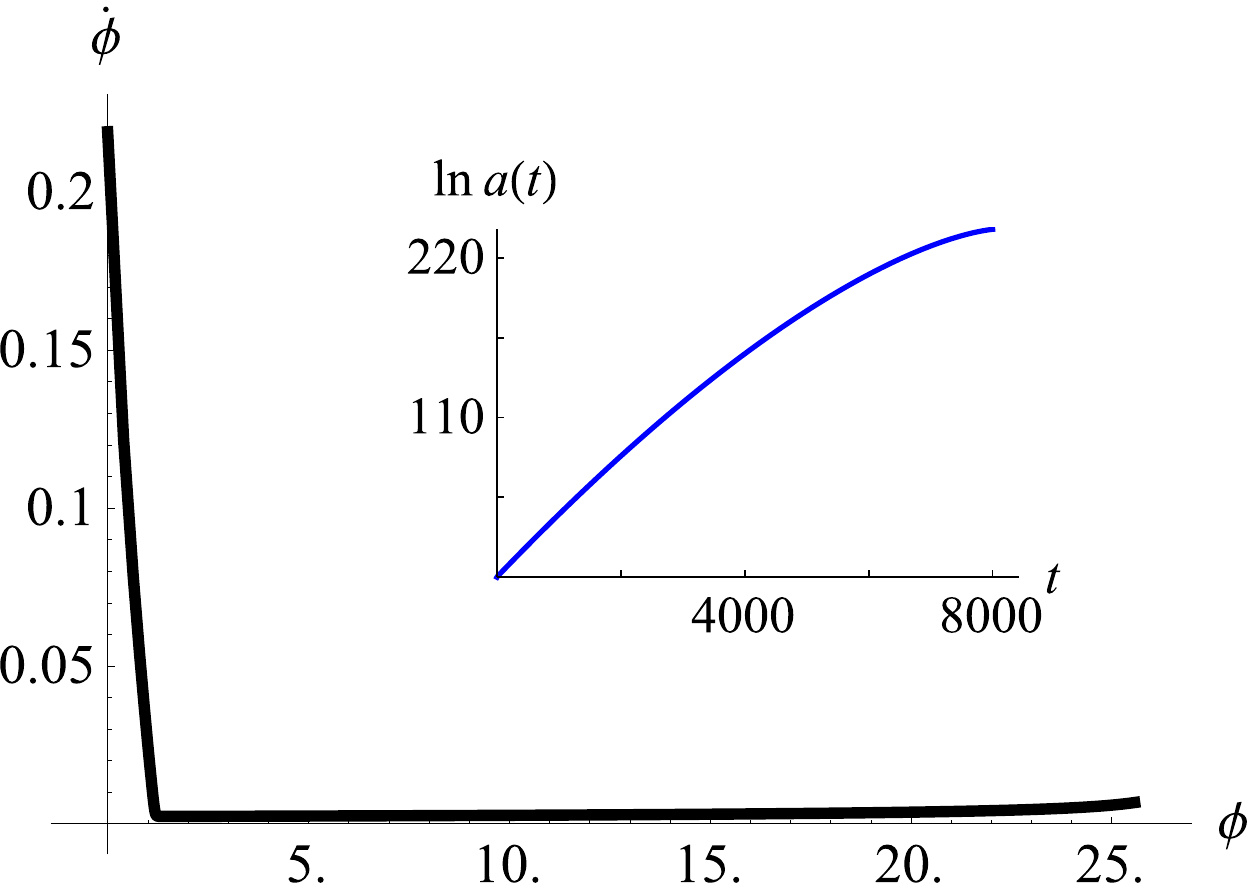}}
\end{minipage}
\begin{minipage}{.32\linewidth}
\subfloat[]{\includegraphics[scale=0.4]{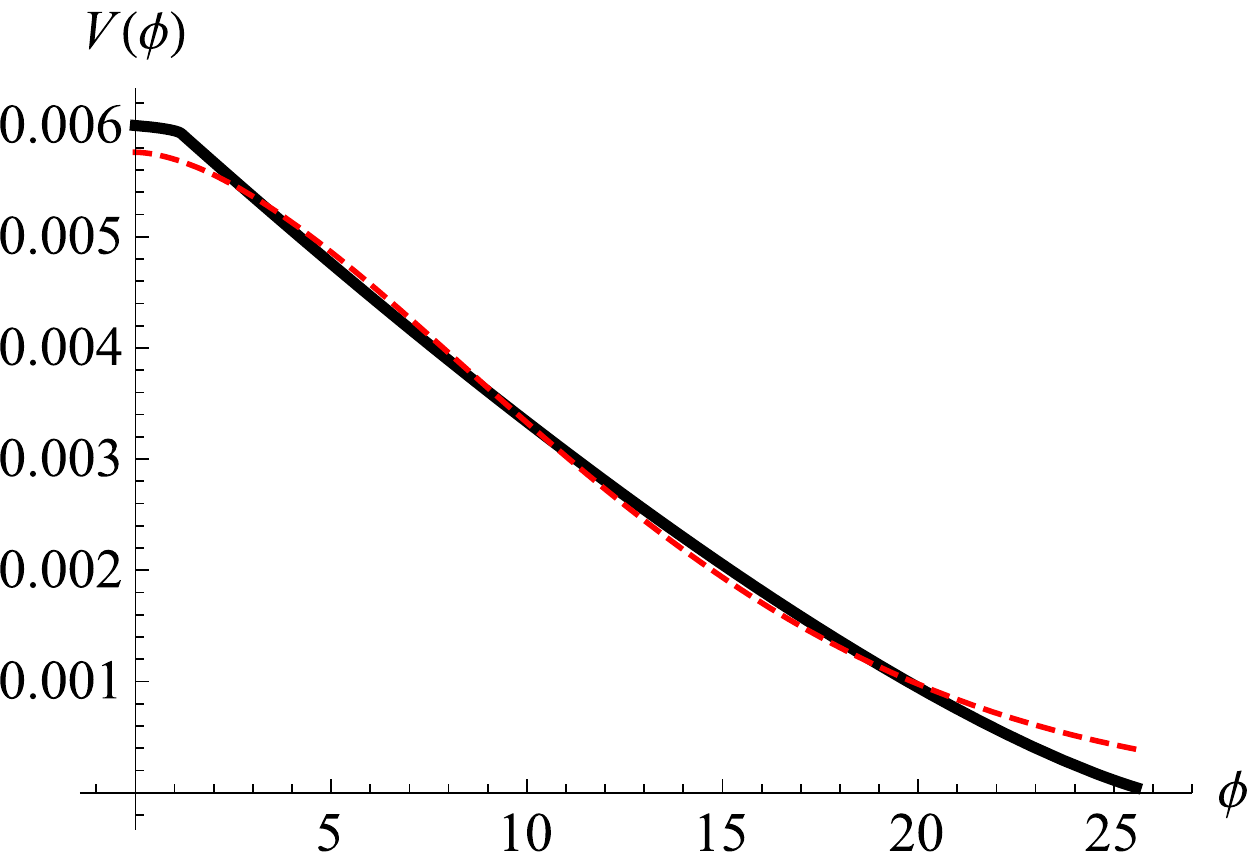}}
\end{minipage}\par\medskip
\begin{minipage}{.32\linewidth}
\subfloat[]{\includegraphics[scale=0.25]{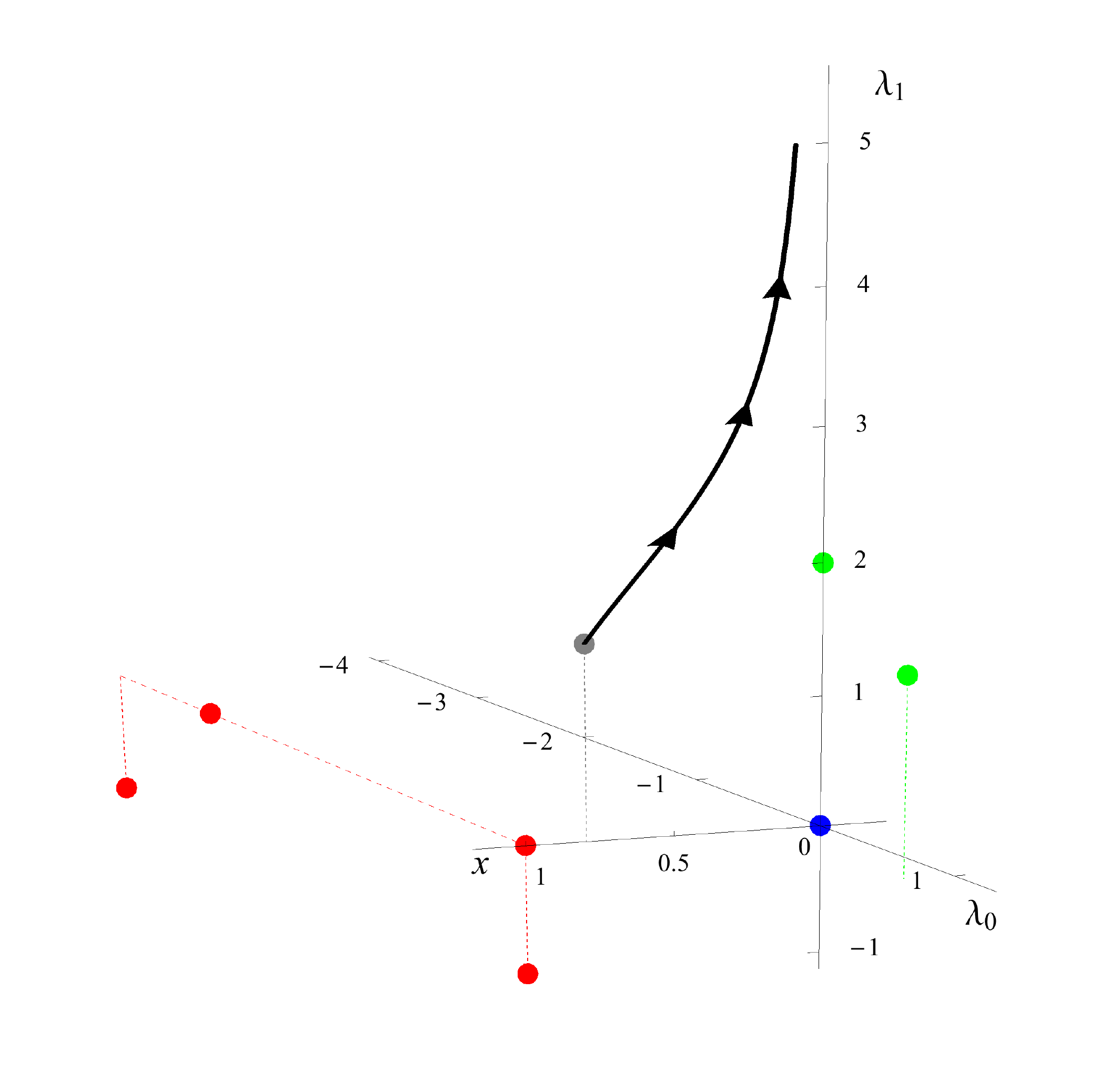}}
\end{minipage}
\begin{minipage}{.32\linewidth}
\subfloat[]{\includegraphics[scale=0.4]{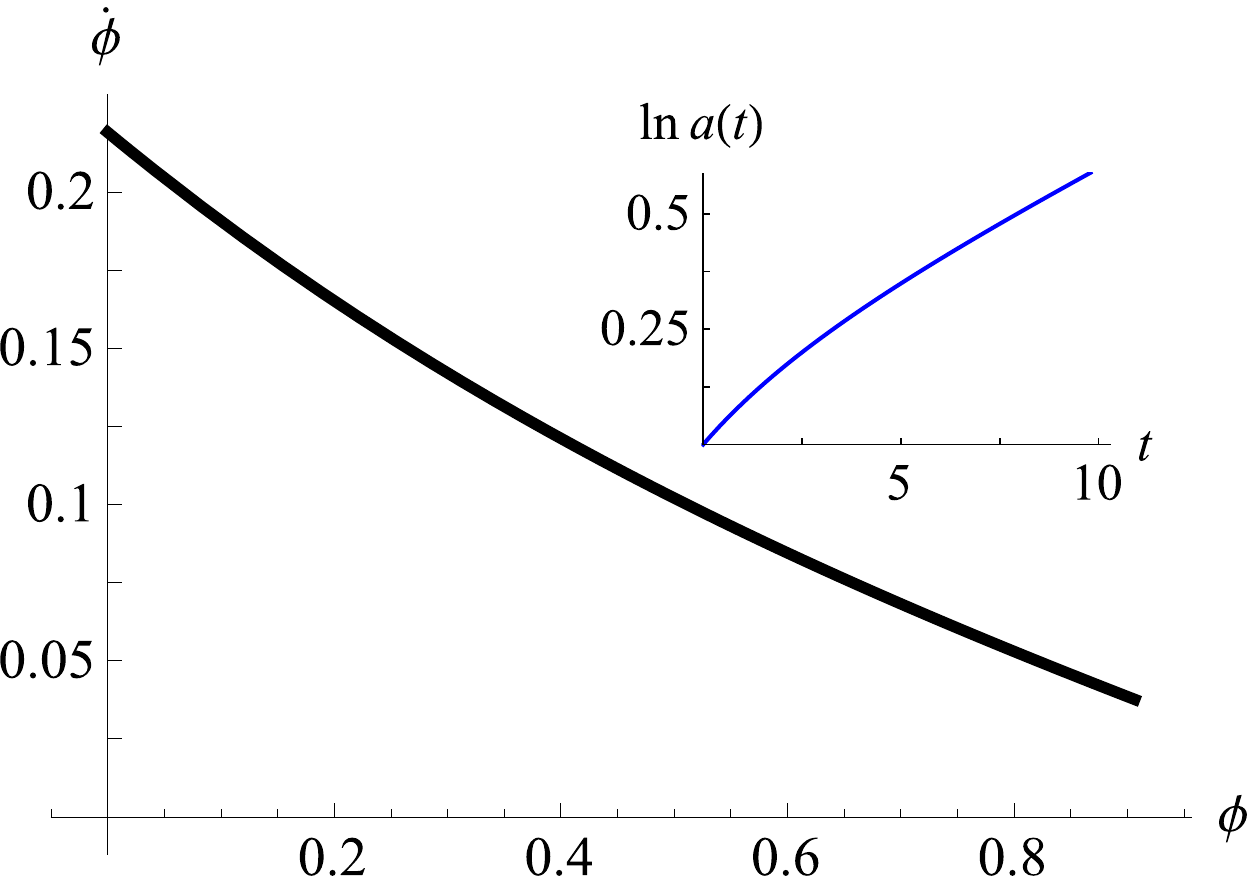}}
\end{minipage}
\begin{minipage}{.32\linewidth}
\subfloat[]{\includegraphics[scale=0.4]{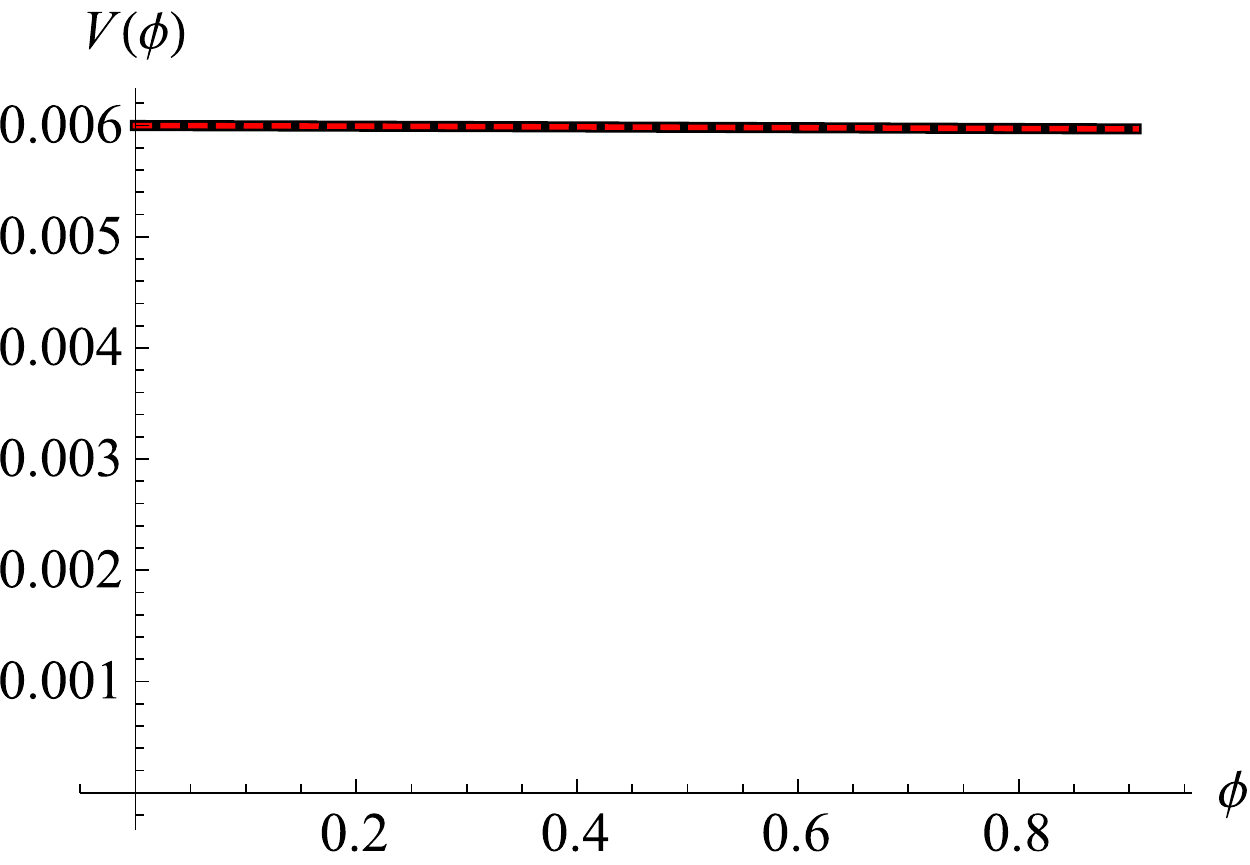}}
\end{minipage}
\caption{Second-order EFT phase space and SSF realizations for $\lambda_{2}=2$ and $H (\tau_i) = 0.1$ (in units of $M_{\rm pl}$), with $(x (\tau_i) , \lambda_0 (\tau_i) , \lambda_1 (\tau_i) ) = (0.8, 0.01, 0.5)$ (top row), $(0.8, 0.01, 1.0)$ (middle row), and $(0.8, 0.01, 1.5)$ (bottom row). In each row, the first plot displays the system's trajectory through the EFT phase space (with the $y$-axis suppressed); inflationary fixed points are displayed as green dots, non-inflationary fixed points are displayed as red dots, a non-hyperbolic fixed point (at the origin of the coordinate system) is displayed as a blue dot, and the starting point of each trajectory is displayed as a gray dot. All dotted lines are drawn to guide the eye. The second plot displays $\dot\phi$ vs.~$\phi$, with $\ln a(t)$ displayed in the inset. The third plot displays $V(\phi)$ as obtained parametrically from the EFT dynamical system (black) and as a fit to the form of $V (\phi)$ in Eq.~(\ref{EQN:FFfit}) (red dashed). Parameters for each fit are given in Table~\ref{TAB:SyTwoCase1a(i)}. In the phase-space plots, the direction of flow along each trajectory is indicated with arrows. In the latter two plots in each row, time evolution flows from left to right. 
}
\label{FIG:SyTwoCase1a(i)}
\end{figure*}

\begin{figure*}
\begin{minipage}{.32\linewidth}
\subfloat[]{\includegraphics[scale=0.25]{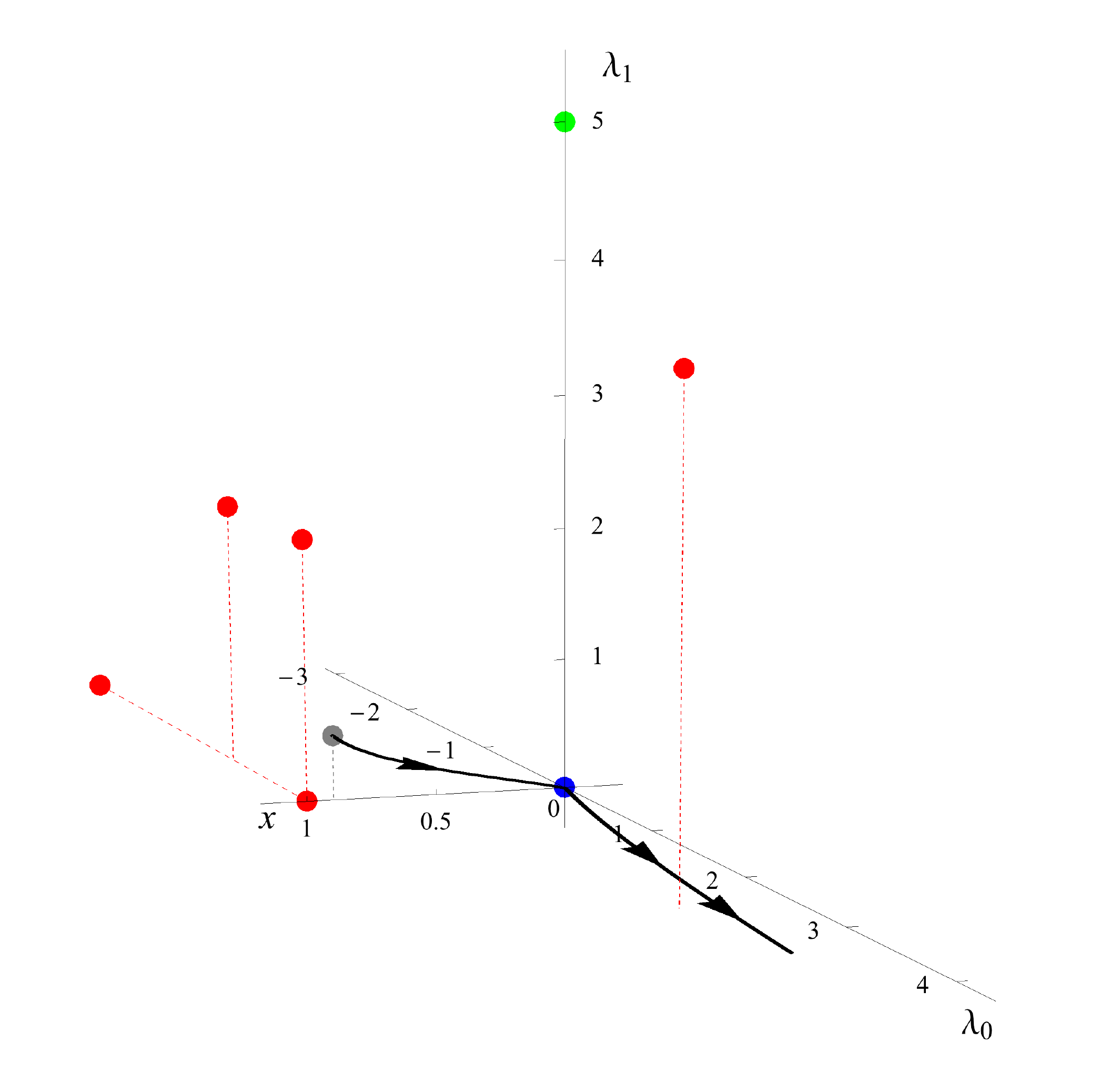}}
\end{minipage}
\begin{minipage}{.32\linewidth}
\subfloat[]{\includegraphics[scale=0.4]{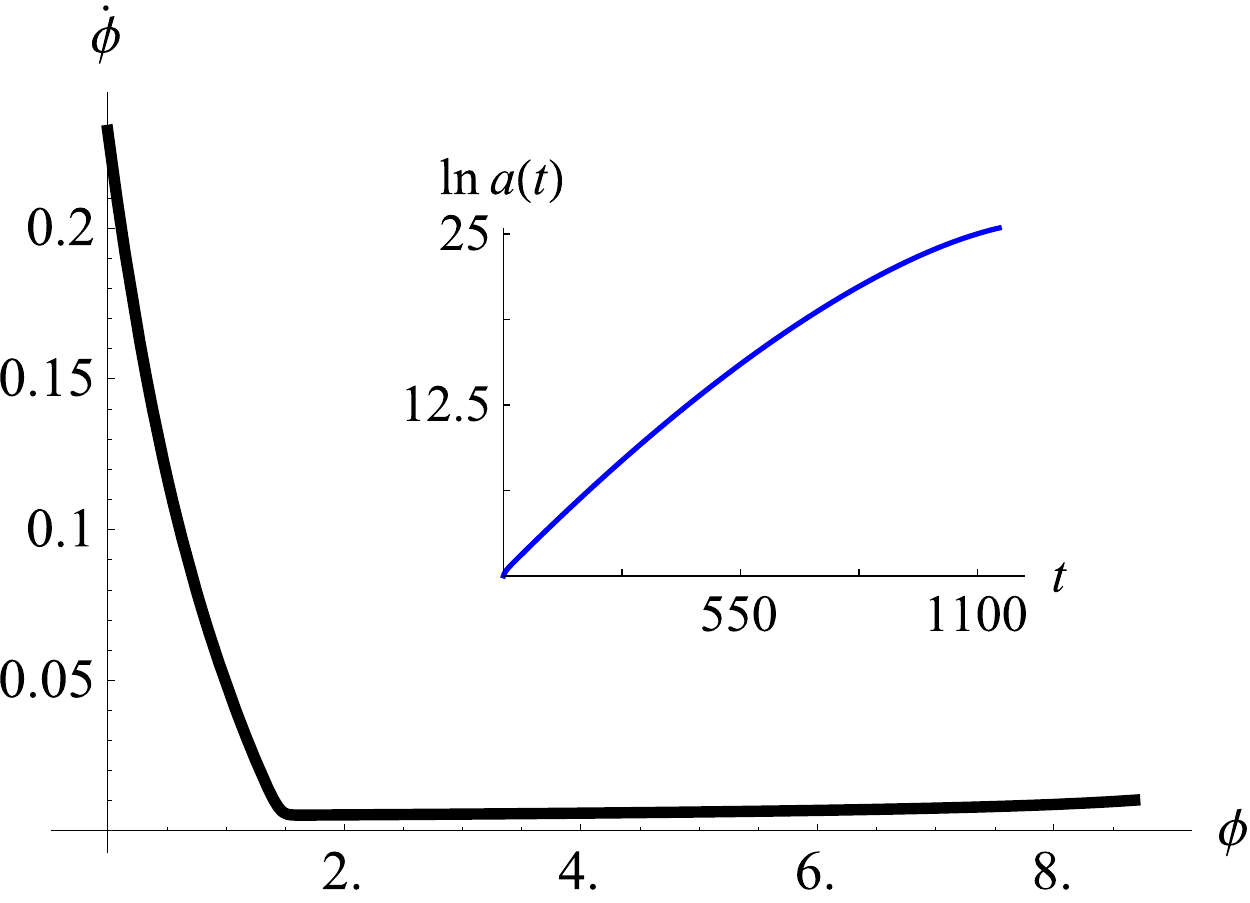}}
\end{minipage}
\begin{minipage}{.32\linewidth}
\subfloat[]{\includegraphics[scale=0.4]{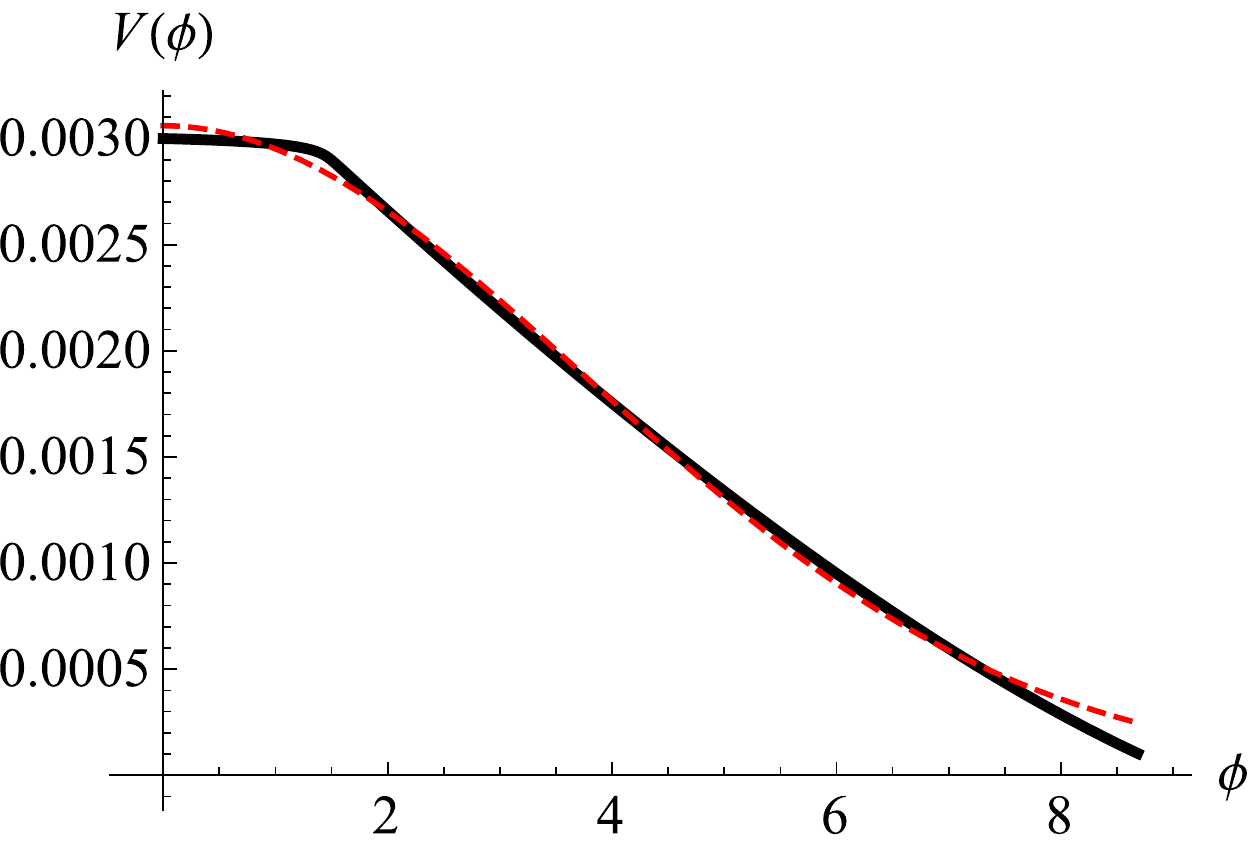}}
\end{minipage}\par\medskip
\begin{minipage}{.32\linewidth}
\subfloat[]{\includegraphics[scale=0.25]{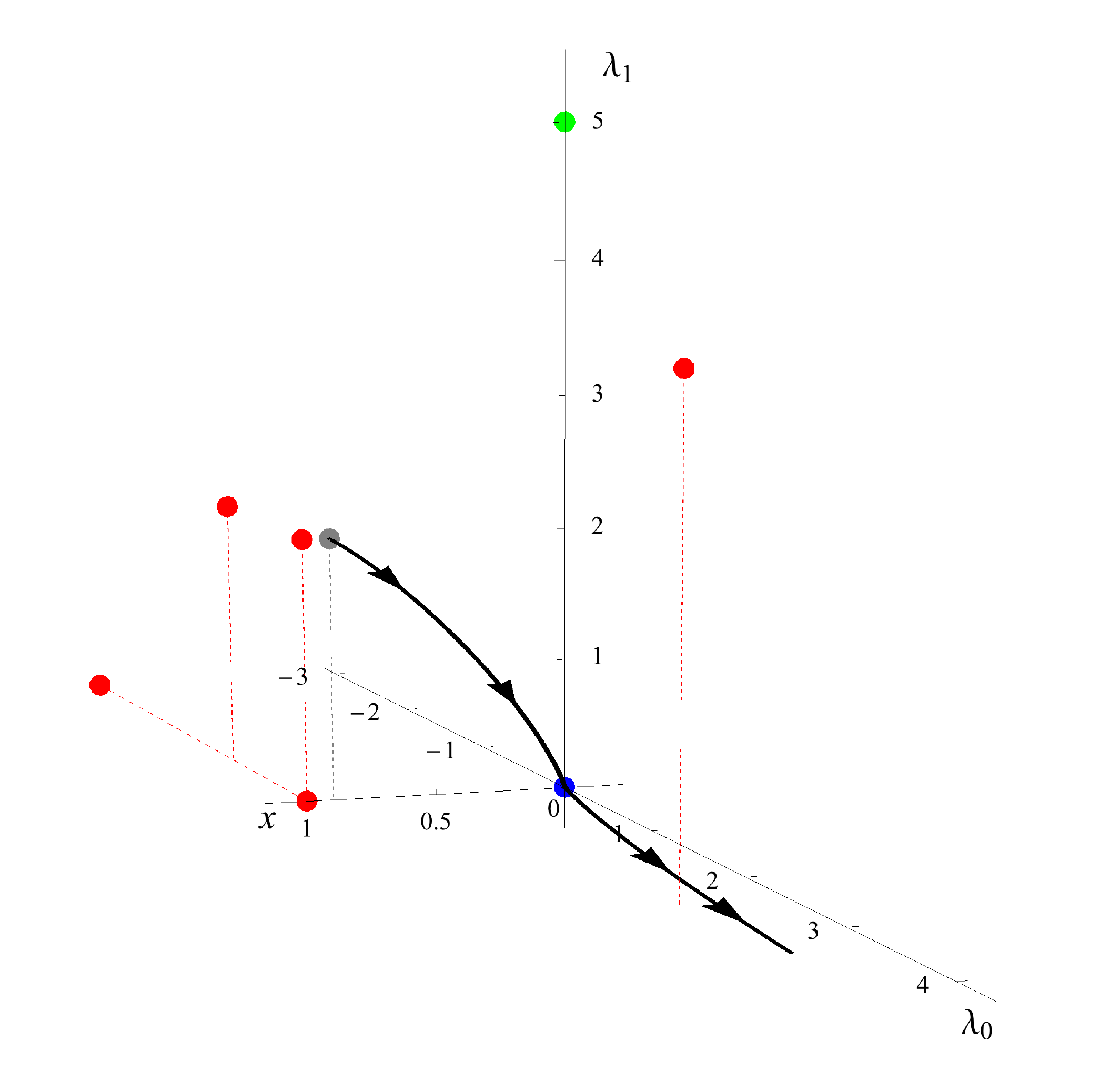}}
\end{minipage}
\begin{minipage}{.32\linewidth}
\subfloat[]{\includegraphics[scale=0.4]{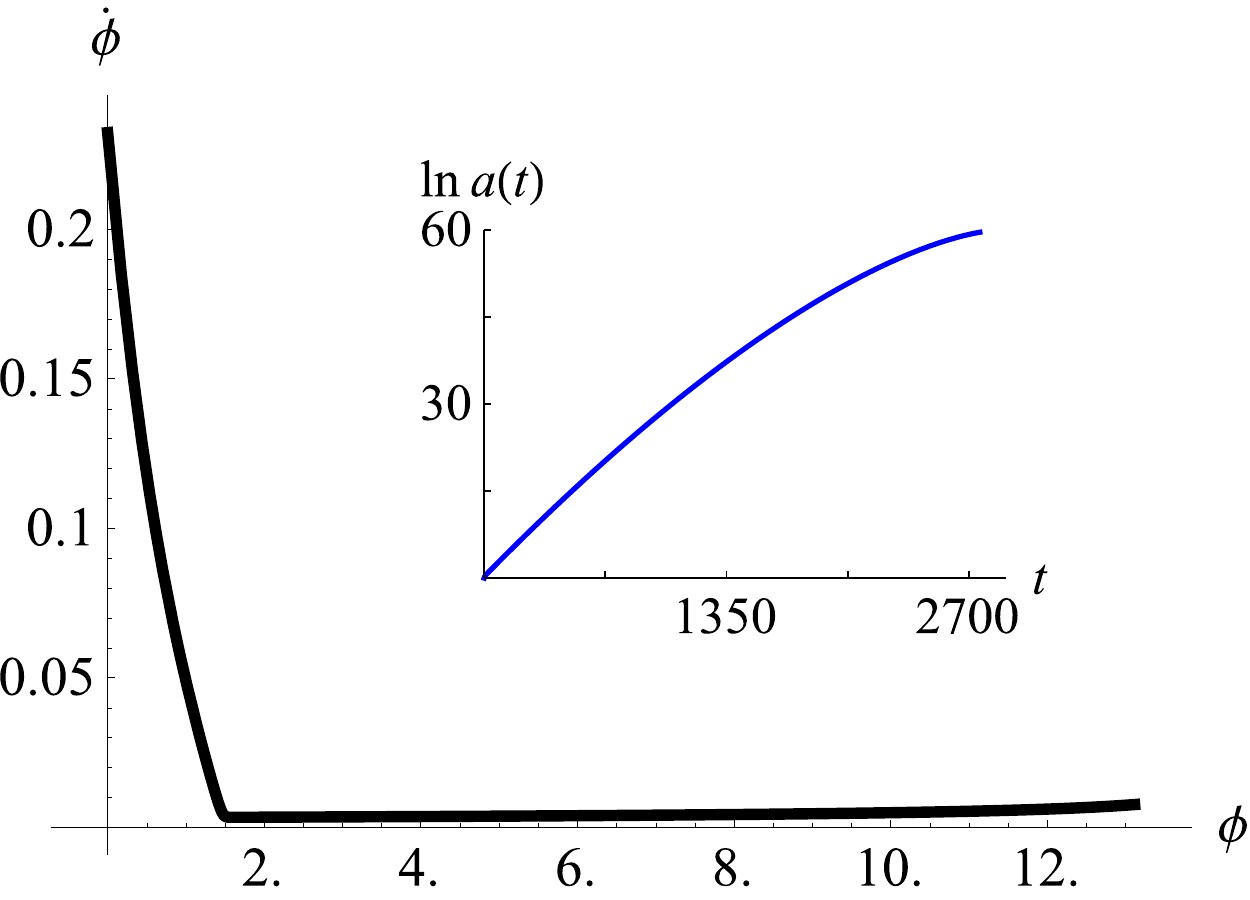}}
\end{minipage}
\begin{minipage}{.32\linewidth}
\subfloat[]{\includegraphics[scale=0.4]{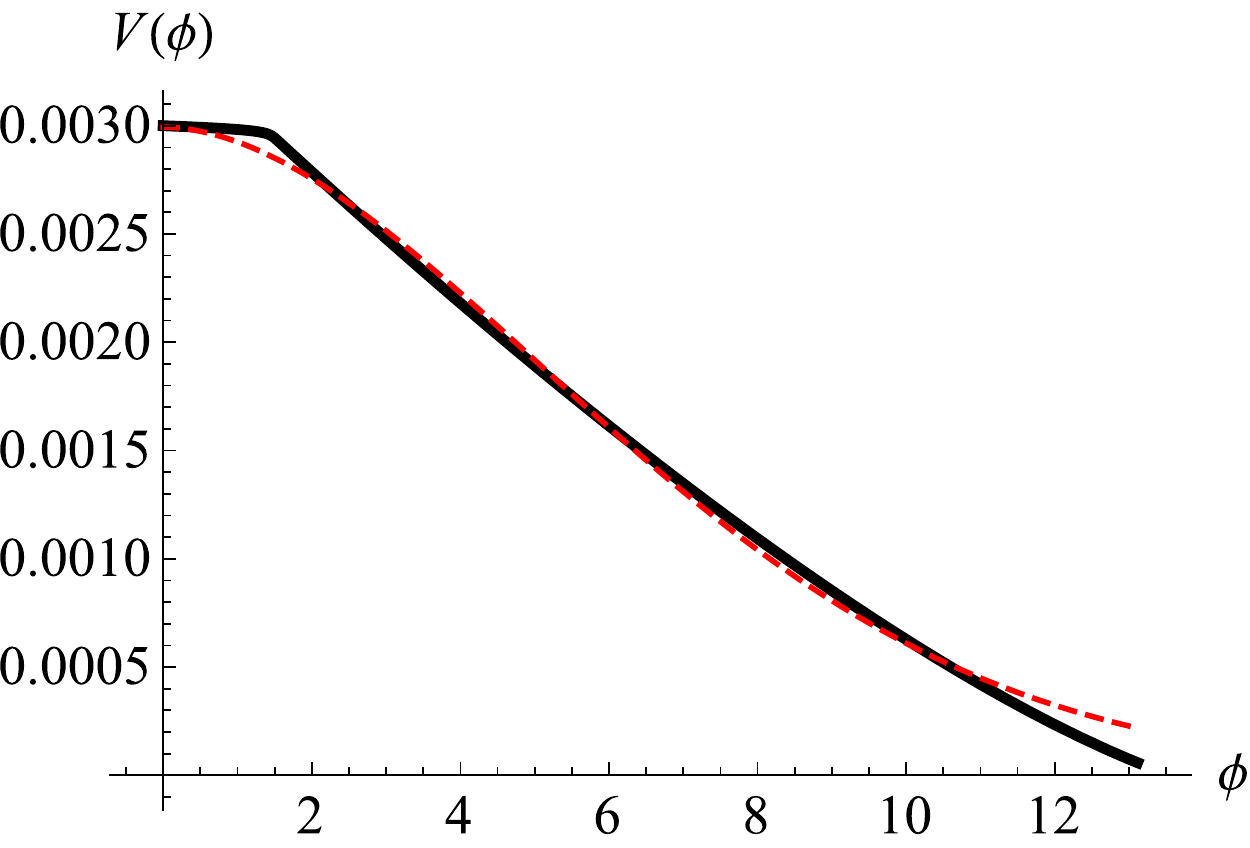}}
\end{minipage}\par\medskip
\begin{minipage}{.32\linewidth}
\subfloat[]{\includegraphics[scale=0.25]{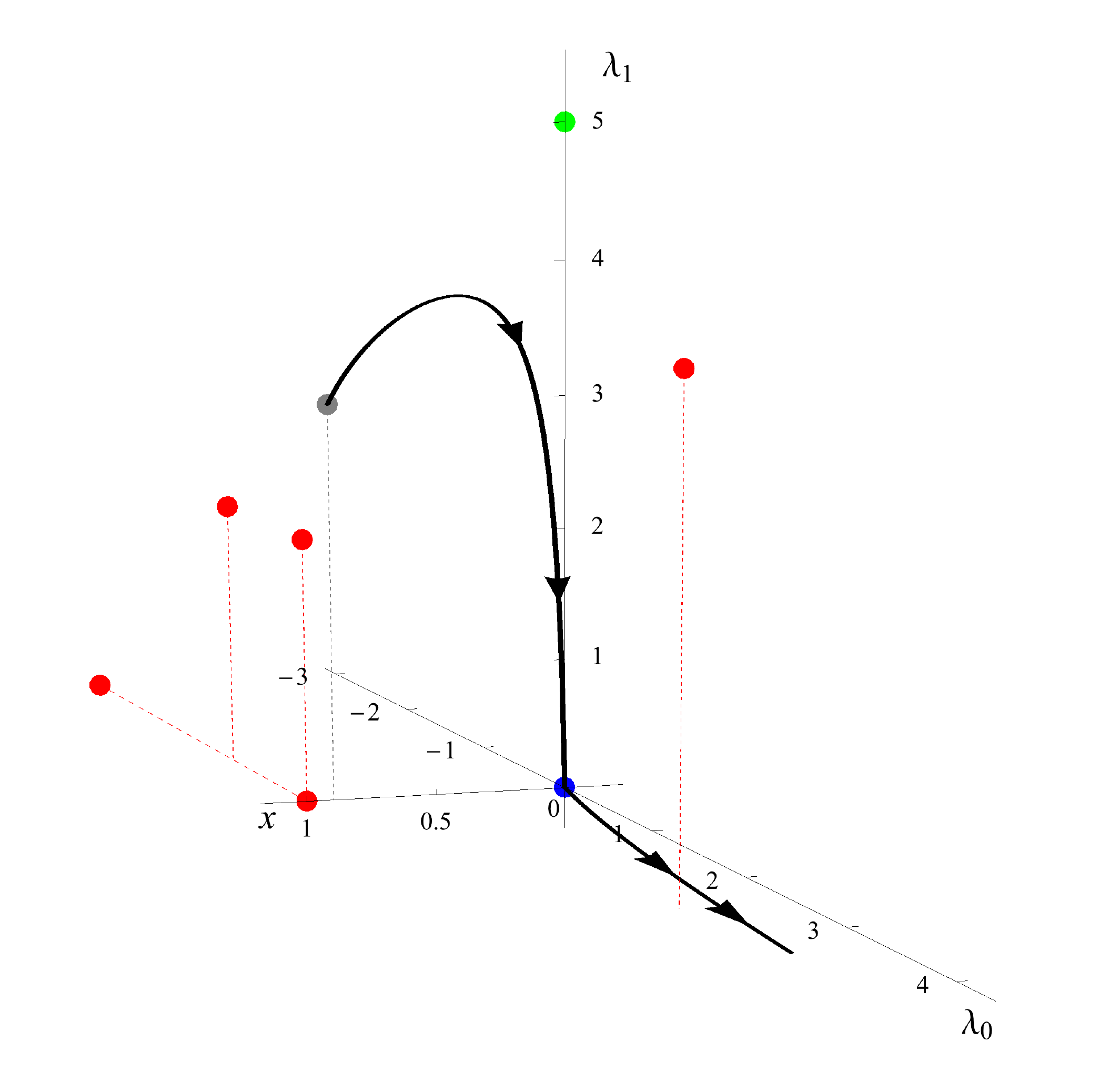}}
\end{minipage}
\begin{minipage}{.32\linewidth}
\subfloat[]{\includegraphics[scale=0.4]{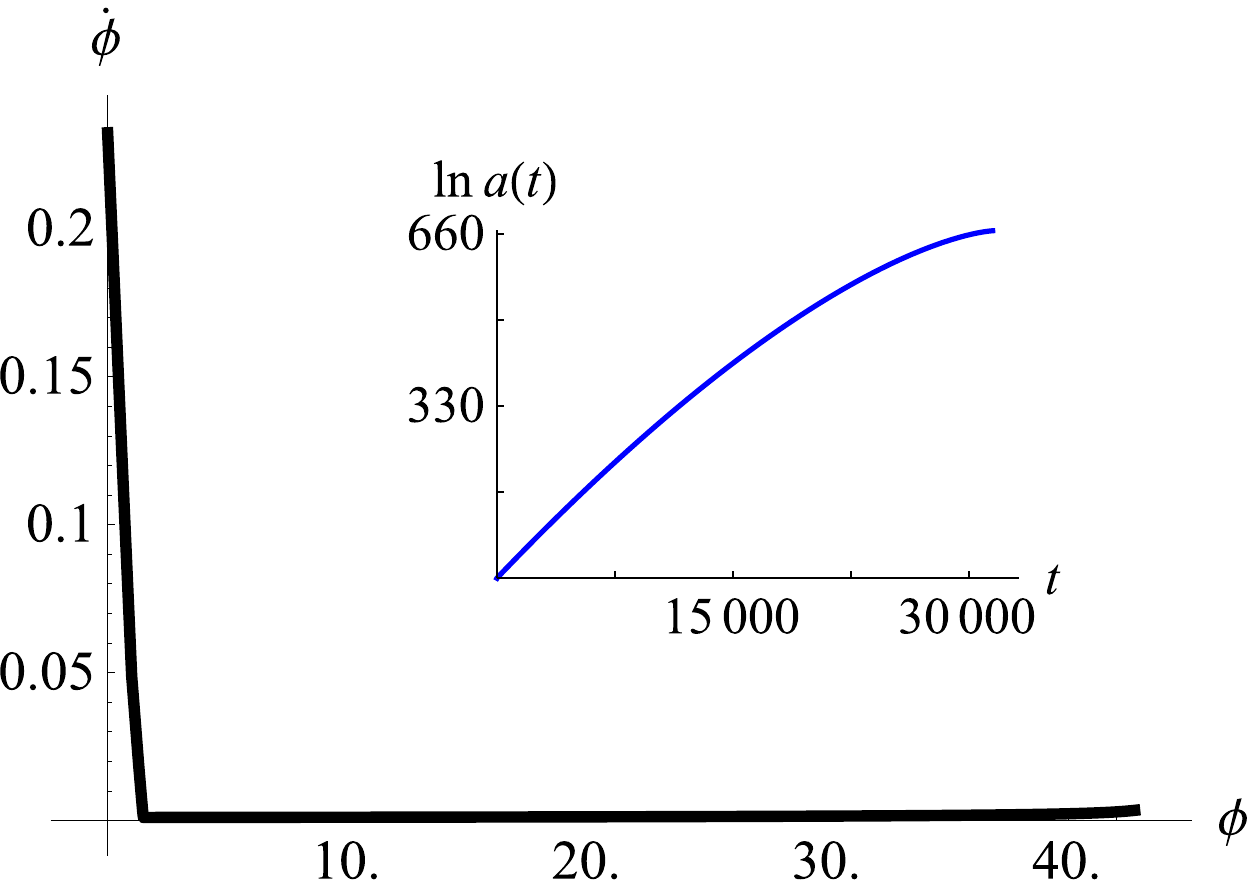}}
\end{minipage}
\begin{minipage}{.32\linewidth}
\subfloat[]{\includegraphics[scale=0.4]{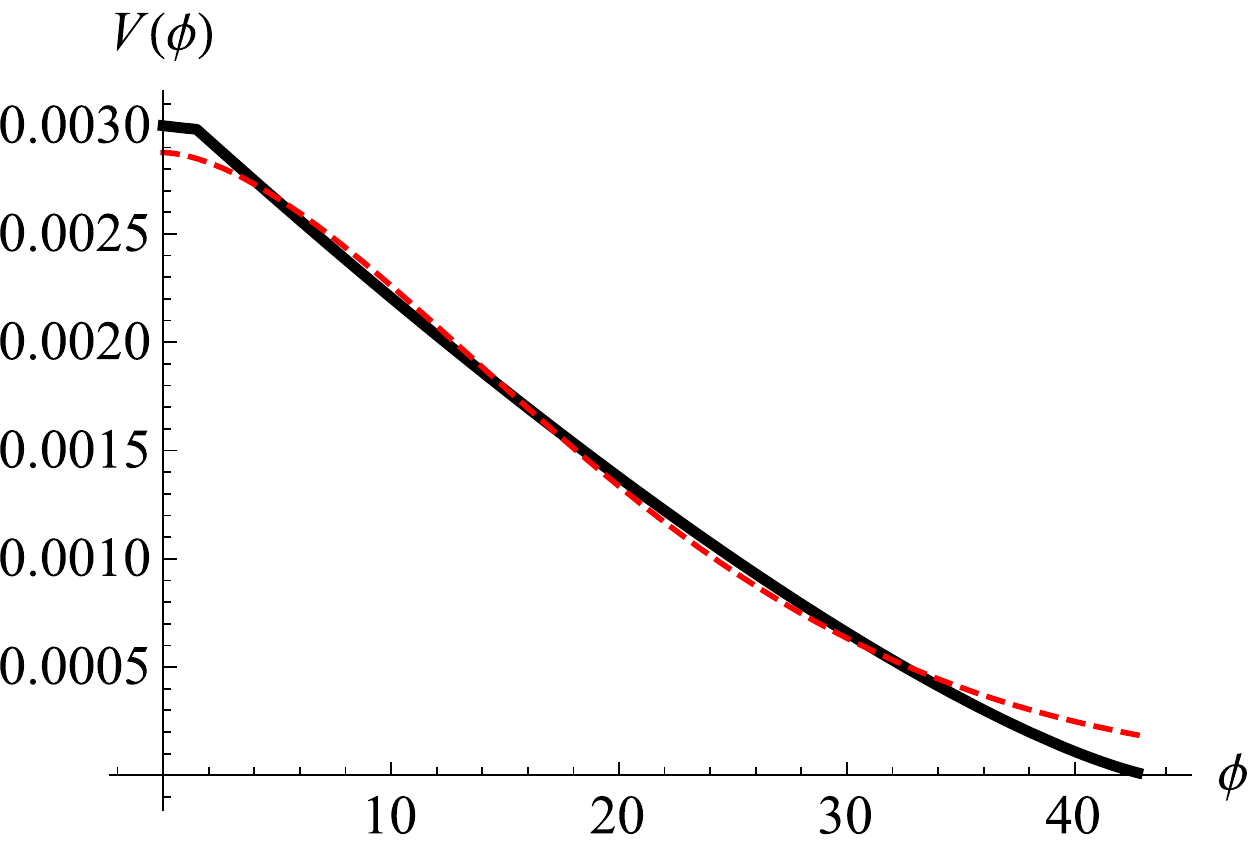}}
\end{minipage}
\caption{Second-order EFT phase space and SSF realization for $\lambda_{2}=5$ and $H(\tau_{i})=0.1$ (in units of $M_{\rm pl})$, with $(x (\tau_i) , \lambda_0 (\tau_i) , \lambda_1 (\tau_i) ) = (0.9, 0.01, 0.5)$ (top row), $(0.9, 0.01, 2)$ (middle row), and $(0.9, 0.01, 3)$ (bottom row). In each row, the first plot displays the EFT phase space (with the $y$-coordinate suppressed); inflationary fixed points are displayed as green dots, non-inflationary fixed points are displayed as red dots, a non-hyperbolic fixed point (at the origin of the coordinate system) is displayed as a blue dot, and the starting point of the trajectory is displayed as a gray dot. All dotted lines are drawn to guide the eye. The second plot displays $\dot\phi$ vs.~$\phi$, with $\ln a(t)$ displayed in the inset. The third plot displays $V(\phi)$ as obtained parametrically from the EFT dynamical system (black) and as a fit to the form of $V (\phi)$ in Eq.~(\ref{EQN:FFfit}) (red dashed). Parameters for the fit are described in Table~\ref{TAB:SyTwoCase1a(i)}. In the phase-space plots, the direction of flow along each trajectory is indicated with arrows. In the latter two plots, time evolution flows from left to right. 
}
\label{FIG:SyTwoCase2b}
\end{figure*}

To find the fixed points of the system, we set the right-hand sides of Eqs.~(\ref{EQN:xFtimeTWO})--(\ref{EQN:lonetimeTWO}) to zero, subject to the constraint of Eq.~(\ref{EQN:ConstraintTWO}). One finds there are at most six hyperbolic fixed points for the system, whose stability properties depend on the value of $\lambda_{2}$. A table summarizing properties of the fixed points at this order is presented in Table~\ref{TAB:SO}. As in the first-order case, there are at most two inflationary fixed points, {\bf FP2a} and {\bf FP2f}. {\bf FP2a}, a saddle point, corresponds to an exact de Sitter background.  {\bf FP2f} is a regularly inflating saddle focus-node for $0<\lambda_{2}<4$. We further note that there exists a non-hyperbolic inflationary fixed point with coordinates $(x,y,\lambda_{0},\lambda_{1})=(0,1,0,0)$. This fixed point appears to play an important role in the dynamics at second order, as one can glean from 
Figs.~\ref{FIG:SyTwoCase1a(i)} and~\ref{FIG:SyTwoCase2b}.

\begin{table*}[htpb!]
\begin{center}
\begin{tabular}{ c | c | c | l }
  \emph{Fixed point} & \emph{Inflationary?}  &\emph{Eigenvalues} & \emph{Stability properties} \\
  $(x, y, \lambda_{0},\lambda_{1} )$ & $(\epsilon= 3x<1?)$&[\emph{Hyperbolic iff}]&\\
  \Xhline{2pt}
  {\bf FP2a} 														&  Yes							& 	$\{-6, -3,-\lambda_{2}, \lambda_{2}\}$ 				& $\lambda_{2}\neq0$: {\small Saddle} \\$(0,1,0,\lambda_{2})$
  																	&			   								&		$[\lambda_{2}\neq 0]$							& \\
  \hline
  {\bf FP2b} 														&  No							& 	$\{9, -3,3,3- \lambda_{2}\}$ 				& $\lambda_{2}\neq3$: {\small Saddle} \\$(1,0,-3,0)$
  																	&			   								&		$[\lambda_{2}\neq 3]$							& \\
  \hline
{\bf FP2c} 														&  No							& 	$\{6, 3,3,3- \lambda_{2}\}$ 				& $\lambda_{2}<3$: {\small Unstable} \\$(1,0,0,0)$
  																	&			   								&		$[\lambda_{2}\neq 3]$							&$\lambda_{2}>3$: {\small Saddle} \\
  \hline
{\bf FP2d} 														&  No							& 	$\{6, 3,6-\lambda_{2},-3+\lambda_{2}\}$ 				& $\lambda_{2}<3$: {\small Saddle} \\$(1,0,0,-3+\lambda_{2})$
  																	&			   								&		$[\lambda_{2}\neq 3,6]$							&$3<\lambda_{2}<6$: {\small Unstable} \\
&&& $\lambda_{2}>6$: {\small Saddle} \\ \hline
{\bf FP2e} 														&  No							& 	$\{3, 12-\lambda_{2},-6+\lambda_{2}, -3+\lambda_{2}\}$ 				& $\lambda_{2}<3$: {\small Saddle} \\$(1,0,-6+\lambda_{2},-3+\lambda_{2})$
  																	&			   								&		$[\lambda_{2}\neq 3,6,12]$							&$3<\lambda_{2}<6$: {\small Saddle} \\
&&& $6<\lambda_{2}<12$: {\small Unstable} \\
&&& $\lambda_{2}>12$: {\small Saddle} \\ \hline
{\bf FP2f} 														&  Yes	 ($\lambda_{2}<4$)						& 	$\{\frac{-6+\lambda_{2}}{2}, *,*,*\}$ 				& $0<\lambda_{2}<4$: {\small Saddle focus} \\$(\frac{\lambda_{2}}{12},1-\frac{\lambda_{2}}{12},\frac{\lambda_{2}}{2},\frac{3\lambda_{2}}{4} )$
  																	&			   								&		$[\lambda_{2}\neq 0,6,12]$							& \\
\end{tabular}
\caption{Hyperbolic second-order fixed points and their stability properties. In the final row, eigenvalues labeled with a * are (different) functions of $\lambda_{2}$, whose precise functional form will not be needed here. For this row, the stability analysis has been carried out numerically, and results in the table are quoted for the domain over which {\bf FP2f} is a regularly inflating hyperbolic fixed point ($0<\lambda_{2}<4$).
We have (numerically) extended the stability analysis beyond these limits (for positive $\lambda_{2}$ only), and this analysis suggests that {\bf FP2f} is not an attractor (or a stable focus-node) for any positive value of $\lambda_{2}$.}\label{TAB:SO}
\end{center}
\end{table*}

We may generate an SSF realization of the second-order EFT dynamical system in a very similar way to the zeroth and first orders (see Sec.~\ref{SEC:GenPotential}). 
Motivated by the analysis for those orders, we first analyze SSF realizations of {\bf FP2a} and {\bf FP2f}, before analyzing SSF realizations of trajectories with kinetic-energy-dominated initial conditions.

Fixed point {\bf FP2a} has coordinates $(x,y, \lambda_{0}, \lambda_{1})=(0,1,0, \lambda_{2})$. Any trajectory that begins at these coordinates, will, of course, remain there for all $t$. Akin to {\bf FP1a} in Sec.~\ref{SEC:FirstOrderSystems}, one can show that evolution at the fixed point {\bf FP2a} corresponds to $\dot{\phi}=\dot{H}=0$: namely, an unending de Sitter phase.

Fixed point {\bf FP2f} corresponds to a particular solution of power-law inflation, akin to fixed point {\bf FP0b} discussed in Sec.~\ref{SEC:ZerothOrderSystems}. One can obtain the relevant equations for the SSF realization at second order by substituting $\lambda_{2}/12$ for $\lambda_{0}/6$ in each relevant equation for {\bf FP0b} in Sec.~\ref{SEC:ZerothOrderSystems}. This is an example of a more general pattern, which we demonstrate in 
Appendix~\ref{APP:Mthorder}: at each order $M\geq 0$, there exists a fixed point whose SSF realization corresponds to a particular solution to power-law inflation.

We now turn to a (restricted) analysis of example trajectories in the EFT phase space and their SSF realizations at second order. We highlight two important points. First, for certain kinetic-energy-dominated initial conditions, one may find trajectories that undergo at least 60 efolds of inflation, even though there does not exist an inflationary attractor in the EFT phase space. Second, the simple functional form for $V (\phi)$ in Eq.~(\ref{EQN:FFfit}) can again be fit to SSF realizations of trajectories.

We present results for two different values of $\lambda_{2}$: $\lambda_{2}=2$ and $\lambda_{2}=5$. These values were chosen so that {\bf FP2f} (see Table~\ref{TAB:SO}) is either an inflationary fixed point ($\lambda_{2}=2$) or a non-inflationary fixed point ($\lambda_{2}=5$). In order to display salient features of each case, we consider illustrative examples, setting $x (\tau_i) = 0.8$ for $\lambda_2 = 2$ and $x (\tau_i) = 0.9$ for $\lambda_2 = 5$, and selecting values of $\lambda_0 (\tau_i)$ and $\lambda_1 (\tau_i)$ that highlight interesting features of the ensuing dynamics. Best-fit parameters for $V (\phi)$ in each case are shown in Table~\ref{TAB:SyTwoCase1a(i)}. Dynamical trajectories are shown in Fig.~\ref{FIG:SyTwoCase1a(i)} ($\lambda_2 = 2$) and Fig.~\ref{FIG:SyTwoCase2b} ($\lambda_2 = 5$). In each set of figures, we suppress the $y$-axis, since all trajectories satisfy the constraint $y = 1 - x$.

\begin{table*}[ht!]
\begin{center}
\begin{tabular}{c | c|c|c|c|c|c|c|c}
  $\lambda_2$ & $x(\tau_{i})$ &$\lambda_{0}(\tau_{i})$&$\lambda_{1}(\tau_{i})$& $V_{0}$ & $\alpha$ & $\beta$ & $N$ & End of integration \\ \Xhline{2pt}
 2 & $0.8$& $0.01$ & $0.5$ & $5.93 \times 10^{-3}$ & 0.0297 & 1.79 & 53 & System stops inflating\\
 2 & $0.8$& $0.01$ & $1.0$ & $5.76 \times 10^{-3}$ & 0.0111 & 1.70 & 240 & System stops inflating \\
 2 & $0.8$& $0.01$ & $1.5$ & $6.00 \times 10^{-3}$ & 0.00575 & 1.11 & 1 & $\lambda_{1} \sim 5$\\
 \hline
 5 & $0.9$& $0.01$ & $0.5$ & $3.06 \times 10^{-3}$ & 0.0363 & 1.96 & 25 & System stops inflating \\
 5 & $0.9$& $0.01$ & $2$ & $2.99 \times 10^{-3}$ & 0.0231 & 1.84 & 60 & System stops inflating \\
 5 & $0.9$& $0.01$ & $3$ & $2.88 \times 10^{-3}$ & 0.00509 & 1.67 & 666& System stops inflating\\
\end{tabular}
\captionsetup{width=\linewidth}
\caption{Best-fit parameters for $V (\phi)$ of Eq.~(\ref{EQN:FFfit}) for second-order systems with $\lambda_2 = 2$ or $\lambda_2 = 5$, and $H (\tau_i) = 0.1$, in units of $M_{\rm pl}$. 
$N$ is the total number of efolds that the system undergoes before integration was stopped, either because inflation ended [with $x (\tau)$ crossing $1/3$ from below, which corresponds to $\epsilon = 1$], or because $\lambda_1 (\tau)$ crossed 5 from below.}
\label{TAB:SyTwoCase1a(i)}
\end{center}
\end{table*}

In sum, there exist trajectories for second-order systems that begin from kinetic-energy-dominated initial conditions and flow into inflationary states, even though for second-order systems --- unlike the zeroth- and first-order cases --- there do not exist inflationary attractors (or inflationary stable focus-nodes) within the EFT phase space. Even in the absence of such fixed-points at second order, inflation for such trajectories can persist for 60 or more efolds, and such trajectories are again well-fit by the simple functional form for $V (\phi)$ of Eq.~(\ref{EQN:FFfit}).

\section{$M$th-order system: $M\geq 3$\label{APP:Mthorder}}

The $M$th-order analysis corresponds to fixing $\lambda_{M}$ = constant. We focus on the case in which $M\geq 3$. To make things transparent, we begin by enumerating the equations that define this order, before outlining the strategy we will use to derive fixed points (which we adapt from Ref.~\cite{frusciante+al_14}). 

For $\lambda_{M}$ = constant, Eqs.~(\ref{EQN:xFtimeM})--(\ref{EQN:ConstraintM}), fall into two blocks. The first block includes (constrained) dynamical equations for $x$, $y$, and $\lambda_{0}$:
\begin{subequations}\label{EQN:blockONE}
\begin{align}
\frac{dx}{d\ln a} &=\lambda_{0} y -3 x +3 x^2-3 x y\label{EQN:xFtimeMapp},\\
\frac{dy}{d\ln a} &=-\lambda_{0} y +3 y +3 x y-3 y^2\label{EQN:yFtimeMapp},\\
\frac{d\lambda_{0}}{d\ln a} &= \left[-\lambda_{1}+\lambda_{0}+\frac{3}{2}(1+x-y)\right]\lambda_{0}\label{EQN:lzerotimeMapp},\\
1&=x+y \label{EQN:ConstraintMapp},
\end{align}
\end{subequations}
where, as in Eq.~(\ref{EQN:Ep}), the slow-roll parameter is given by $\epsilon=3x$. The second block of equations comprises dynamical equations for the $\lambda_{m}$ with $1\leq m \leq M-1$:
\begin{subequations}\label{EQN:blockTWO}
\begin{align}
\frac{d\lambda_{1}}{d\ln a} &= \left[-\lambda_{2}+\lambda_{1}+\frac{3}{2}(1+x-y)\right]\lambda_{1}\label{EQN:lonetimeMapp},\\
\frac{d\lambda_{2}}{d\ln a} &= \left[-\lambda_{3}+\lambda_{2}+\frac{3}{2}(1+x-y)\right]\lambda_{2}\label{EQN:ltwotimeMapp},\\
&\vdots\nonumber\\
\frac{d\lambda_{M-1}}{d\ln a} &= \left[-\lambda_{M}+\lambda_{M-1}+\frac{3}{2}(1+x-y)\right]\lambda_{M-1}\label{EQN:lMminusonetimeMapp}.
\end{align}
\end{subequations}
Comparing the two blocks of equations, we see that, for $m \geq 2$, the variables $x$, $y$, and $\lambda_{0}$ depend on $\lambda_{m}$ only via their dependence on $\lambda_1$. Thus, adapting the procedure outlined in Ref.~\cite{frusciante+al_14}, we proceed by deriving fixed points for the entire system of equations by first finding fixed points of the first block of equations as functions of $\lambda_{1}$, and then solving for fixed points for the second block of equations after assuming all possible values for the critical value of $\lambda_{1}$, which we denote $\lambda_{1,c}$. (For clarity, we append the subscript `$c$' for `critical' values of the phase-space variables.) This strategy sounds onerous, but things simplify dramatically because we need only  consider cases in which $\lambda_{1,c}=0$ and $\lambda_{1,c}\neq 0$, to solve for fixed points for the entire system of equations in full generality.

We begin by indicating how to derive all possible fixed points, hyperbolic and non-hyperbolic alike. We will focus specifically on inflationary fixed points, and will carry out a stability analysis only for the hyperbolic inflationary fixed points. We have already solved the first block of equations for fixed points, as displayed in Table~\ref{TAB:FO}. We reproduce those in Table~\ref{TAB:initsegFO}, relabeling what now amount to `initial segments' of higher-dimensional fixed points, by which we mean the first few elements of the full set of $M+2$ phase-space variables $\{ x, y, \lambda_0, \lambda_1, \dots , \lambda_{M-1} \}$. In what follows, we consider the two general cases that will allow us to find critical values for each of the $M+2$ phase-space variables, namely, $\lambda_{1,c} = 0$ and $\lambda_{1,c} \neq 0$.

\begin{table}[htbp!]
\begin{center}
\begin{tabular}{c|c||c|c|c}
 Solution label & Inflationary? & $x_{c}$ & $y_{c}$ & $\lambda_{0,c}$ \\ \Xhline{2pt}
$\alpha$ & Yes & 0 & 1& 0 \\
$\beta$ & No & 1 & 0 & 0 \\
$\gamma$ & No & 1 & 0& $-3+\lambda_{1}$ \\
$\delta$ &Yes $(\lambda_{1}<3)$ &$\frac{\lambda_{1}}{9}$&$1-\frac{\lambda_{1}}{9}$& $\frac{2 \lambda_{1}}{3}$
\end{tabular}
\caption{Initial segments from the first block of equations: Eq.~(\ref{EQN:blockONE}).  }
\label{TAB:initsegFO}
\end{center}
\end{table}

\subsection{Case {\bf A}: $\lambda_{1,c}=0$}\label{SEC:CaseA}

For $\lambda_{1,c} = 0$, there are only 3 distinct initial segments, since $\delta\to\alpha$ in Table~\ref{TAB:initsegFO}, only one of which ($\alpha$) is inflationary (with $\epsilon = 3x = 0$). Hence we obtain the initial segments displayed in Table~\ref{TAB:initseg0}.
\begin{table}[htbp!]
\begin{center}
\begin{tabular}{c|c||c|c|c|c}
 Solution label & Inflationary? & $x_{c}$ & $y_{c}$ & $\lambda_{0,c}$& $\lambda_{1,c}$ \\ \Xhline{2pt}
$\alpha^{\prime}$ & Yes & 0 & 1& 0 & 0 \\
$\beta^{\prime}$ & No & 1 & 0 & 0 & 0 \\
$\gamma^{\prime}$ & No & 1 & 0& $-3$ & 0
\end{tabular}
\caption{Distinct initial segments derived from the first block of equations, Eq.~(\ref{EQN:blockONE}), with $\lambda_{1,c}=0$.}\label{TAB:initseg0}
\end{center}
\end{table}

Next we find fixed points of the second block of equations subject to the initial segments displayed in Table~\ref{TAB:initseg0}. There exist three general cases:
\begin{itemize}
\item[(i)] $\lambda_{m,c}=0$ $\textrm{for all}\;2\leq m\leq M-1$;
\item[(ii)] $\lambda_{m,c}\neq0$ $\textrm{for all}\;2\leq m\leq M-1$; and
\item[(iii)] $\lambda_{m,c}=0$ for just some of the $2\leq m\leq M-1$.
\end{itemize}
These cases just list all the ways one can distribute 0's among all (critical values of) phase-space variables that have yet to be determined. Having distributed 0's in this way, the structure of the equations that appear in the second block allows one to determine the nonzero critical values in a straightforward way, as we will now show.\\

\noindent\underline{Case {\bf A}(i)}: ($\lambda_{m,c}=0$ $\textrm{for all}\;2\leq m\leq M-1$)
\\
\\
This is the most straightforward case, as the initial segments displayed in Table~\ref{TAB:initseg0} are appended with 0's for $\lambda_{2}$ though to $\lambda_{M-1}$. One thus derives three fixed points in total, only one of which is inflationary (which we label {\bf IA}(i), with `I' for `inflationary'). We find
\begin{align}\label{EQN:InfAi}
{\bf IA}\textrm{(i)}: & \>\>(x_{c}, y_{c}, \lambda_{0,c}, \lambda_{1,c}, \lambda_{2,c},\dots, \lambda_{M-1,c}) \nonumber\\
&\hspace{1in}= (0,1,0,0,0,\dots, 0).
\end{align}
There are no restrictions on the value of $\lambda_{M}$.\\

\noindent\underline{Case {\bf A}(ii)}: ($\lambda_{m,c}\neq0$ $\textrm{for all}\;2\leq m\leq M-1$)
\\
\\
In this case, for each of the initial segments displayed in Table~\ref{TAB:initseg0}, we can solve the second block of equations for fixed points by starting at the top of the tower of the $\lambda_{m}$'s, namely, at $\lambda_{M-1}$, and working our way down. So, the fixed point $\lambda_{M-1,c}$ can first be determined from Eq.~(\ref{EQN:lMminusonetimeMapp}) by noting that the term in square brackets must vanish, which yields
\begin{equation}
\lambda_{M-1,c}=\lambda_{M}-\frac{3}{2}(1+x_{c}-y_{c})\equiv \lambda_{M}-\epsilon(x_{c}).
\end{equation}
Recall from Eq.~(\ref{EQN:Ep}) that $\epsilon(x) = 3x$. We can then proceed up the second block of equations, sequentially determining $\lambda_{m,c}$ for $m = M-2, \dots, 2$. In general, we find:
\begin{equation}\label{EQN:generallambdaAii}
\lambda_{m,c} = \lambda_{M} - (M-m)\epsilon(x_{c}),
\end{equation}
for $m= 2,3,\dots, M-1$. Each initial segment in Table~\ref{TAB:initseg0} is thus appended with Eq.~(\ref{EQN:generallambdaAii}), giving the corresponding $(M+2)$-dimensional fixed point. Note that the inflationary initial segment ($\alpha^{\prime}$) has $\epsilon(x_{c}) = 0$, and therefore we have a new inflationary fixed point, which we label {\bf IA}(ii), only when $\lambda_{M}\neq 0$, in which case
\begin{align}\label{EQN:InfAii}
{\bf IA}\textrm{(ii)}:& \>\>(x_{c}, y_{c}, \lambda_{0,c}, \lambda_{1,c}, \lambda_{2,c},\dots, \lambda_{M-1,c}) \nonumber\\
&\hspace{0.8in}= (0,1,0,0,\lambda_{M},\dots, \lambda_{M}).
\end{align}
Similarly, this case gives two new non-inflationary fixed points only when $\lambda_{M} \neq (M-m)\epsilon(x_{c})= 3(M-m)$ (for all $m=2,3,\dots, M-1)$.\\

\noindent\underline{Case {\bf A}(iii)}: ($\lambda_{m,c}=0$ for just some $2\leq m\leq M-1$)
\\
\\
In this final case (which can only provide new fixed points when $M>3$), there are, in principle, $2^{M-2}-2$ new ways to distribute 0's among the $\lambda_{m,c}$'s, since $m=2,3,\dots, M-1$, and we have subtracted cases {\bf A}(i) and {\bf A}(ii) from the total number of ways of distributing 0's among $M-2$ variables. Note, however, that not all of these different ways are consistent with the initial segments displayed in Table~\ref{TAB:initseg0}. We illustrate our procedure by considering consistent extensions of the inflating case, $\alpha^\prime$.

The total number of possibilities for consistent extensions of $\alpha^\prime$ simplifies dramatically, because the second block of equations, Eq.~(\ref{EQN:blockTWO}), does not allow for a solution in which a nonzero critical value somewhere in the tower is followed by a critical value that is zero. First consider that this were {\it not} the case. That is, assume that there is some $j\in \{2,3,\dots,M-2\}$ for which $\lambda_{j,c}\neq0$, but for which $\lambda_{j+1,c}= 0$. Then the second block of equations would yield the following equation:%
\begin{equation}
 \left[-\lambda_{j+1,c}+\lambda_{j,c}+\epsilon(x_{c})\right]\lambda_{j,c}=0.
\end{equation}
But then noting that for $\alpha^{\prime}$, $\epsilon(x_{c})=0$, we find 
\begin{equation}
\lambda_{j+1,c}= 0\implies \lambda_{j,c}= 0,
\end{equation}
contradicting the original assumption that $\lambda_{j,c}\neq0$. 

This argument leaves just $M-3$ new cases, namely, the cases in which there exist a string of $k-1$ zeros starting from $\lambda_{2,c}=0$ up to and including some $\lambda_{k,c}=0$, with $\lambda_{j,c}\neq 0$ for $k+1\leq j\leq M-1$. One generates all $M-3$ possibilities by considering, in turn, $k=2,3,\dots, M-2$. Having chosen some initial sequence of $k-1$ zeros, it is straightforward to show that the remaining nonzero terms are given simply by $\lambda_{j,c}=\lambda_{M}$.
Thus only for $\lambda_{M}\neq 0$ do we find $M-3$ new, inflationary solutions, which we refer to as
\begin{widetext}
\begin{equation}\label{EQN:InfAiii}
{\bf IA}\textrm{(iii)}_{k}: \>\> (x_{c}, y_{c}, \lambda_{0,c}, \lambda_{1,c}, \lambda_{2,c},\dots,\lambda_{k,c},\lambda_{k+1,c},\dots,\lambda_{M-1,c}) = (0,1,0,0,0,\dots, 0, \lambda_{M}, \dots, \lambda_{M}),
\end{equation}
\end{widetext}
for $k = 2,3,\dots, M-2$. Note, again, that for $M=3$, ${\bf IA}\textrm{(iii)}_{k}$ provides no new fixed points.


Similar arguments may be used to derive extensions for the non-inflationary cases, $\beta^\prime$ and $\gamma^\prime$, of Table \ref{TAB:initseg0}. Next we consider the second general case, where $\lambda_{1,c} \neq 0$.

\subsection{Case {\bf B}: $\lambda_{1,c}\neq 0$}\label{SEC:CaseB}

In this case, we have 4 distinct initial segments, as displayed in Table~\ref{TAB:initsegFO}, which we reproduce and relabel for clarity in Table~\ref{TAB:initsegFOnonzero}.
\begin{table}[H]
\begin{center}
\begin{tabular}{c|c||c|c|c|c}
 Solution label & Inflationary? & $x_{c}$ & $y_{c}$ & $\lambda_{0,c}$ & $\lambda_{1,c}$ \\ \Xhline{2pt}
$\alpha^{\prime\prime}$ & Yes & 0 & 1& 0 &$\neq 0$ \\
$\beta^{\prime\prime}$ &No & 1 & 0 & 0&$\neq 0$ \\
$\gamma^{\prime\prime}$ & No& 1 & 0& $-3+\lambda_{1}$&$\neq 0$ \\
$\delta^{\prime\prime}$ &Yes $(\lambda_{1}<3)$ &$\frac{\lambda_{1}}{9}$&$1-\frac{\lambda_{1}}{9}$& $\frac{2 \lambda_{1}}{3}$&$\neq 0$
\end{tabular}
\caption{Initial segments from the first block of equations, Eq.~(\ref{EQN:blockONE}), with $\lambda_{1,c}\neq 0$. }\label{TAB:initsegFOnonzero}
\end{center}
\end{table}
As for Case A, we may find fixed points for the second block of equations, Eq.~(\ref{EQN:blockTWO}), subject to the initial segments in Table \ref{TAB:initsegFOnonzero}, by invoking three general cases, depending on which $\lambda_{m,c}$ vanish. We again work through these cases in turn. For $\lambda_{1,c} \neq 0$, the two blocks of equations in Eqs.~(\ref{EQN:blockONE}) and (\ref{EQN:blockTWO}) are not as independent as for the case $\lambda_{1,c} = 0$, which introduces only modest additional complications. \\

\noindent\underline{Case {\bf B}(i)}: ($\lambda_{m,c}=0$ $\textrm{for all}\;2\leq m\leq M-1$)
\\
\\
We again focus on inflationary fixed points, which can only correspond to extensions of $\alpha^{\prime \prime}$ and $\delta^{\prime\prime}$. For $\alpha^{\prime\prime}$, there does not exist any extension, because if 
$\lambda_{2,c}=0$, then Eq.~(\ref{EQN:lonetimeMapp}) yields
\begin{equation}\label{EQN:Intermed}
\left[-\lambda_{2,c}+\lambda_{1,c}+\epsilon (x_{c})\right]\lambda_{1,c}=0 .
\end{equation}
But since $\epsilon (x_c) = 0$ for $\alpha^{\prime\prime}$, we find $\lambda_{1,c} = 0$, which contradicts the defining assumption of Case B. We find a similar result for $\delta^{\prime\prime}$. In that case, $\epsilon (x_c) = \lambda_1 / 3$, which, together with Eq.~(\ref{EQN:Intermed}), yields $\lambda_{1,c} = - \lambda_{1,c} / 3$, whose only solution is $\lambda_{1,c} = 0$, again yielding a contradiction for Case B. Hence we find no new inflationary solutions in this case. (Similar manipulations indicate that there exist two new non-inflationary solutions for $\beta^{\prime\prime}$ and $\gamma^{\prime\prime}$, with no constraints on $\lambda_M$.)
 \\

\noindent\underline{Case {\bf B}(ii)}: ($\lambda_{m,c}\neq0$ $\textrm{for all}\;2\leq m\leq M-1$)
\\
\\
The solutions in this case mirror those of {Case {\bf A}(ii)} except that now, $\lambda_{1,c}$ is also nonzero. Hence the appropriate generalization of Eq.~(\ref{EQN:generallambdaAii}) is
\begin{equation}\label{EQN:generallambdaBii}
\lambda_{m,c} = \lambda_{M} - (M-m)\epsilon(x_{c}),
\end{equation}
for $m= 1,2,\dots, M-1$. Aside from two new non-inflationary solutions (extending $\beta^{\prime\prime}$ and $\gamma^{\prime\prime}$, with certain restrictions on the value of $\lambda_{M}$ that we will not enumerate here), we now have two new inflationary solutions. 

We first consider {\color{blue} the} extension of $\alpha^{\prime\prime}$, where, noting $\epsilon (x_{c})=0$, we find (for $\lambda_{M}\neq 0$),
\begin{widetext}
\begin{equation}\label{EQN:InfBiialpha}
{\bf IB}\textrm{(ii)}\alpha^{\prime\prime}: (x_{c}, y_{c}, \lambda_{0,c}, \lambda_{1,c}, \lambda_{2,c},\dots, \lambda_{M-1,c}) = (0,1,0,\lambda_{M},\lambda_{M},\dots, \lambda_{M}).
\end{equation}
\end{widetext}
The second inflationary fixed point corresponds to the extension of $\delta^{\prime\prime}$. It can be found by noting that for $m=1$, Eq.~(\ref{EQN:generallambdaBii}), in combination with the fact that $\epsilon(x_{c})={\lambda_{1}}/{3}$, yields
\begin{equation}
\lambda_{1,c}= \frac{3\lambda_{M}}{M+2}.
\end{equation}
Thus we can compute the extension to $\delta^{\prime\prime}$, which yields (for $\lambda_{M}\neq 0$)
\begin{widetext}
\begin{align}\label{EQN:InfBiidelta}
{\bf IB}\textrm{(ii)}\delta^{\prime\prime}: \>\> (x_{c}, y_{c}, \lambda_{0,c}, \lambda_{1,c}, \{\lambda_{j,c}\}_{j=2}^{M-1}) = \left(\frac{\lambda_{M}}{3(M+2)}, 1-\frac{\lambda_{M}}{3(M+2)},\frac{2\lambda_{M}}{M+2}, \frac{3\lambda_{M}}{M+2}, \left\{\left(\frac{j+2}{M+2}\right)\lambda_{M}\right\}_{j=2}^{M-1} \right).
\end{align}
\end{widetext}
Note that in this case, the original condition for inflation, namely, $\lambda_{1}<3$, translates to a condition on $\lambda_{M}$: $\lambda_{M}<M+2$. In addition, on both sides of Eq.~(\ref{EQN:InfBiidelta}), one can set the lower limit of the term in braces to $j=0$, thereby consistently subsuming the two terms preceding the term in braces.\\

\noindent\underline{Case {\bf B}(iii)}: ($\lambda_{m,c}=0$ $\textrm{for just some}\;2\leq m\leq M-1$)
\\
\\
For this case, there are no new inflationary fixed points. Consider $\alpha^{\prime\prime}$ first. If there were to be a consistent extension of $\alpha^{\prime\prime}$, there would need to exist some first $\lambda_{j,c} = 0$ for some $j=2,3,\dots, M-1$, before which all $\lambda_{j',c}\neq0$, for $j' < j$. Then, from Eq.~(\ref{EQN:blockTWO}), the righthand side of the relevant dynamical equation for $\lambda_{j - 1}$ would take the form $[ - \lambda_j + \lambda_{j - 1} + \epsilon (x)] \lambda_{j - 1}$. Setting this expression to zero and solving for $\lambda_{j-1,c}$ would yield $\lambda_{j - 1,c} = - \epsilon (x_c)$, but for $\alpha^{\prime\prime}$, we have $\epsilon (x_c) = 0$, thus contradicting the assumption that $\lambda_{j,c}$ (rather than $\lambda_{j-1,c}$) is the first such zero critical value. A similar argument indicates that there does not exist a consistent extension of $\delta^{\prime\prime}$ for this case, either, though consistent extensions of $\beta^{\prime\prime}$ and $\gamma^{\prime\prime}$ may be found.

To summarize: at any order $M \geq 3$, there are at most $M + 1$ inflationary fixed points. These correspond to Eqs.~(\ref{EQN:InfAi}),~(\ref{EQN:InfAii}),~(\ref{EQN:InfAiii}),~(\ref{EQN:InfBiialpha}), and~(\ref{EQN:InfBiidelta}), which we reproduce in Table~\ref{TAB:MthordersummaryINF}. Note that in the third row of results, we display a representative fixed point for ${{\bf IA}\textrm{(iii)}}_{k}$ from Eq.~(\ref{EQN:InfAiii}).  However, as we next demonstrate, not all of these inflationary fixed points are hyperbolic.

\begin{table*}[htpb!]
\begin{center}
\begin{tabular}{l||c||c|c|c|c|c|c|c|c|c|c}
 Case& $\epsilon(x_{c})$ &$x_{c}$ & $y_{c}$ & $\lambda_{0,c}$ &$\lambda_{1,c}$&$\lambda_{2,c}$&\dots&$\lambda_{k,c}$&$\lambda_{k+1,c}$&\dots&$\lambda_{M-1,c}$\\ \Xhline{2pt}
 {\bf IA}(i)&0&0&1&0&0&0&\dots&0&0&\dots&0\\ \hline
 {\bf IA}(ii)&0&0&1&0&0&$\lambda_{M}$&\dots&$\lambda_{M}$&$\lambda_{M}$&\dots&$\lambda_{M}$\\ \hline
 ${{\bf IA}\textrm{(iii)}}_{k}$&0&0&1&0&0&0&\dots&0&$\lambda_{M}$&\dots&$\lambda_{M}$\\ \hline
 {\bf IB}(ii)$\alpha^{\prime\prime}$&0&0&1&0&$\lambda_{M}$&$\lambda_{M}$&\dots&$\lambda_{M}$&$\lambda_{M}$&\dots&$\lambda_{M}$\\ \hline
 {\bf IB}(ii)$\delta^{\prime\prime}$&$\frac{\lambda_{M}}{M+2}$&$\frac{\lambda_{M}}{3(M+2)}$&$1-\frac{\lambda_{M}}{3(M+2)}$&$\frac{2\lambda_{M}}{M+2}$&$\frac{3\lambda_{M}}{M+2}$&$\frac{4\lambda_{M}}{M+2}$&\dots&$\frac{(k+2)\lambda_{M}}{M+2}$&$\frac{(k+3)\lambda_{M}}{M+2}$&\dots&$\frac{(M+1)\lambda_{M}}{M+2}$\\ 
 \end{tabular}
\caption{Maximal set of inflationary fixed points (realized for $\lambda_{M}\neq 0$). At any order $M\geq 3$, there are at most $M+1$ such fixed points. The subscript $k$ in ${{\bf IA}\textrm{(iii)}}_{k}$ runs over $k=2,3,\dots,M-2$. Note that ${{\bf IA}\textrm{(iii)}}_{k}$ only provides new fixed points for $M>3$.}\label{TAB:MthordersummaryINF}
\end{center}
\end{table*}

\subsection{Stability analysis of $M$th-order inflationary fixed points}\label{SEC:PrelimStabM}

To investigate the stability properties of various fixed points, we consider the eigenvalues of the Jacobian. For the two blocks of equations
listed in Eqs.~(\ref{EQN:blockONE}) and~(\ref{EQN:blockTWO}), the Jacobian takes a somewhat simple form. One can use this fact to determine the stability properties of any (hyperbolic) fixed point of interest. As above, we will focus solely on the fixed points that can be inflationary. We find that only fixed points {\bf IB}(ii)$\alpha^{\prime\prime}$ and {\bf IB}(ii)$\delta^{\prime\prime}$ are hyperbolic: they comprise a saddle point and, as a numerical analysis reveals, a saddle focus-node, respectively.

The Jacobian is an $(M+2) \times (M+ 2)$ matrix, given by
\begin{widetext}
\begin{equation}
\resizebox{0.9\columnwidth}{!}{%
$
\left(
\begin{array}{cccccccc}
 6 x-3 y-3 & \lambda_{0} -3 x & y& 0&0&0&\dots& 0 \\
 3 y & 3 x-6 y-\lambda_{0} +3 & -y &0&0&0&\dots&0 \\
 \frac{3 \lambda_{0}}{2} & -\frac{3\lambda_{0}}{2} &-\lambda_{1} +2 \lambda_{0}+\epsilon &-\lambda_{0}&0&0&\dots&0\\
\frac{3 \lambda_{1}}{2} & -\frac{3\lambda_{1}}{2} &0&-\lambda_{2} +2 \lambda_{1}+\epsilon &-\lambda_{1}&0&\dots&0\\
\frac{3 \lambda_{2}}{2} & -\frac{3\lambda_{2}}{2} &0&0&-\lambda_{3} +2 \lambda_{2}+\epsilon &-\lambda_{2}&\dots&0\\
  \vdots&  \vdots&\vdots & \vdots&\vdots&\vdots&\ddots& \vdots \\
\frac{3 \lambda_{M-1}}{2} & -\frac{3\lambda_{M-1}}{2} &0&0&0&0&\dots&-\lambda_{M} +2 \lambda_{M-1}+\epsilon\label{EQN:JACOBIAN}\\ \\
\end{array} 
\right) ,
$}
\end{equation}
\end{widetext}
where, as usual, $\epsilon \equiv {3 \over 2} (1 + x - y)$. For the inflationary fixed points in Table~\ref{TAB:MthordersummaryINF}, we find the following stability properties:
\begin{itemize}
\item {\bf IA}(i): Substituting the first row of results in Table~\ref{TAB:MthordersummaryINF}  into the Jacobian of Eq.~(\ref{EQN:JACOBIAN}), one finds that the Jacobian is block diagonal, with the first block (a $3\times 3$ matrix in the upper lefthand corner) given by 
\begin{equation}\label{EQN:zeroEgm}
\left(
\begin{array}{ccc}
-6 & 0 & 1 \\
 3  & -3 & -1 \\
 0 & 0 &0  \\
\end{array}
\right).
\end{equation}
The matrix in Eq.~(\ref{EQN:zeroEgm}) may itself be further decomposed as a lower block-diagonal ($2 \times 2)$ matrix, consisting of the first two rows and columns of Eq.~(\ref{EQN:zeroEgm}), nested within an upper block-diagonal matrix. 
The eigenvalues are just the diagonal entries. Thus there exists a zero eigenvalue for the (full) Jacobian in this case, and hence {\bf IA}(i) is not hyperbolic.
\item {\bf IA}(ii), {\bf IA}$\textrm{(iii)}_{k}$: Analogous arguments to the previous one hold in all of these cases. Although the full Jacobian is not block diagonal, it is
lower block-diagonal in all cases. 
In particular, one finds that the same matrix in Eq.~(\ref{EQN:zeroEgm}) occupies the upper left block in all cases. Thus, as for fixed point {\bf IA}(i), there exists a zero eigenvalue in all of these cases, rendering these fixed points not hyperbolic. 
\item {\bf IB}(ii)$\alpha^{\prime\prime}$: Direct substitution of the fourth row of results in Table~\ref{TAB:MthordersummaryINF} into Eq.~(\ref{EQN:JACOBIAN}) shows that the Jacobian is a lower block-diagonal matrix, with an upper block-diagonal $3\times 3$ matrix in the upper lefthand corner, and an $(M-1)\times (M-1)$ upper triangular block in the lower righthand corner. The eigenvalues of the full Jacobian thus lie along the diagonal. One can easily show that they are ($-6$,$-3$,$-\lambda_{M}$, $\lambda_{M}$, $\lambda_{M}$, \dots, $\lambda_{M}$). Thus, this fixed point is hyperbolic and is a saddle point. 
\item {\bf IB}(ii)$\delta^{\prime\prime}$: The stability analysis is not as straightforward in this case because there is ostensibly no way to simplify the Jacobian as in the previous cases. We have thus performed a numerical analysis (for $3\leq M\leq 10$) that reveals, for $0<\lambda_{M}< M+2$ --- for which this fixed point is inflationary with $\epsilon > 0$
--- that this fixed point is a saddle focus-node. 
\end{itemize}

To summarize: at all orders $M$, there exist cosmologically viable fixed points. 
In particular, at any order $M \geq 0$, there exist at most two hyperbolic, inflationary fixed points. Moreover, at each order $M \geq 1$, one of the hyperbolic fixed points corresponds to exact de Sitter evolution, whereas the other corresponds to quasi-de Sitter evolution (for appropriate values of $\lambda_M$). Furthermore, it appears that the latter fixed point is not an attractor.

\subsection{SSF realization of hyperbolic inflationary fixed points}\label{SEC:SSFMthInfl}

We conclude our analysis by focussing on SSF realizations of hyperbolic inflationary fixed points for the $M$th-order system for $M\geq 3$. This analysis will also allow us to tie together some results generated for $M=0,1,$ and $2$. 

As Appendix~\ref{SEC:PrelimStabM} revealed, there are at most two hyperbolic inflationary fixed points at any order $M \geq 3$, namely, the points {\bf IB}(ii)$\alpha^{\prime\prime}$ and {\bf IB}(ii)$\delta^{\prime\prime}$ in Table~\ref{TAB:MthordersummaryINF}. (In fact, as we have seen above, this statement holds for any $M\geq 0$.) The point {\bf IB}(ii)$\alpha^{\prime\prime}$ has $x$-coordinate $x_{c}=0$, so we immediately see --- in direct analogy to the cases of fixed points {\bf FP1a} at first order (see Sec.~\ref{SEC:FirstOrderSystems}) and {\bf FP2a} at second order (see Appendix~\ref{APP:secondorder}) --- that the SSF realization of a trajectory that begins at {\bf IB}(ii)$\alpha^{\prime\prime}$ will remain there and undergo de Sitter evolution forever.

We will now show (by construction) that a trajectory that begins at fixed point {\bf IB}(ii)$\delta^{\prime\prime}$ corresponds to the particular solution to power-law inflation that we have already discussed for {\bf FP0b} at zeroth order (Sec.~\ref{SEC:ZerothOrderSystems}), {\bf FP1d} at first order (Sec.~\ref{SEC:FirstOrderSystems}), and {\bf FP2f} at second order (Appendix~\ref{APP:secondorder}). We will refer to this particular solution of power-law inflation as `PLI' --- as governed by Eqs.~(\ref{EQN:VPLI})--(\ref{EQN:aPLI}).

Our construction proceeds in two steps. First we demonstrate that PLI will indeed behave like an SSF realization of the EFT dynamical system at $M$th order. Then we demonstrate that PLI corresponds to the fixed point {\bf IB}(ii)$\delta^{\prime\prime}$.

We first demonstrate that PLI corresponds to an SSF realization of the EFT dynamical system at \emph{any} order. That is, if we fix $\lambda_{M}$ for some $M\geq 0$ (and make a judicious choice of $p$), PLI can be made to satisfy the defining relationship for that order, namely
\begin{equation}\label{EQN:defrel}
V^{(M)}(t)\propto \left[ a(t) \right]^{-\lambda_{M}},
\end{equation}
where, as in Eq.~(\ref{EQN:VM}), $V^{(M)}(t)$ is the $M$th derivative of the scalar potential with respect to cosmic time. In particular, it is straightforward to show (via direct integration) that for PLI,
\begin{equation}
V_{\textrm{\tiny{PLI}}}^{(M)}(t)\propto \left[ a(t)\right]^{-\left(M+2 \right) /p}.
\end{equation}
Upon setting
\begin{equation}\label{EQN:MPPLI}
p\equiv\frac{M+2}{\lambda_{M}},
\end{equation}
we find that PLI satisfies Eq.~(\ref{EQN:defrel}), and hence one should be able to locate it somewhere as an SSF realization of the EFT dynamical system at $M$th order.

Next we demonstrate that PLI corresponds to evolution at a fixed point. Notice that if we compute $\lambda_{m}$ for PLI from its definition in Eq.~(\ref{EQN:lSF}), we find 
\begin{equation}\label{EQN:genPPLI}
\lambda_{m}=\frac{m+2}{p}.
\end{equation}
So if we assume we are analyzing the EFT dynamical system at $M$th order and have fixed $p$ according to Eq.~(\ref{EQN:MPPLI}), then we can (easily) sequentially construct lower-order $\lambda_{m}$'s, as well as $x$ and $y$, for PLI. Upon using Eq.~(\ref{EQN:genPPLI}) and proceeding down the tower of $\lambda_{m}$'s, we find
%
\begin{align}
\lambda_{M-1}=\frac{(M-1)+2}{p}&=\frac{M+1}{M+2}\lambda_{M},\label{EQN:Seqone}\\
\lambda_{M-2}&=\frac{M}{M+2}\lambda_{M},\\
&\vdots\nonumber\\
\lambda_{1}&=\frac{3}{M+2}\lambda_{M},\\
\lambda_{0}&=\frac{2}{M+2}\lambda_{M}.
\end{align}
In addition, using Eqs.~(\ref{EQN:phiPLI}) and (\ref{EQN:aPLI}) for PLI, one can show that 
\begin{equation}
    (x,y)= \left(\frac{1}{3p}, 1-\frac{1}{3p} \right). 
\end{equation}
Thus
\begin{align}
x&=\frac{1}{3p}=\frac{\lambda_{M}}{3(M+2)},\\
y&=1-\frac{1}{3p}=1-\frac{\lambda_{M}}{3(M+2)}.\label{EQN:Seqlast}
\end{align}
That is, all the lower order dynamical variables for PLI are constants (determined by $M$ and $\lambda_{M}$). So PLI corresponds to a fixed point at order $M$. In fact, if we compare the sequence of equations in Eqs.~(\ref{EQN:Seqone})--(\ref{EQN:Seqlast}) to the table summarizing inflationary fixed points at order $M$, Table~\ref{TAB:MthordersummaryINF}, we see that the fixed point corresponding to PLI is precisely {\bf IB}(ii)$\delta^{\prime\prime}$ in that table. (Although the entries in Table~\ref{TAB:MthordersummaryINF} were constructed for $M \geq 3$, the expression for this particular fixed point is valid for $M = 1,2$ as well --- and likewise covers the case $M = 0$ with appropriate modifications --- and hence the argument here includes the case of $M = 2$, as alluded to in Appendix~\ref{APP:secondorder}.)

Thus, we have shown that all hyperbolic inflationary fixed points, at any order $M\geq 0$, have SSF realizations that come in one of two varieties: pure de Sitter evolution, or quasi-de Sitter evolution consistent with power-law inflation.

\section*{Acknowledgements} It is a pleasure to thank John Barrow, Jeremy Butterfield, Larry Ford, Alan Guth, Mark Hertzberg, Evan McDonough, Mohammad Hossein Namjoo, Christopher Smeenk, Vincent Vennin, and Alexander Vilenkin for helpful comments and suggestions. This work was conducted in part in the Center for Theoretical Physics at MIT, and has been supported in part by the U.S. Department of Energy under grant Contract Number DE-SC0012567. FA acknowledges support from the Wittgenstein Studentship in Philosophy at Trinity College, University of Cambridge; and the Black Hole Initiative at Harvard University, which is funded through a grant from the John Templeton Foundation.


%

\end{document}